\documentclass[]{emulateapj_table}

\usepackage{graphicx}
\usepackage{subfigure}

\usepackage[colorlinks=true,
		linkcolor=red,
		linkbordercolor={1 0 0},
		anchorcolor=magenta,
		citecolor=blue,
		citebordercolor={0 1 0},
		filecolor=magenta,
		filebordercolor={0 .5 .5},
		menucolor=magenta,
		menubordercolor={1 0 0},
		runcolor=magenta,
		urlcolor=magenta,
		urlbordercolor={0 1 1},
		frenchlinks=false]{hyperref}

\usepackage{multirow}
\usepackage{nicefrac}

\usepackage{longtable}
\usepackage{lscape}

\slugcomment{Accepted for publication in ApJ, December 17, 2019}
\shorttitle{multiple kinematic components}\shortauthors{Catal\'an-Torrecilla et al.}

\begin{document}

\title{Spatially-resolved analysis of neutral winds, stars and ionized gas kinematics with MEGARA/GTC: new insights on the nearby galaxy UGC 10205}





\author{C. Catal\'an-Torrecilla\altaffilmark{1,2}}
\author{\'A. Castillo-Morales\altaffilmark{1}}
\author{A. Gil de Paz\altaffilmark{1}}
\author{J. Gallego\altaffilmark{1}}
\author{E. Carrasco\altaffilmark{3}}
\author{J. Iglesias-P\'aramo\altaffilmark{4,5}}
\author{R. Cedazo\altaffilmark{6}}
\author{M. Chamorro-Cazorla\altaffilmark{1}}
\author{S. Pascual\altaffilmark{1}}
\author{M.L. Garc\'ia-Vargas\altaffilmark{7}}
\author{N. Cardiel\altaffilmark{1}}
\author{P. G\'omez-Alvarez\altaffilmark{7}}
\author{A. P\'erez-Calpena\altaffilmark{7}}
\author{I. Mart\'inez-Delgado\altaffilmark{7}}
\author{B. T. Dullo\altaffilmark{1}}
\author{P. Coelho\altaffilmark{8}}
\author{G. Bruzual\altaffilmark{9}}
\author{S. Charlot\altaffilmark{10}}


\affil{\altaffilmark{1}Departamento de F\'isica de la Tierra y Astrof\'isica, Instituto de F\'isica de Part\'iculas y del Cosmos IPARCOS, Universidad Complutense de Madrid, E-28040 Madrid, Spain \email{ccatalan@ucm.es}}
\affil{\altaffilmark{2}Centro de Astrobiolog\'ia (CAB, CSIC-INTA), Carretera de Ajalvir km 4, E-28850 Torrej\'on de Ardoz, Madrid, Spain}
\affil{\altaffilmark{3}Instituto Nacional de Astrof\'isica, \'Optica y Electr\'onica, Luis Enrique Erro No. 1, C.P. 72840, Tonantzintla, Puebla, Mexico}
\affil{\altaffilmark{4}Instituto de Astrof\'isica de Andaluc\'ia-CSIC, Glorieta de la Astronom\'ia s/n, E-18008, Granada, Spain}
\affil{\altaffilmark{5}Estaci\'on Experimental de Zonas \'Aridas - CSIC, Ctra. de Sacramento s/n, 04120 Almer\'ia, Spain} 
\affil{\altaffilmark{6}Universidad Polit\'ecnica de Madrid, Madrid, Spain}
\affil{\altaffilmark{7}FRACTAL S.L.N.E. C/Tulip\'an 2, p13, 1A. E-28231, Las Rozas de Madrid, Spain}
\affil{\altaffilmark{8}Universidade de S\~ao Paulo, Instituto de Astronomia, Geof\'isica e Ci\^encias Atmosf\'ericas, Rua do Mat\~ao 1226, 05508-090, S\~ao Paulo, Brazil}
\affil{\altaffilmark{9}Instituto de Radioastronom{\'i}a y Astrof{\'i}sica, UNAM, Campus Morelia, Michoacan, M{\'e}xico, C.P. 58089, M{\'e}xico}
\affil{\altaffilmark{10}Sorbonne Universit\'e, CNRS, UMR7095, Institut d'Astrophysique de Paris, F-75014, Paris, France}



\begin{abstract} 

We present a comprehensive analysis of the multi-phase structure of the interstellar medium (ISM) and the stellar kinematics in the edge-on nearby galaxy UGC 10205 using integral field spectroscopy (IFS) data taken with MEGARA at the GTC. We explore both the neutral and the ionized gas phases using the interstellar \mbox{Na {\small I} D} doublet absorption (LR$-$V set-up, R $\sim$ 6000) and the H$\alpha$ emission line (HR$-$R set-up, R $\sim$ 18000), respectively. The high-resolution data show the complexity of the H$\alpha$ emission line profile revealing the detection of up to three kinematically distinct gaseous components. Despite of this fact, a thin disk model is able to reproduce the bulk of the ionized gas motions in the central regions of UGC 10205. The use of asymmetric drift corrections is needed to reconciliate the ionized and the stellar velocity rotation curves. We also report the detection of outflowing neutral gas material blueshifted by $\sim$ 87 km\,s$^{-1}$. The main physical properties that describe the observed outflow are a total mass M$_{out}$ $=$ (4.55 $\pm$ 0.06) $\times$ 10$^{7}$ M$_{\sun}$ and a cold gas mass outflow rate $\dot{M}$$_{out}$ $=$ 0.78 $\pm$ 0.03 M$_{\sun}$ yr$^{-1}$. This work points out the necessity of exploiting high-resolution IFS data to understand the multi-phase components of the ISM and the multiple kinematical components in the central regions of nearby galaxies.

\end{abstract}


\section{Introduction}

Being a fundamental physical process for the galaxy evolution models, a vast observational evidence have probed the ubiquity of galaxy-scale outflows across all cosmic epochs \citep[see reviews by][]{Veilleux_2005, Rupke_2018}. Stellar winds, supernovae or active galactic nuclei (AGNs) have been postulated as the potential mechanisms to drive these phenomena to which we will broadly refer as Galactic Winds (GWs).

GWs are candidates to contribute to the metal enrichment of the intergalactic medium (IGM) \citep{Nath_1997,Dalcanton_2007}. The gas entrained in the wind may escape the halo potential \citep{Heckman_2000} or it could fall back down on to the galaxy in a process referred to as {\it galactic fountain} \citep{Shapiro_1976}. GWs are essential for understanding the feedback mechanisms and their impact on the host galaxy. While they might produce a positive effect in which the compression of the insterstellar and circumgalactic medium can potentially boost the star formation \citep{Nayakshin_2012,Zubovas_2013}, they can also be considered as a plausible mechanism to suppress star formation within galaxies \citep{Granato_2004, King_2015}. 



Given their intrinsic multiphase nature, different tracers (X-rays, optical nebular emission lines, ISM absorption lines, CO transitions) ought to be used to characterize their different phases (hot, warm, cold, molecular). While most of the studies have traditionally focused on the study of GWs on extreme objects such as luminous and ultraluminous infrared galaxies (LIRGs and ULIRGs), mergers and strong starburst \citep{Martin_2005,Rupke_2005,Rupke_2005_2, Arribas_2014,Cazzoli_2014,Cazzoli_2016}, the presence of relatively strong winds in less active galaxies have remain mostly unexplored \citep{Castillo_Morales_2007, Jimenez_Vicente_2007, Chen_2010, Roberts_Borsani_2019}. 
Within this scenario, the use of new instrumental capabilities to explore the existence, frequency and strength of winds in relatively quiet galaxies seems the necessary step to follow. 






The advent of the Multi-Espectr\'ografo en GTC de Alta Resoluci\'on para Astronom\'ia (MEGARA, Gil de Paz et al$.$ in prep$.$) will improve the detectability of the GWs helping to separate them from their hosts with exquisite detail. These high-quality Integral Field Spectroscopy (IFS) data are fundamental to kinematically discriminate both the ionized medium and the cold interstellar gas phase of winds and their geometry. That is why, the characterization of GWs in the central regions of nearby galaxies is one of the main scientific objectives of the MEGARA instrument \citep{Carrasco_2018SPIE}.

In this context, UGC 10205 was selected as part of the night-time commissioning observations due to the presence of residual \mbox{Na {\small I} D} interestelar absorption previously identified in the CALIFA survey \citep{Sanchez_2012}.  In the optical range, neutral gas motions are commonly studied using this interstellar medium (ISM) \mbox{Na {\small I} D} absorption line doublet at 5889.95\,\AA\, and 5895.92\,\AA. The lower value of the ionization potential for the \mbox{Na {\small I} D} (5.14 eV) in comparison with the one for the hydrogen, makes it an ideal candidate to trace the cool (T $<$ 10$^{4}$\,K) neutral gas.

Besides, by examining the multiple components of the ISM in galaxies such as the case of UGC 10205, where mergers might have played an important role \citep{Reshetnikov_1999}, one could provide important clues on the impact of these events on their evolution. Past long-slit spectroscopy studies limited to a few directions only \citep[see][]{Vega_1997} have revealed the complexity of this object. The use of high-resolution IFS data is essential to provide a comprehensive view on the dynamics of this galaxy. In this work, we perform a reconstruction of the stellar kinematics and gas morphology in the central regions of this nearby galaxy. Thus, UGC 10205 is intended to provide a exploratory study to define the final selection for the optimal exploitation of the MEGARA capabilities both in terms of the stellar and gas kinematics.

  
The structure of this paper is as follows: Section~\ref{ugc10205_properties} describes the general properties of the UGC 10205 galaxy, Section~\ref{observations} presents MEGARA observations and data reduction, Section~\ref{analysis} outlines the various steps in our analysis, Section~\ref{results} describes the main results of this study including details on the kinematics of the ionized gas and the stellar component and the properties of the cold gas. Finally, we discuss the proposed scenario for this object in Section~\ref{scenario} and we summarize our conclusions in Section~\ref{final_conclusions}. Unless otherwise stated, throughout this paper we adopt a a flat $\Lambda$CDM cosmology with H$_{0}$ $=$ 70 km\,s$^{-1}$\,Mpc$^{-1}$, $\Omega$$_{m}$ $=$ 0.3 and $\Omega$$_{\Lambda}$ $=$ 0.7.

\section{General Properties of UGC 10205} \label{ugc10205_properties}

\begin{figure}
\centering
\includegraphics[trim={0.9cm 0.2cm 0.2cm 0.2cm},width=75mm]{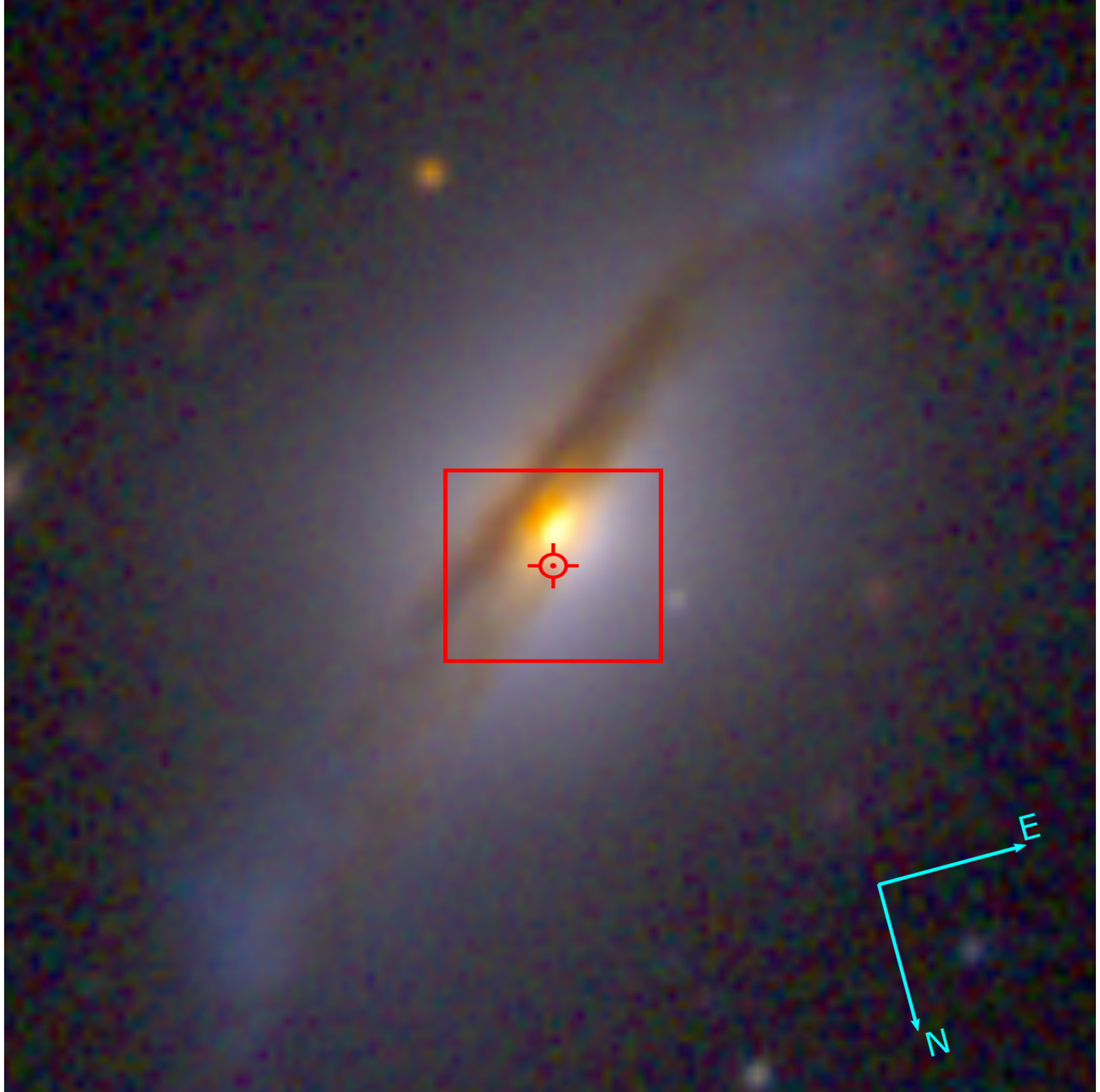}
\includegraphics[trim={1.5cm 0cm 0cm 0cm},width=100mm]{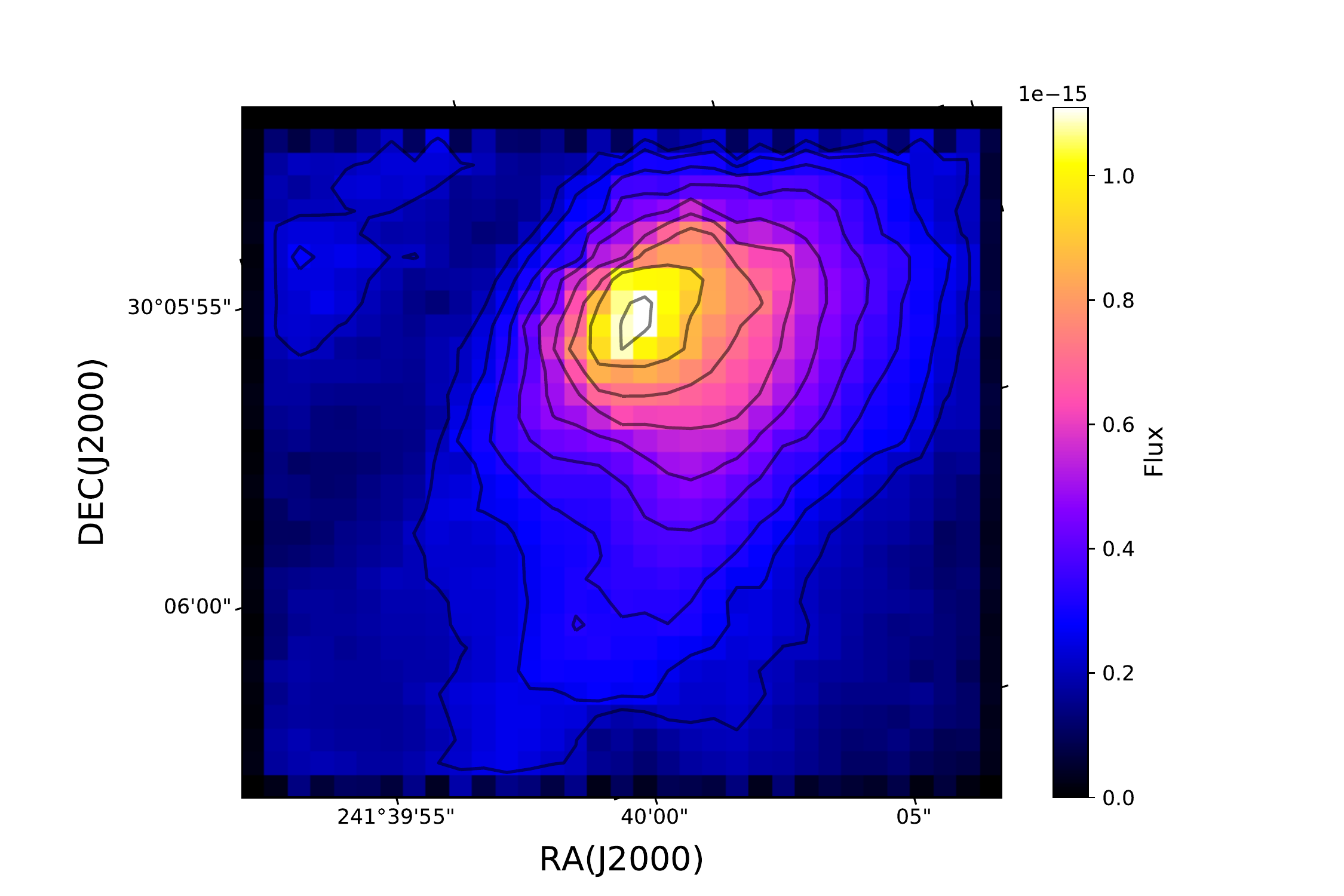}\\
\caption{Top panel: SDSS composite image of UGC 10205. Overlaid on the image is the footprint of the MEGARA IFU field-of-view (red rectangle) covering a region of 5.1 $\times$ 5.7 kpc$^{2}$. The orientation is indicated in the bottom-right corner. Bottom panel: Distribution of the continuum emission of UGC 10205 derived from the MEGARA IFS datacube in the spectral range 5600-5700\,\AA. The flux units are erg s$^{-1}$cm$^{-2}$pixel$^{-2}$\AA$^{-1}$. Pixel size is 0.4'' in this case. The area of these synthetic spaxels was chosen as the best compromise between the seeing of the observing nights and the routine used to create the datacubes (see more details in Section~\ref{recover_halpha_emission}).\\}  
\label{ugc10205_megara_fov}
\end{figure}

UGC 10205 is an edge-on disk galaxy classified as an Sa/S0 galaxy by \citet{Nilson_1973,Tarenghi_1994} and \citet{Nair_2010}, respectively, with an inclination of {\it i} $=$ 84$^{\circ}$ \citep{Rubin_1985}. A SDSS composite image for this galaxy appears in Figure~\ref{ugc10205_megara_fov} showing the presence of dust lanes roughly parallel to the major axis of the galaxy. A shell  structure  in  the  outer  regions of this galaxy was also discovered by the early studies of \citet{Rubin_1987}.

The main properties of UGC 10205 are summarized in Table~\ref{tab:table_1} and briefly described below. A systemic heliocentric radial velocity of 6556 km$^{-1}$ was calculated in \citet{Theureau_1998} using measurements of the 21 cm neutral hydrogen line carried out with the meridian transit Nan\c{c}ay radiotelescope. \citet{Walcher_2014} used \citet{Bruzual_Charlot_2003} stellar population models with a \citet{Chabrier_2003} stellar IMF to construct UV to NIR SED to estimate a total stellar mass of log(M$_{\star}$/M$_{\sun}$) $=$ 10.997 $\pm$ 0.111. A set of panchromatic luminosity measurements can be also found in the literature for this galaxy. In particular, a far-infrared luminosity of log({\it L$_{FIR}$}/L$_{\sun}$) $=$ 10.10 was reported by \citet{Vlahakis_2005} using the Submillimetre Common-User Bolometer Array (SCUBA) data and a log({\it L$_{TIR}$/L$_{\sun}$}) $=$ 10.35 by \citet{Willmer_2009} combining Spitzer Space Telescope observations with SCUBA 850 $\mu$m data. UGC 10205 was observed as part of the CALIFA survey \citep{Sanchez_2012}. In this context, \citet{Catalan_Torrecilla_2015} derived a global {\it L$_{H\alpha,obs}$} $=$ (10.35 $\pm$ 0.57) $\times$ 10$^{40}$ erg s$^{-1}$ and a {\it L$_{FUV,obs}$} $=$ (2.71 $\pm$ 0.67) $\times$ 10$^{42}$ erg s$^{-1}$ from the Galaxy Evolution Explorer \citep[GALEX,][]{Martin_2005}.

Questions arise about the complexity of this object as has been already manifested by several authors. The numerical simulations run by \citet{Reshetnikov_1999} show the capture and tidal disruption of a companion (a small type E/S0). Other authors have considered this galaxy as a possible polar ring candidate \citep{Whitmore_1990,van_Driel_2000}. A profound study done by \citet{Vega_1997} using long-slit spectroscopy data revealed the presence of up to three kinematically distinct gaseous components. The use of IFS data will help to discriminate the spatial distribution of these components and to overcome limitations associated with the use of one-dimensional velocity curves. At a distance of 150 Mpc \citep[H$_{0}$ = 75.2 $\pm$ 3.3 km s$^{-1}$ Mpc$^{-1}$,][]{Sorce_2014}, the central 5.7 $\times$ 5.1 kpc$^{2}$ of UGC 10205 can be mapped with MEGARA. These observations have the potential to shed some light on the multi-phase components of the ISM to understand the current state and future evolution of this galaxy.


    \begin{table}
    \centering
    \small

    \caption{Global Properties of UGC 10205}
    \begin{tabular}{ccc}
    \hline\hline                 
    Property  &  UGC 10205 & References \\    

    \hline
    Morphology & Sa/S0 &  (1)(2)/(3) \\   
   z &  0.02187 &  (4)     \\
   RA(J2000) & 16h06m40.180s & (4)  \\
   DEC(J2000) & +30d05m56.70s  &  (4)   \\
   PA (deg) & 135 &  (4)   \\ 
   {\it r$-$SDSS} diameter ('') & 105.3 & (4)  \\ 
   Inclination (deg) & 84  &  (5)   \\ 
   Log(M$_{\star}$/M$_{\sun}$) & 10.997 &  (6) \\
   Log({\it L$_{TIR}$/L$_{\sun}$}) & 10.35 & (7) \\
   L$_{H\alpha}$ (erg s$^{-1}$) & 10.35 $\times$ 10$^{40}$  & (8) \\
   L$_{FUV}$ (erg s$^{-1}$) & 2.71 $\times$ 10$^{42}$  & (8) \\
   \hline
    \vspace{-0.5cm}
    \tablenotetext{0}{Notes: (1): \citet{Nilson_1973}, (2): \citet{Tarenghi_1994}, (3): \citet{Nair_2010} , (4): NASA/IPAC Extra-galactic Database , (5): \citet{Rubin_1985}, (6): \citet{Walcher_2014}, (7): \citet{Willmer_2009}, (8): \citet{Catalan_Torrecilla_2015},} \\
    
    \\

    \end{tabular}
    \label{tab:table_1}
    \end{table}


\section{Observations: MEGARA Integral Field Spectroscopy data} \label{observations}

The spectrophotometric observations of UGC 10205 were carried out with the {\it Multi-Espectr\'ografo en GTC de Alta Resoluci\'on para Astronom\'ia} (MEGARA, Gil de Paz et al. in prep.) on the 10.4m GTC telescope at La Palma Observatory. MEGARA is composed of an Integral Field Unit (IFU) and a Multi-Object Spectroscopy (MOS) mode. For the purpose of this work, we used only IFU data taken in 2017 June 30th and July 1st as part of the science verification of its commissioning phase. The IFU unit provides a field-of-view (FoV) of 12.5'' x 11.3'' using 567 hexagonal spaxels of 0.62'' in size which makes it ideal for studying the central regions of nearby galaxies. Due to the reasons explained later in Section~\ref{recover_halpha_emission}, the analysis of the data will be performed using squared spaxels. A recent example showing the potential of MEGARA data can be found in \citet{Dullo_2019}. 

The values of the seeing during the observations ranged between approximately 0.93'' and 1.11'' as recorded by the differential image motion monitor (DIMM) close to the Telescopio Nazionale Galileo (TNG). Four 1200 s science exposures were combined for the LR$-$V data set while in the HR$-$R set-up a total integration time of 2200s were made in two exposures. These data provide a unique view into the cold and warm phases of the interstellar medium. In particular, the LR$-$V grating is optimal to analyze the ISM Na I $\lambda$$\lambda$ 5890,5896 doublet (Na\,{\small I} D) as it yields a wavelength coverage from 5143\,\AA\, to 6164\,\AA\, with a spectral resolution of FWHM $=$ 0.937\,\AA\, at central wavelength $\lambda$$_{c}$ $=$ 5695\,\AA. The HR$-$R grating provides a wavelength coverage from 6445\,\AA\, to 6837\,\AA\, (FWHM $=$ 0.355\,\AA\, at $\lambda$$_{c}$ $=$ 6646\,\AA) that allow us to characterize the ionized gas by means of the H$\alpha$ emission line. The reciprocal dispersion is 0.27\,\AA/pix and 0.097\,\AA/pix respectively for LR$-$V and HR$-$R.

Additional images were acquired to help in the data reduction process. In particular, the spectrophotometric standard star BD$+$33 2642 was observed and used to create a response function to calibrate the absolute flux scale. Arc calibration lamp frames and twilight sky flat-field/lamp flat-field were obtained at the beginning and at the end of each night, respectively. Finally, the reduction of the raw data comprises the standard procedures of bias subtraction, cosmic-ray removal, flat fielding, tracing and extraction of the spectra, arc calibration solutions, sky-subtraction and flux calibration. It was performed applying the standard MEGARA pipeline \citep{sergio_pascual_2018_2206856,nicolas_cardiel_2018_2270518}.

\begin{figure*}
\centering
\includegraphics[trim={2cm 0cm 2cm 0cm},width=130mm]{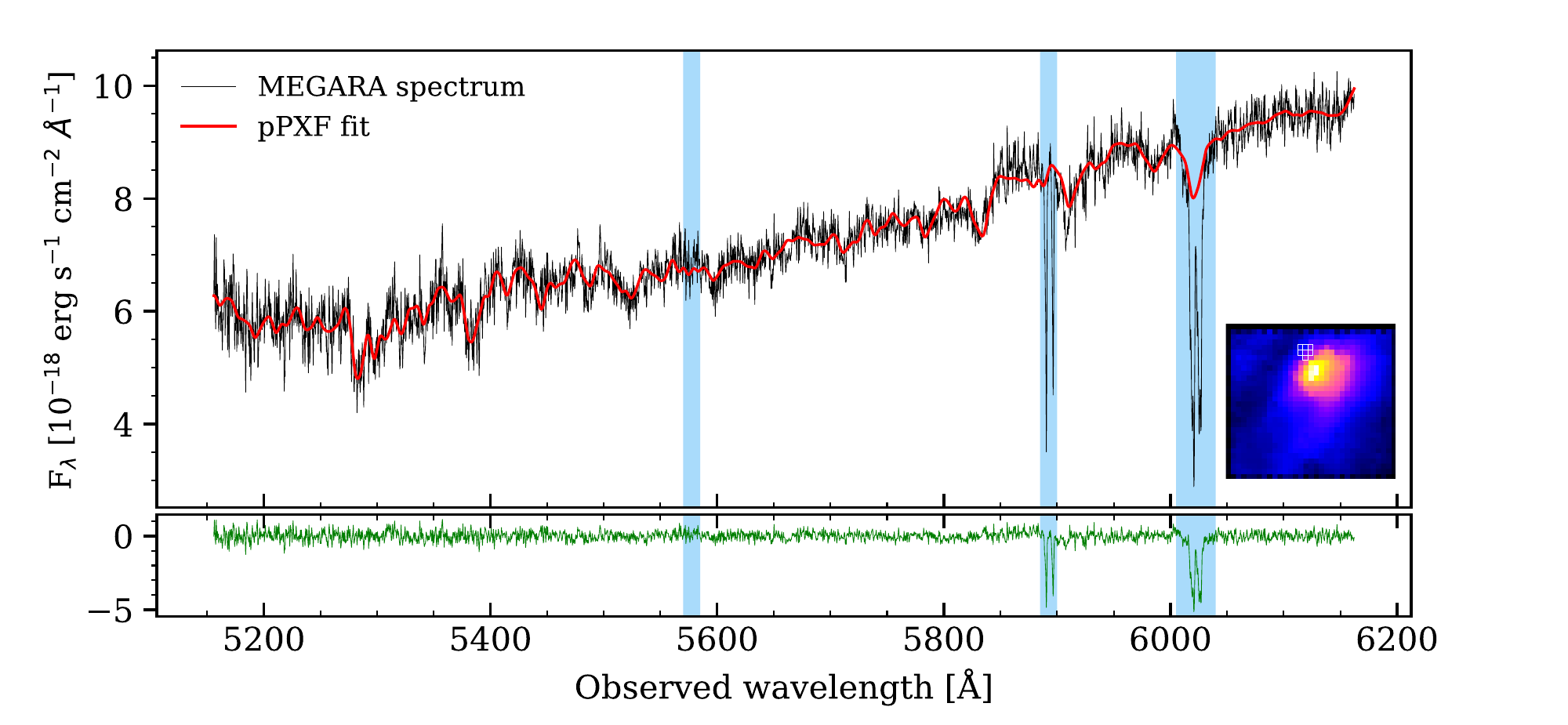} 
\caption{Spectrum for the LR$-$V MEGARA datacube is shown in black while its best fit from pPXF appears in red. The green spectrum at the bottom represents the fit residuals (note the different vertical scale). The blue shaded areas correspond to the masked regions (ISM features and sky line residuals). The inset shows the position of the Voronoi cell corresponding to this spectrum on the same continuum image of the galaxy as in Figure~\ref{ugc10205_megara_fov}.\\}
\label{spectra_ppxf}
\end{figure*}

\section{Analysis} \label{analysis}

\subsection{Galaxy continuum subtraction and interstellar \mbox{Na {\small I} D} detection} \label{LR-V grating}

In this section, we describe the main characteristics of the process followed to model the galaxy continuum for the case of the LR$-$V grating. As commented before, the LR$-$V covers the range from 5143\,\AA\, to 6164\,\AA\, with a spectral resolution of FWHM = 0.937\,\AA\, at $\lambda$$_{c}$ $=$ 5695\,\AA. 

Firstly, a Voronoi binning algorithm \citep{Cappellari_Copin_2003} is applied to increase the signal-to-noise ratio (S/N) of the spectra in the individual spaxels, reaching a S/N $\sim$ 20 over the continuum. This process allows a more homogeneous distribution of the S/N within the datacube. Then, we model the galaxy continuum in the spatially-binned MEGARA datacube using a Penalized Pixel-Fitting method \citep[pPXF,][]{Cappellari_2004, Cappellari_2017}. The aim of pPXF is to build the stellar template that will best reproduce the observed spectrum in each Voronoi cell. For that purpose, we used the latest theoretical stellar library by \citet{Coelho_2014} and P. Coelho, G. Bruzual $\&$ S. Charlot in preparation, which represents an expansion of the previous stellar population modelling in \citet{Coelho_2005,Coelho_2007} and \citet{Walcher_2009}. The resolution of these models, R $=$ 20000, is optimal for all the MEGARA set-ups and especially for the highest resolution VPHs (R $\approx$ 20000). We selected a total of 660 models with different ages ranging from 30 Myr to 14 Gyr and 3 different metallicities ([Fe/H] = $-$1.0,0.0,0.2) at both scaled-solar and $\alpha$-enhanced mixtures. The stellar continuum was fitted over the spectral range (5150 $-$ 6160)\,\AA\, which contains multiple stellar absorption lines. We carefully masked the ISM features as they are the result of both the stellar and the ISM contribution. Also, the He\,{\small I} emission line at 5875.67\,\AA\, was masked due to its proximity to the \mbox{Na {\small I} D} line. Finally, the best stellar model was subtracted from the observed spectra to detect the presence of \mbox{Na {\small I} D} interstellar absorption. To assess the robustness of the pPXF fits within the region measured and its impact on the results, we have applied a standard bootstrap analysis \citep[see][]{wu1986} using 100 simulations to each of the Voronoi spectra. First, a redistribution of the best-fit model residuals is done. Then, we fitted the resampled spectrum using pPXF. For the two first LOSVD moments (V and $\sigma$) we obtained dispersions of up to 9 km\,s$^{-1}$ and 10 km\,s$^{-1}$ after running the 100 repetitions, respectively. We also found mean differences of only 7 km\,s$^{-1}$ (5 km\,s$^{-1}$) between the original velocities (stellar velocity dispersions) values and the ones obtained after applying the bootstrapping technique.

Figure~\ref{spectra_ppxf} shows a representative fit for one of the Voronoi cells. The position of this particular Voronoi cell within the galaxy can be seen in the inset of the figure. The black line is the observed spectrum while the red line is the fit for the stellar component. The vertical blue shaded regions show the ISM features and sky-lines that have been masked in the fitting process. The residuals of the fitting appear in green color at the bottom. There is a clear detection of ISM \mbox{Na {\small I} D} absorption in this spectrum.

The analysis done along this Section allows us (i) to derive fundamental parameters of the galaxy kinematics such as the first four order moments of the line of sight velocity distribution (V, $\sigma$, h$_{3}$ and h$_{4}$) for each Voronoi cell of the MEGARA datacube (Section~\ref{resolved_stellar_kinematics}) and (ii) to explore the presence of interstellar \mbox{Na {\small I} D} absorption in UGC 10205 and its connection with neutral gas outflows (Section~\ref{nature_sodium}).

\subsection{Recovering the H$\alpha$ emission} \label{recover_halpha_emission}

In the following paragraphs we describe the analysis performed to obtain a reliable measure of the H$\alpha$ emission flux for the case of the HR$-$R grating. This set-up is characterized by a wavelength coverage from 6445\,\AA\, to 6837\,\AA\, and a FWHM $=$ 0.355\,\AA\, at $\lambda$$_{c}$ $=$ 6646\,\AA. The HR$-$R observations were taken on different nights, June 30th and July 1st (2017), and the pointing was slightly offset between them (we remind the reader that the data were taken as part of the Commissioning Time). Thus, the first step was to combine both observations to create a new datacube taking into account the relative offsets between them.

The second step was to remove the stellar continuum of the galaxy. However, due to the lack of absorption features in the wavelength range covered by the HR$-$R set-up, the use of codes such as pPXF is not recommended. For that reason, the method applied for the HR$-$R case slightly differs from the one used in the LR$-$V grating (Section~\ref{LR-V grating}). The approach followed here to estimate the H$\alpha$ absorption-corrected emission flux is to derive (i) the equivalent width of the stellar H$\alpha$ absorption (EW$_{H\alpha,abs}$) and (ii) the stellar velocity corresponding to each spaxel in the datacube by anchoring the stellar populations models to the ones obtained for the LR$-$V grating. The latest calculation requires to reference the HR$-$R observations to the LR$-$V ones. Due to pointing effects, the spatial coverage of the galaxy is slightly different in both set-ups so we have created datacubes with a pixel scale of 0.4 arcsec in each of the spatial axes. The area of these new synthetic spaxels, on which we will perform the subsequent analysis, will therefore be 0.4$\times$0.4\,arcsec$^2$ and will be simply referred as spaxels. The value of the spaxel size was chosen as the best compromise between the seeing of the observing nights and the routine used to create the datacubes. To estimate the EW$_{H\alpha,abs}$ we used the CALIFA spectrum (3700$-$7000\,\AA) integrated in the same area as the one in our MEGARA FoV datacube. To recover the best model of the stellar continuum, we run the pPXF code. Finally, we have obtained the gaussian profile corresponding to the stellar absorption that is added to the H$\alpha$ emission flux in the HR$-$R datacube. The main assumption in this approach is that the model that best reproduce the stellar continuum is the same for all the spaxels in the MEGARA datacube. As the mean value of the recovered stellar flux at the H$\alpha$ wavelength is around 9\,$\%$ of the total H$\alpha$ flux, we applied this correction assuming that it represents a small second order correction for the estimation of the total H$\alpha$ flux. As done in the case of the LR$-$V, we also increase the S/N ratio making a spatial binning of the datacube. In this case, we have defined the S/N as the ratio between the H$\alpha$ emission line and the nearest continuum and imposed a S/N $>$ 15.

The high spectral resolution of MEGARA allow us to distinguish double and triple-peaked H$\alpha$ emission line profiles in this galaxy. The finding of up to three kinematically distinct gaseous components is in accordance with the results from \citet{Vega_1997}. Some examples showing the complexity of the H$\alpha$ emission line profiles will be shown and examined in the following sections.

\section{Results} \label{results}

\subsection{Kinematics of the Emitting Gas as revealed by the H$\alpha$ emission line} \label{gas_kinematics_section}

To investigate the gas kinematics, we make use of the H$\alpha$ emission line. A previous long-slit spectroscopic study of the kinematics of the gas in UGC 10205 was performed by \citet{Vega_1997} using the Intermediate Dispersion Spectrograph (IDS) at the Isaac Newton Telescope (INT) obtaining a mean value of FWHM $=$ 0.86\AA\, (\mbox{$\sigma$ $\sim$ 17 km\,s$^{-1}$}). These authors found line splitting for three kinematically distinct gaseous components in the position-velocity diagram along the major axis of the galaxy in the inner $\pm$ 13''. A double-peaked H$\alpha$ emission appeared in the region $-$13''$\leq$ r $<$ 3'' and a third H$\alpha$ component is found at distances in the range 3'' $\leq$ r $\leq$ 13''. Later, \citet{Garcia_Lorenzo_2015} presented a comprehensive analysis of the gas velocity fields for 177 galaxies among which UGC 10205 is present. The galaxies were observed as part of the CALIFA survey using the Potsdam Multi-Aperture Spectrophotometer \citep[PMAS,][]{Roth_2005} in the PPak mode \citep{Kelz_2006} which provides a nominal resolution (R $\sim$ 850 at $\lambda$ $\sim$ 5000\,\AA) that is not sufficient to probe the presence of multiple components in this object.

Here, the high-resolution MEGARA IFS data improves the ability to map the kinematics of the line-emitting gas in a spatially-resolved manner, combining the information on the morphology and the kinematics simultaneously. In the next section, we attempt an interpretation for the kinematic properties of the ionized-gas phase by fitting a rotating exponential thin disk model to the observations.

\begin{figure*}
\centering

\includegraphics[trim={0.35cm 0.05cm 0.1cm 0.35cmm}, width=44mm]{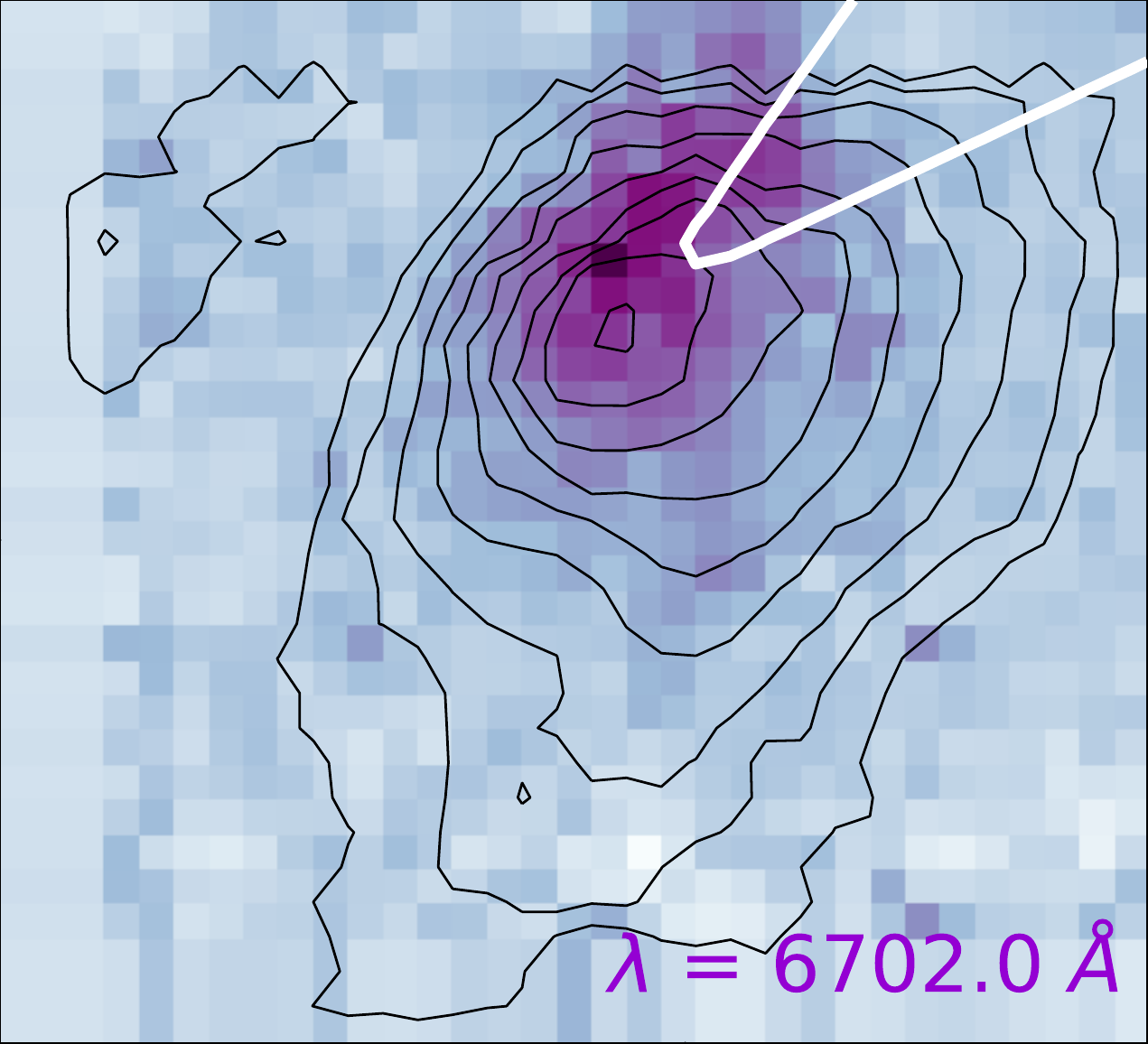} 
\includegraphics[trim={0.35cm 0.05cm 0.1cm 0.35cmm}, width=44mm]{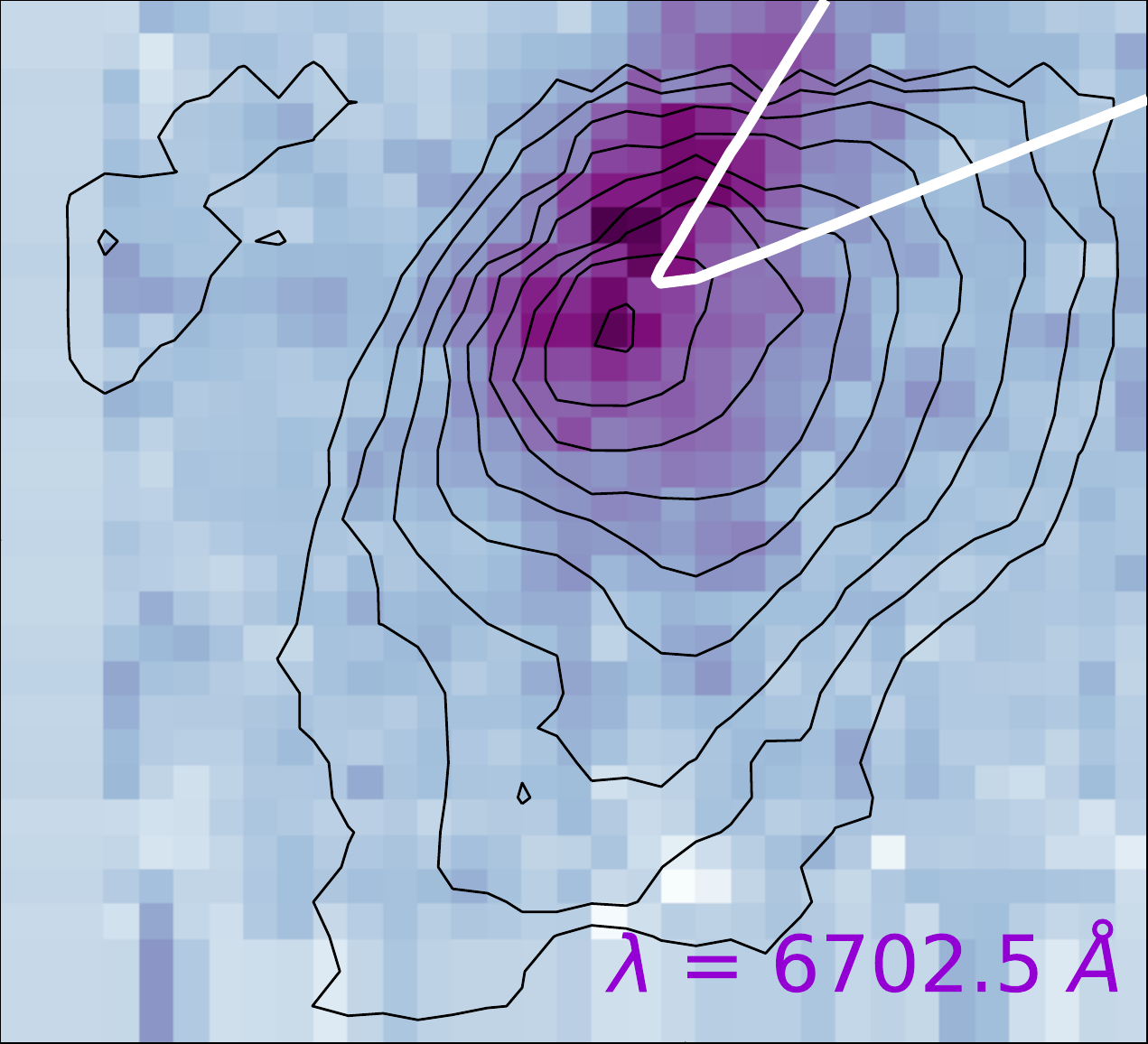} 
\includegraphics[trim={0.35cm 0.05cm 0.1cm 0.35cmm}, width=44mm]{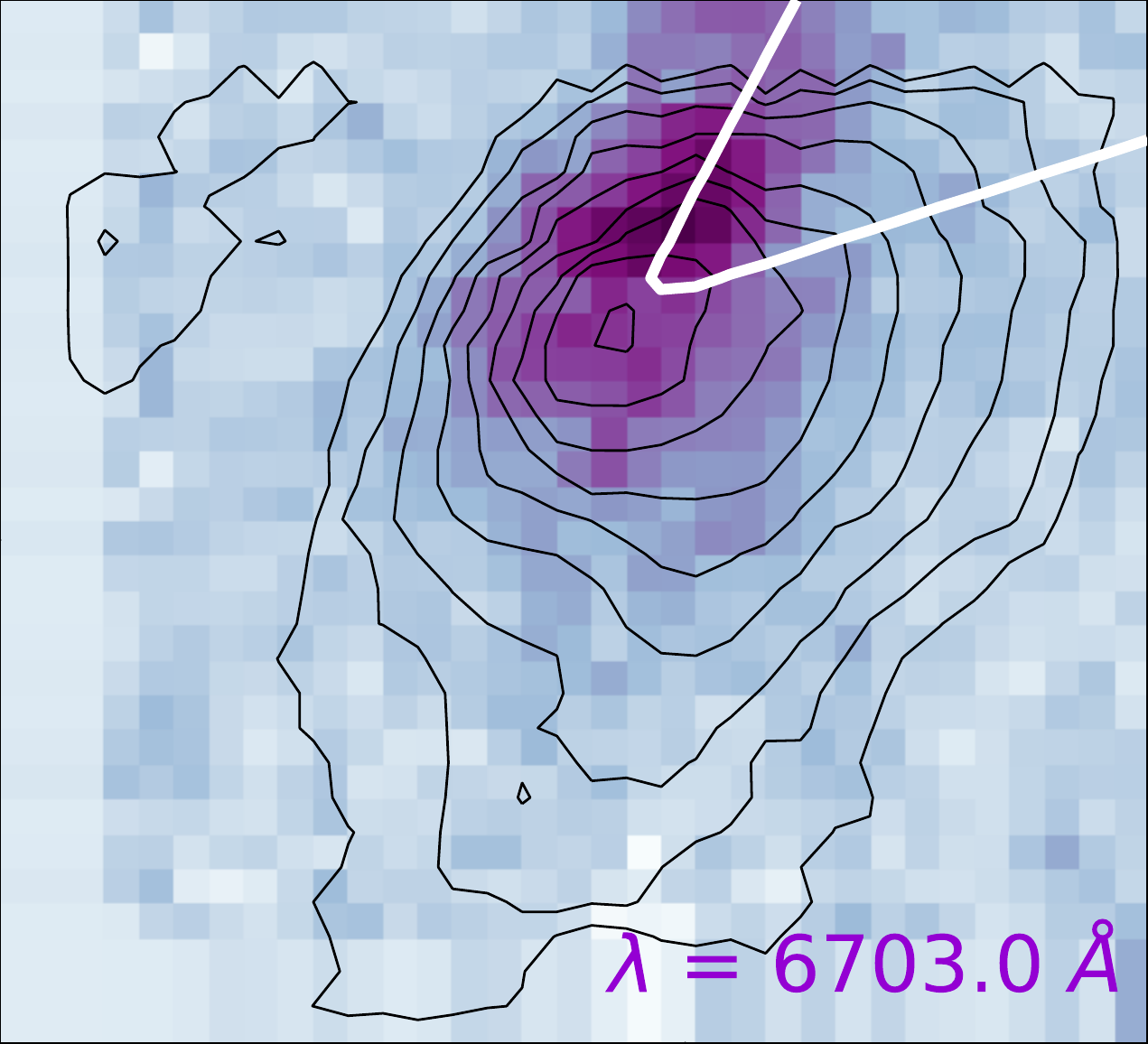}
\includegraphics[trim={0.35cm 0.05cm 0.1cm 0.35cmm}, width=44mm]{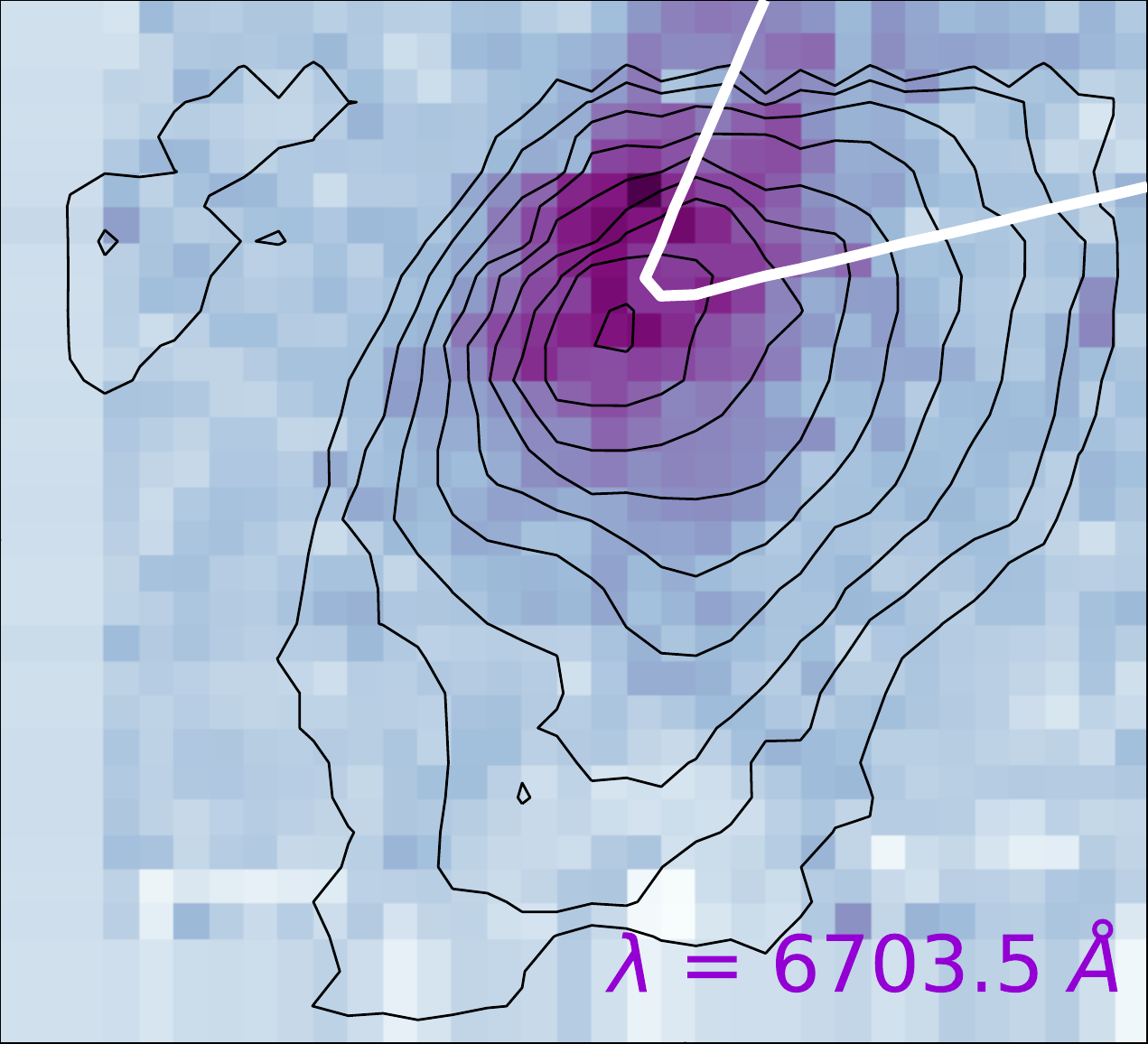} \\

\includegraphics[trim={0.35cm 0.05cm 0.1cm 0.35cmm}, width=44mm]{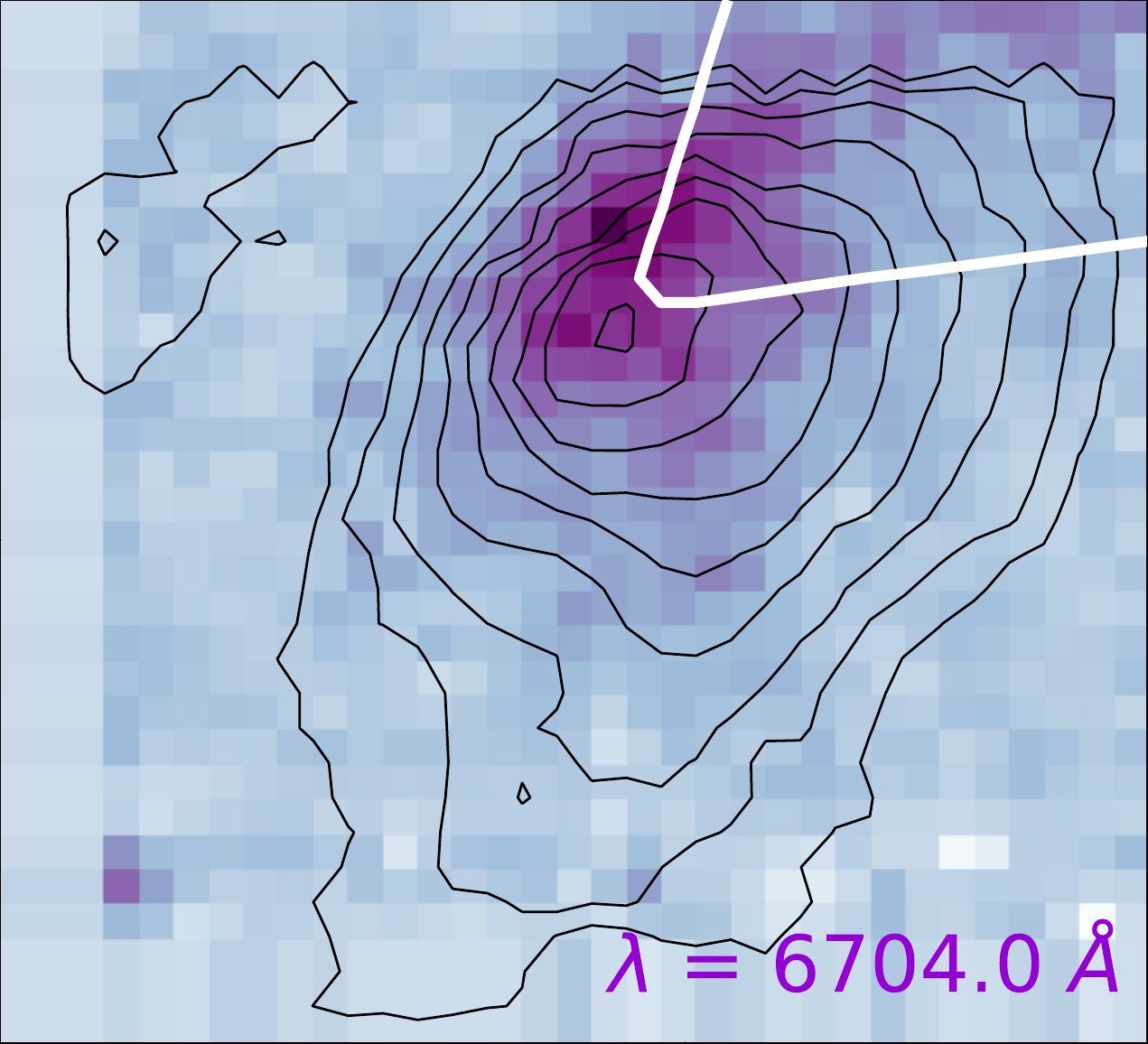}
\includegraphics[trim={0.35cm 0.05cm 0.1cm 0.35cmm}, width=44mm]{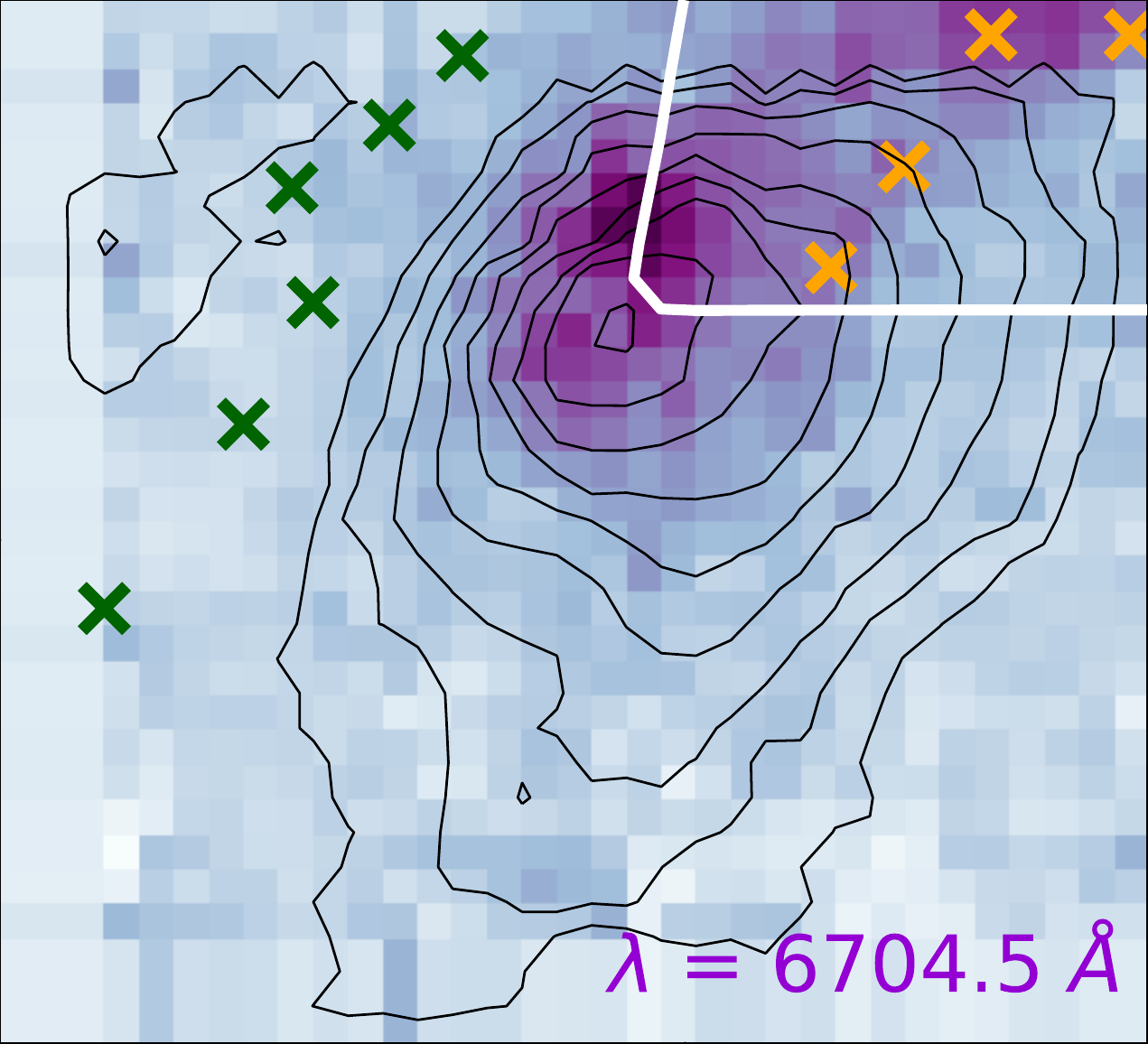}
\includegraphics[trim={0.35cm 0.05cm 0.1cm 0.35cmm}, width=44mm]{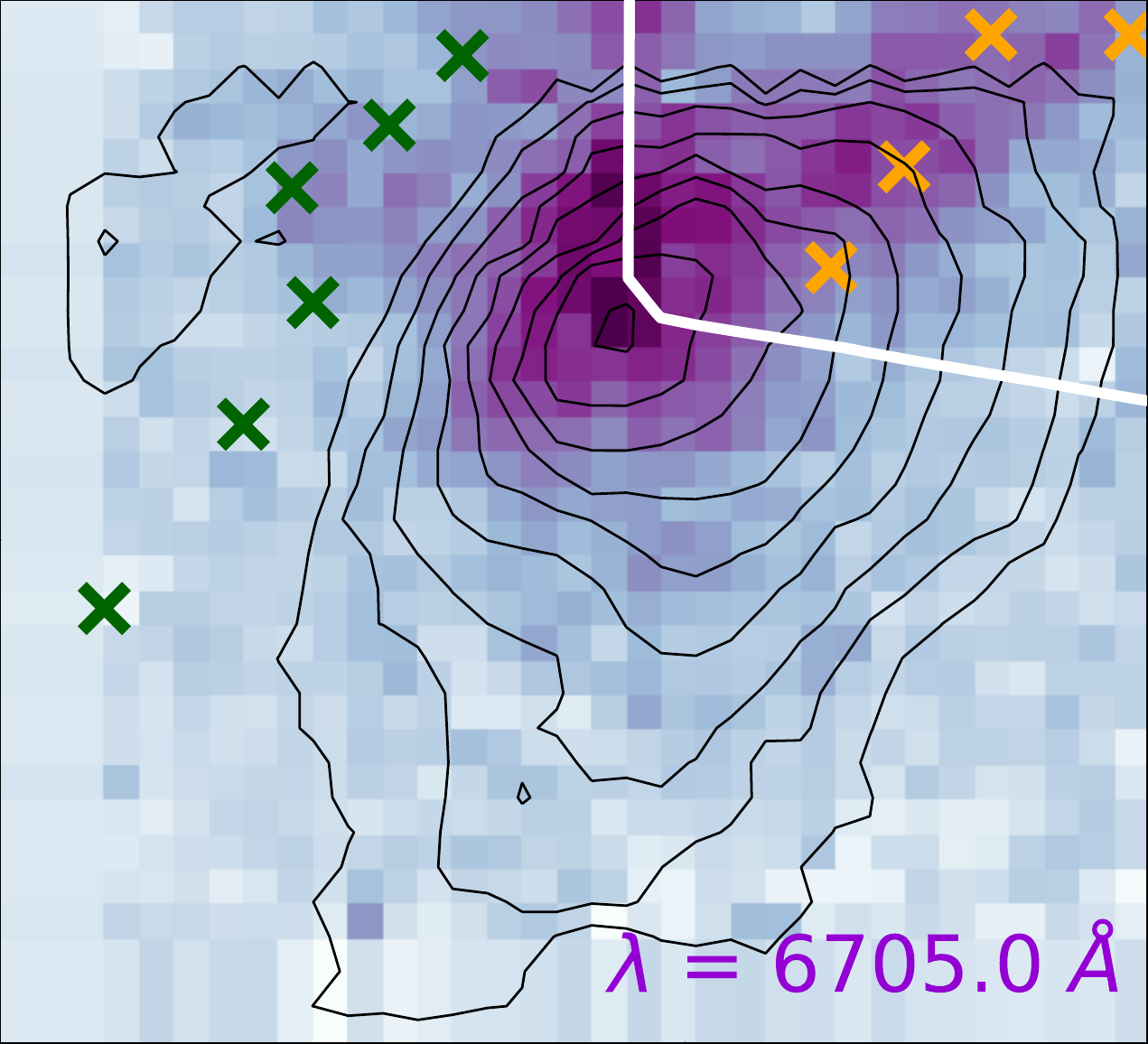} 
\includegraphics[trim={0.35cm 0.05cm 0.1cm 0.35cmm}, width=44mm]{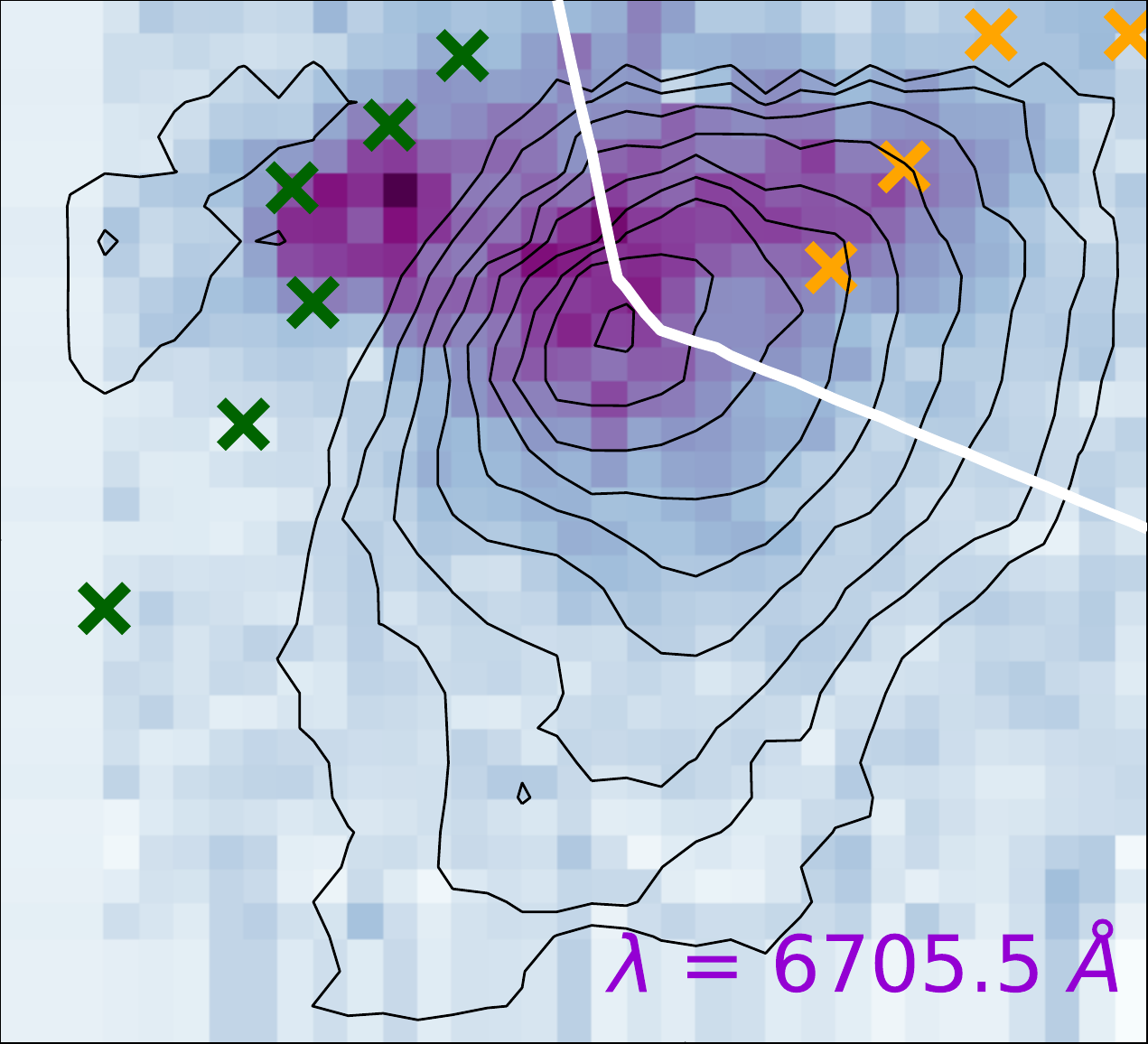} \\

\includegraphics[trim={0.35cm 0.05cm 0.1cm 0.35cmm}, width=44mm]{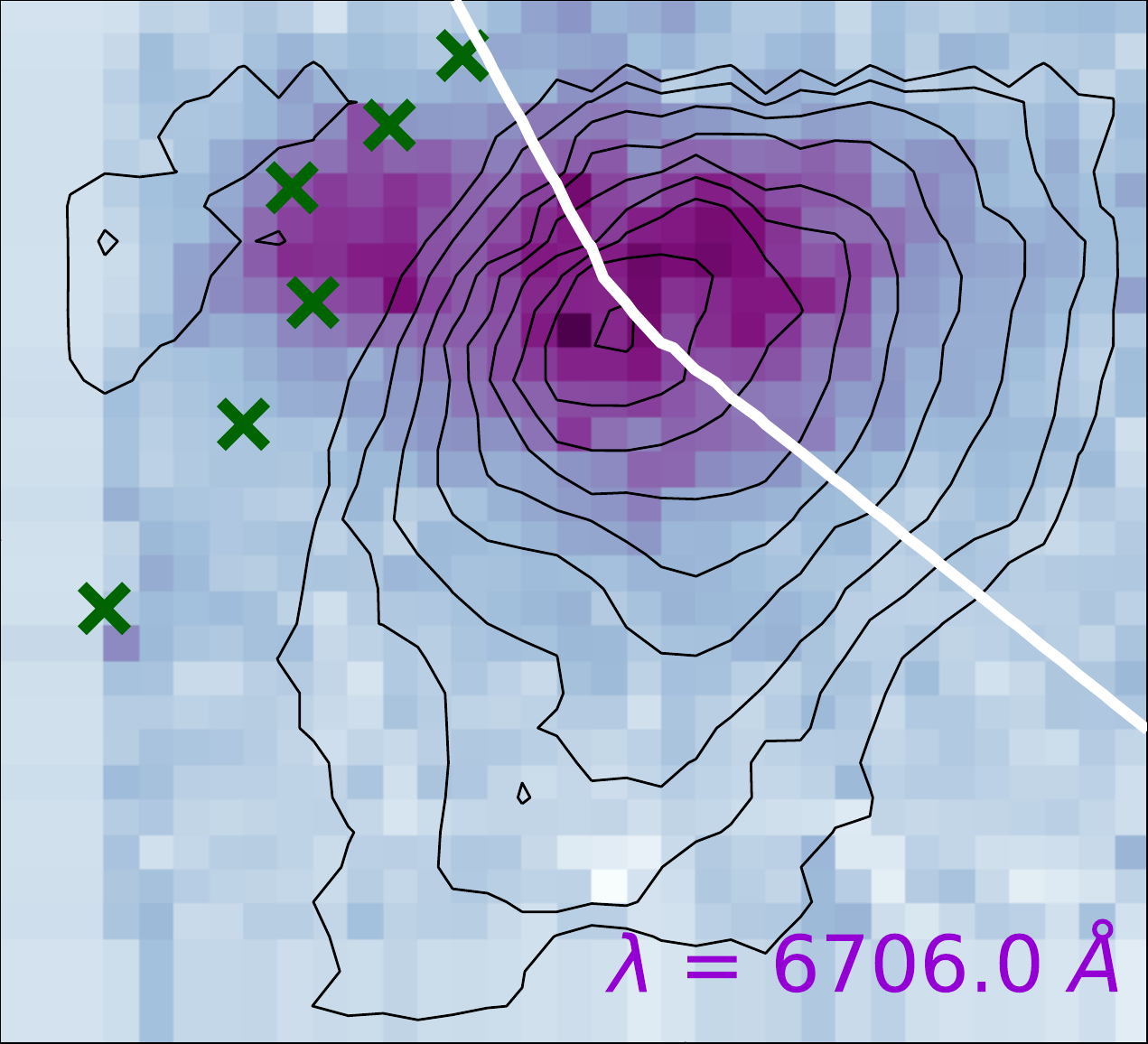}
\includegraphics[trim={0.35cm 0.05cm 0.1cm 0.35cmm}, width=44mm]{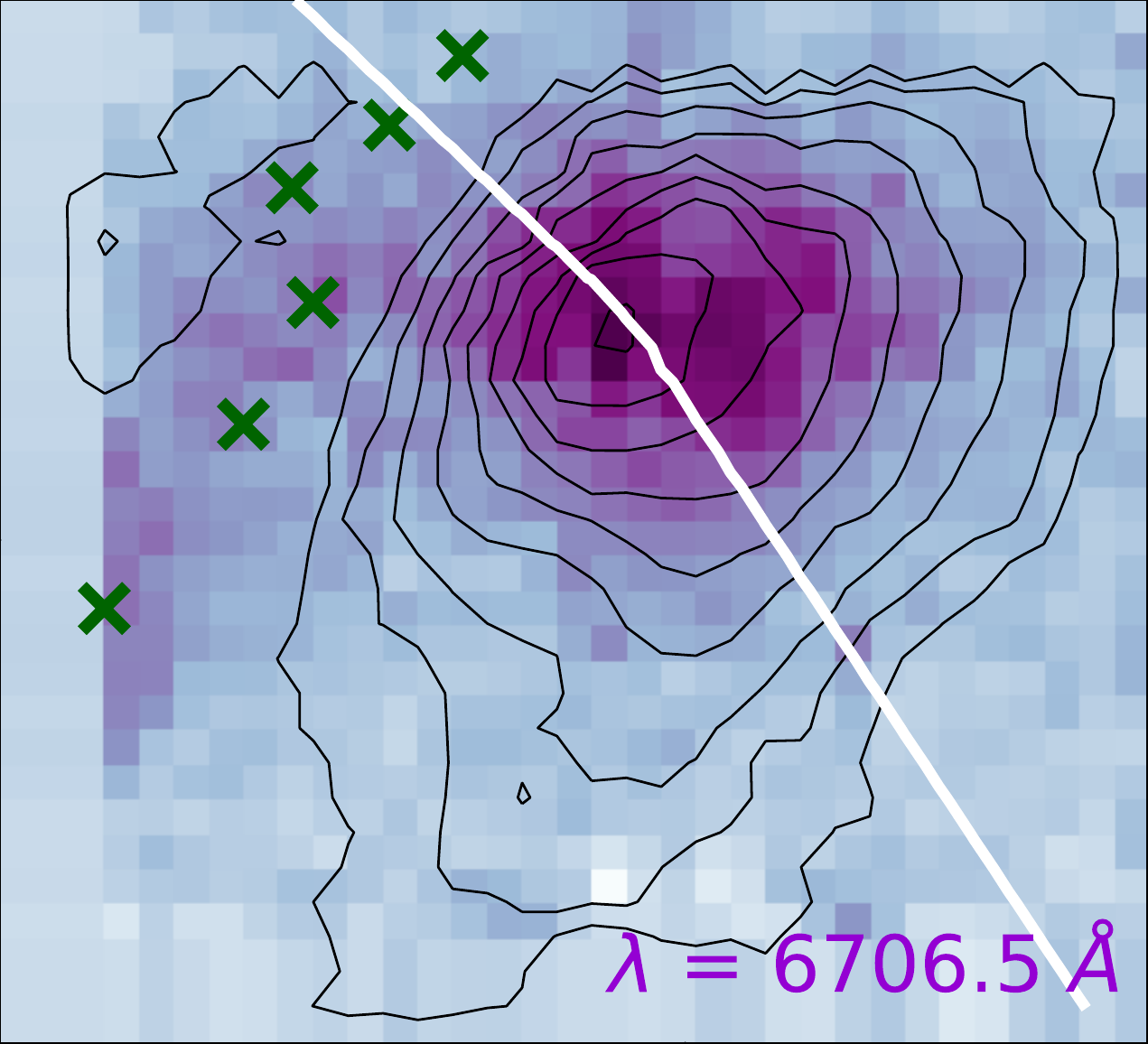}
\includegraphics[trim={0.35cm 0.05cm 0.1cm 0.35cmm}, width=44mm]{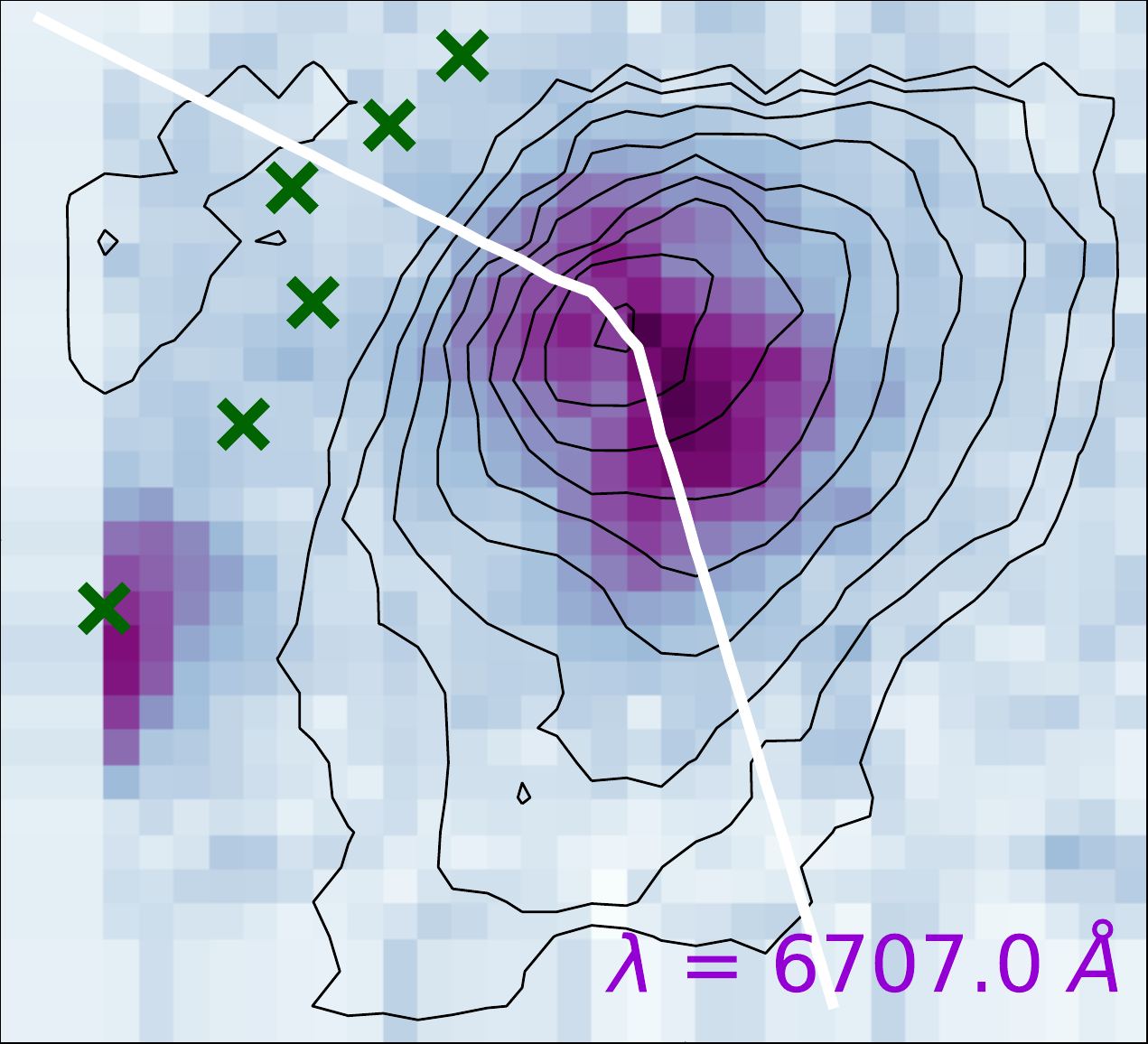} 
\includegraphics[trim={0.35cm 0.05cm 0.1cm 0.35cmm}, width=44mm]{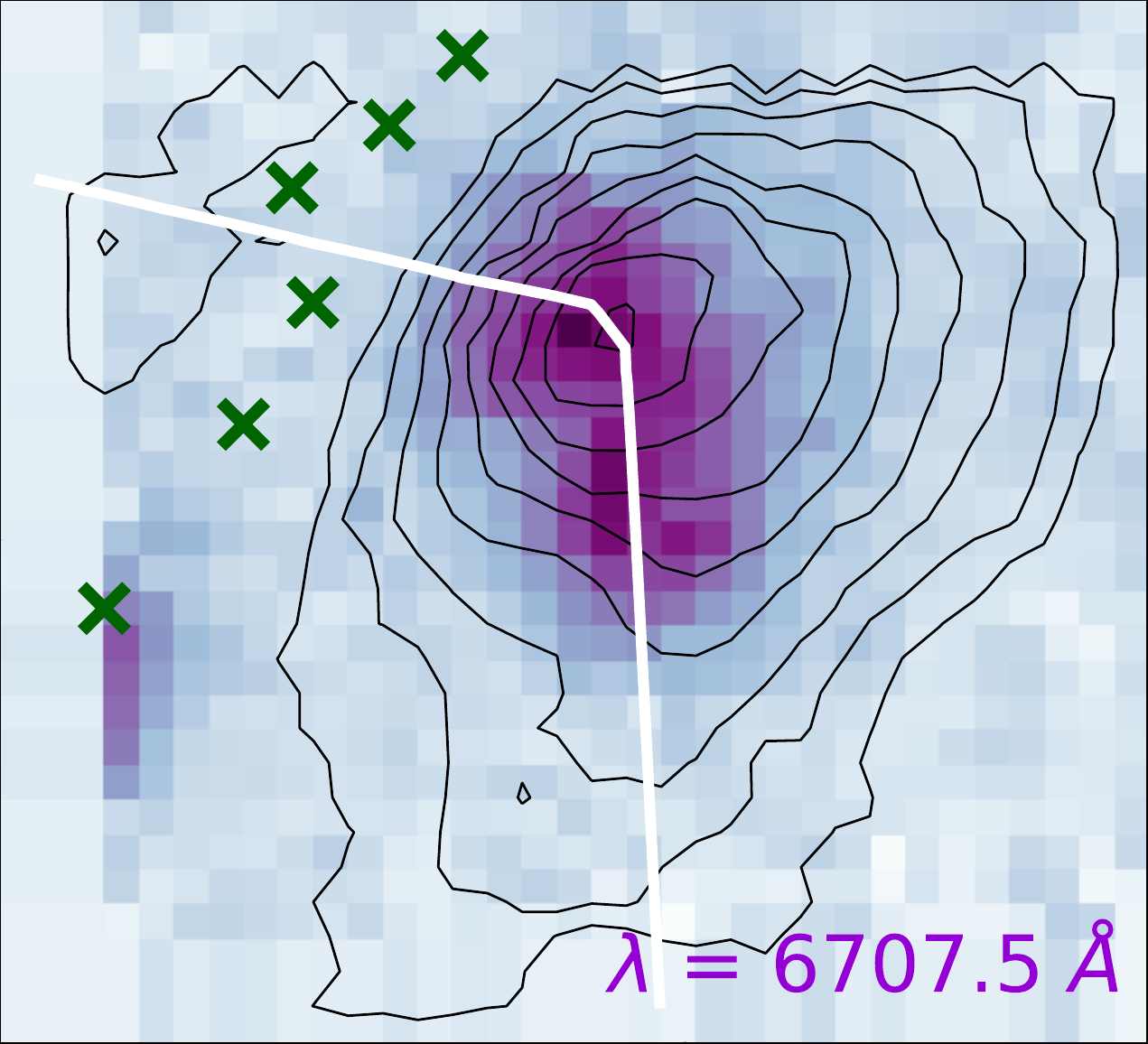} \\
 
\includegraphics[trim={0.35cm 0.05cm 0.1cm 0.35cmm}, width=44mm]{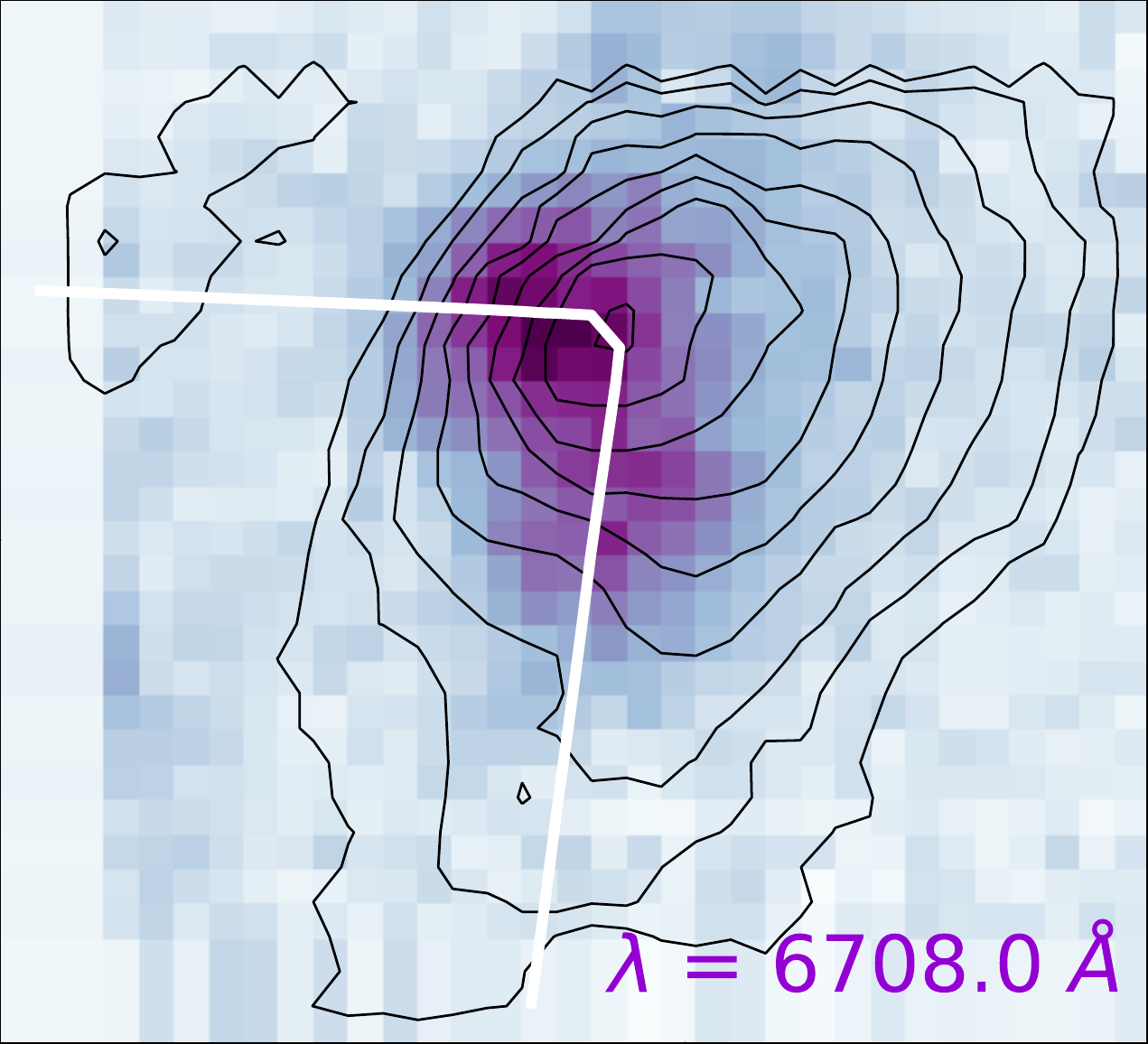}
\includegraphics[trim={0.35cm 0.05cm 0.1cm 0.35cmm}, width=44mm]{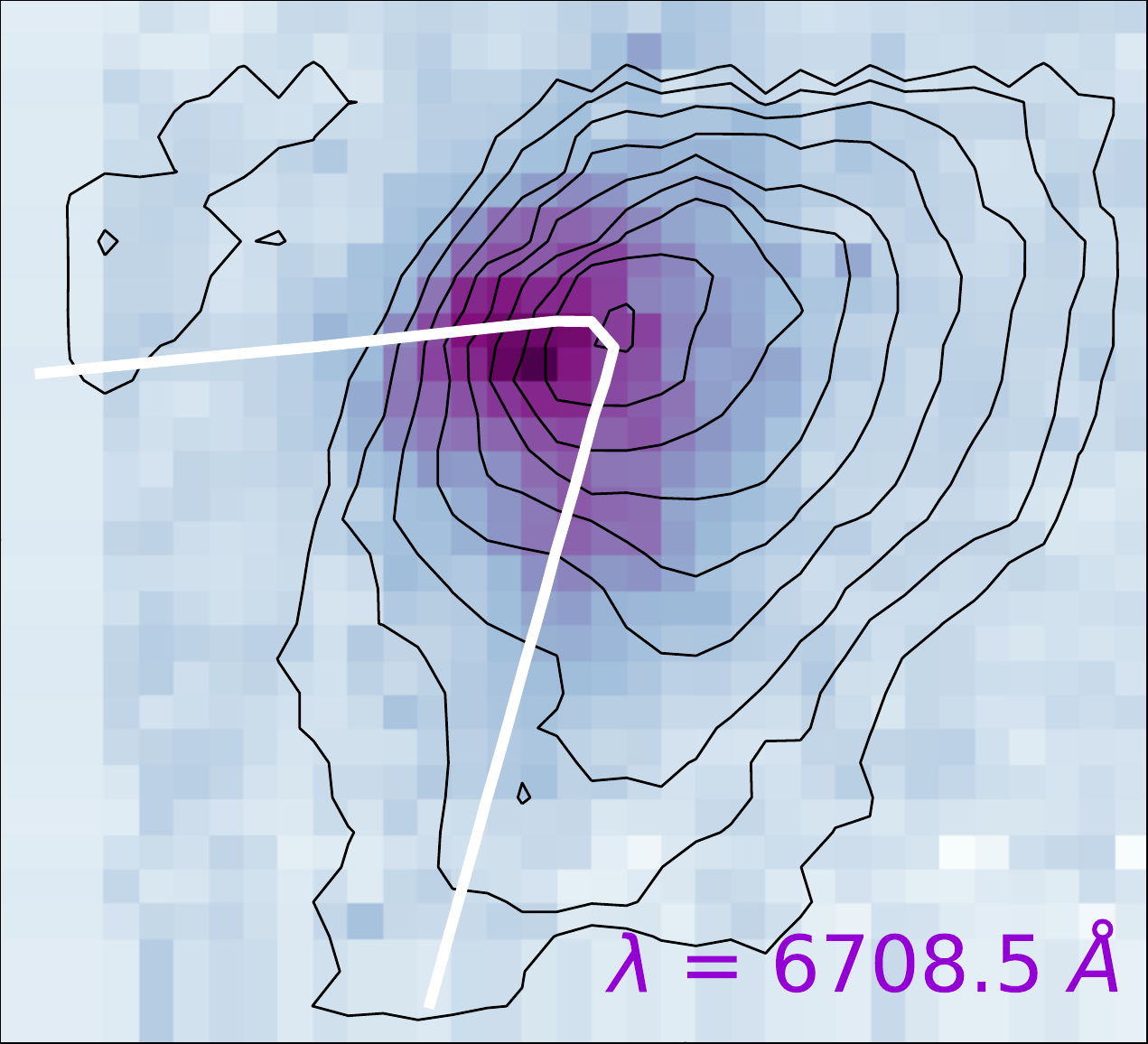}
\includegraphics[trim={0.35cm 0.05cm 0.1cm 0.35cmm}, width=44mm]{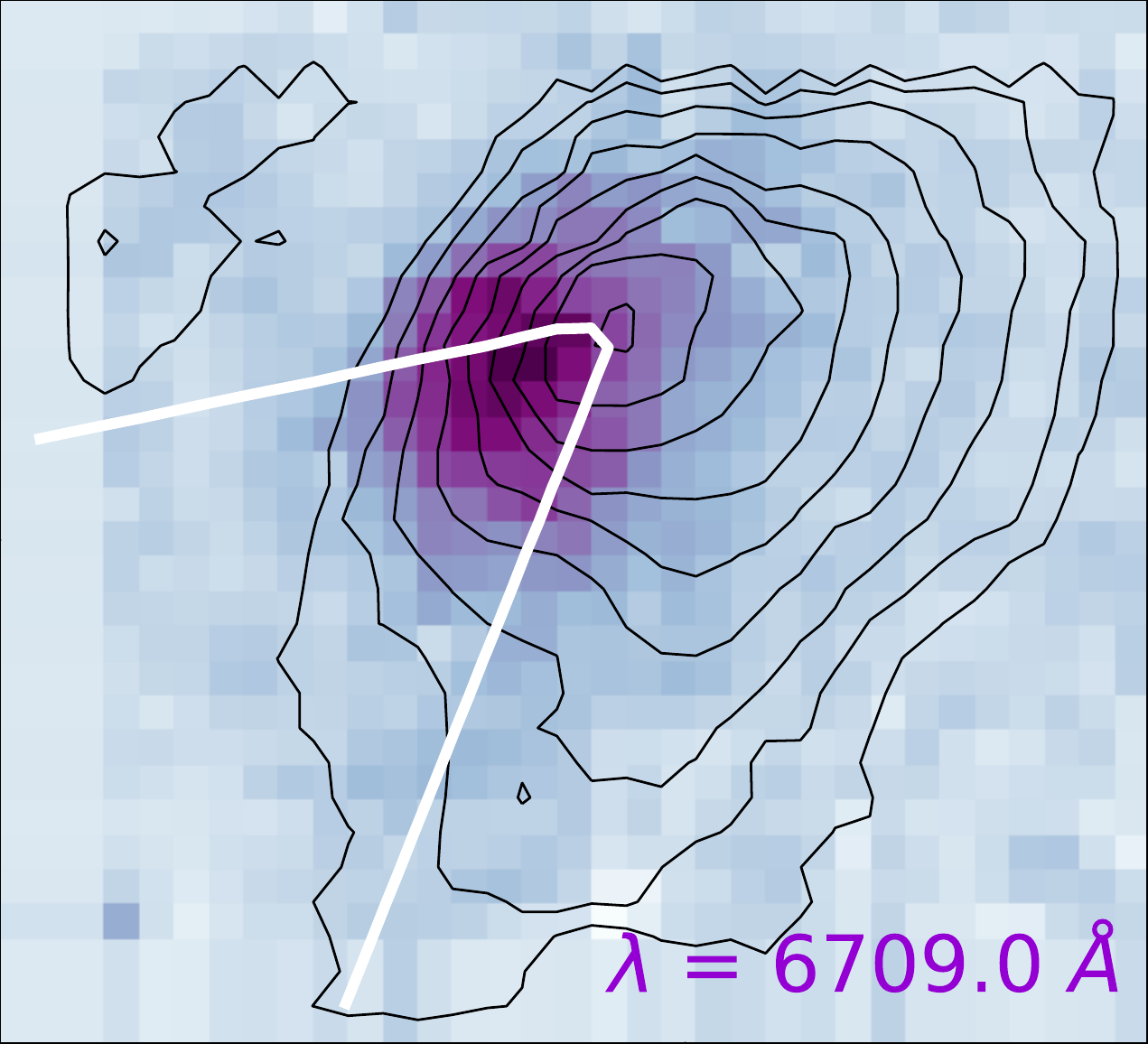} 
\includegraphics[trim={0.35cm 0.05cm 0.1cm 0.35cmm}, width=44mm]{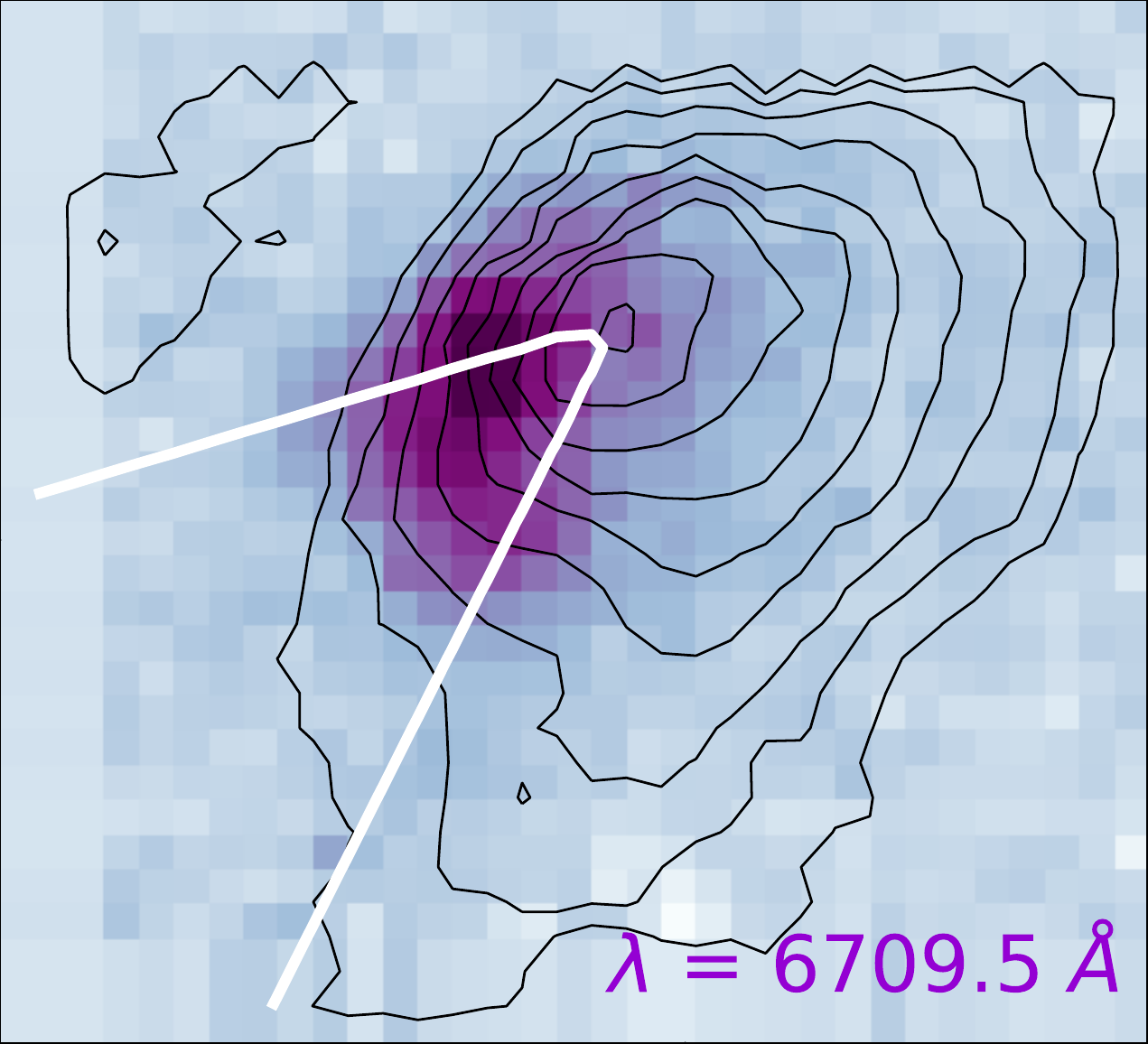} \\

\includegraphics[trim={0.35cm 0.05cm 0.1cm 0.35cmm}, width=44mm]{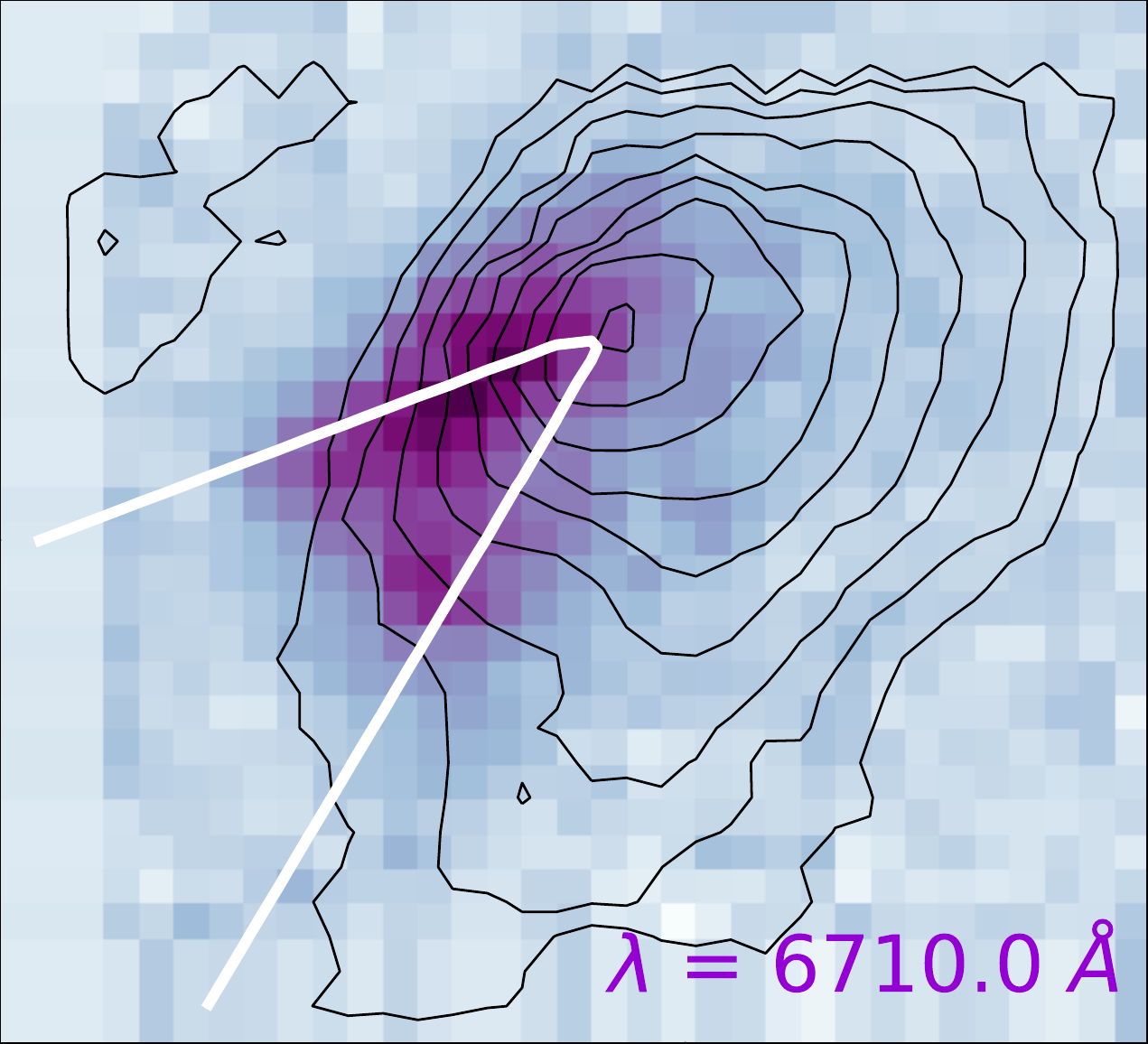}
\includegraphics[trim={0.35cm 0.05cm 0.1cm 0.35cmm}, width=44mm]{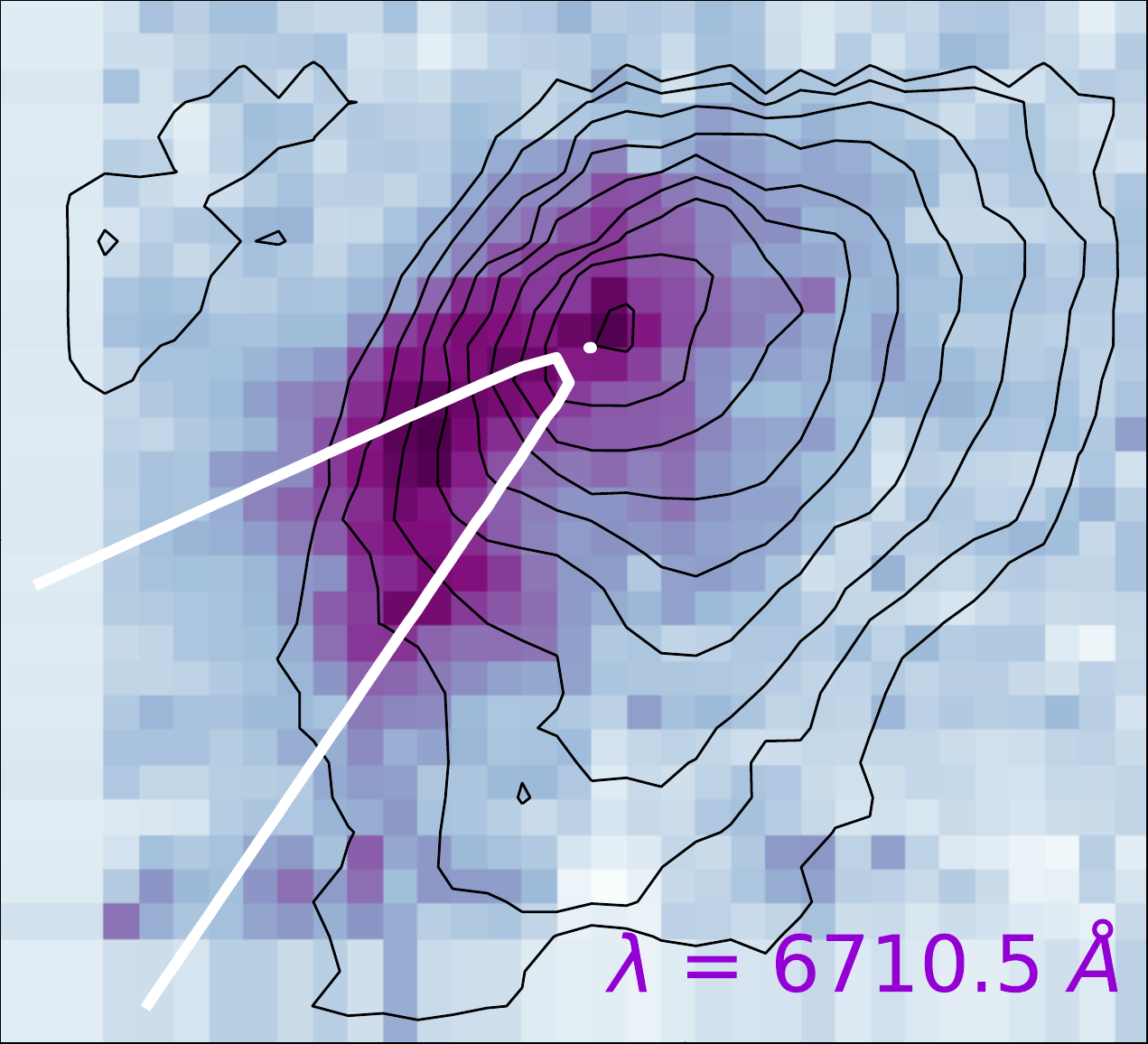}
\includegraphics[trim={0.35cm 0.05cm 0.1cm 0.35cmm}, width=44mm]{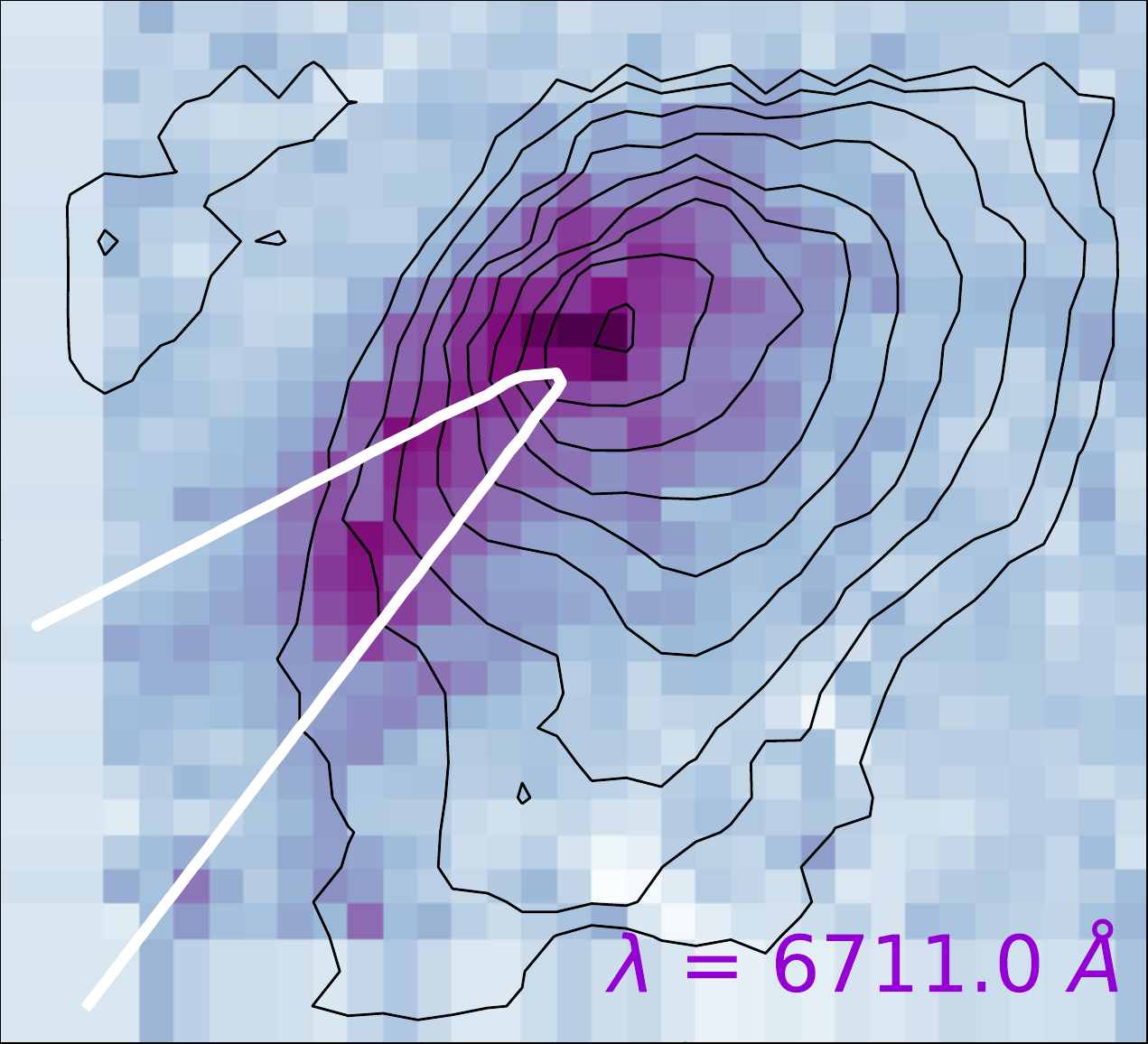} 
\includegraphics[trim={0.35cm 0.05cm 0.1cm 0.35cmm}, width=44mm]{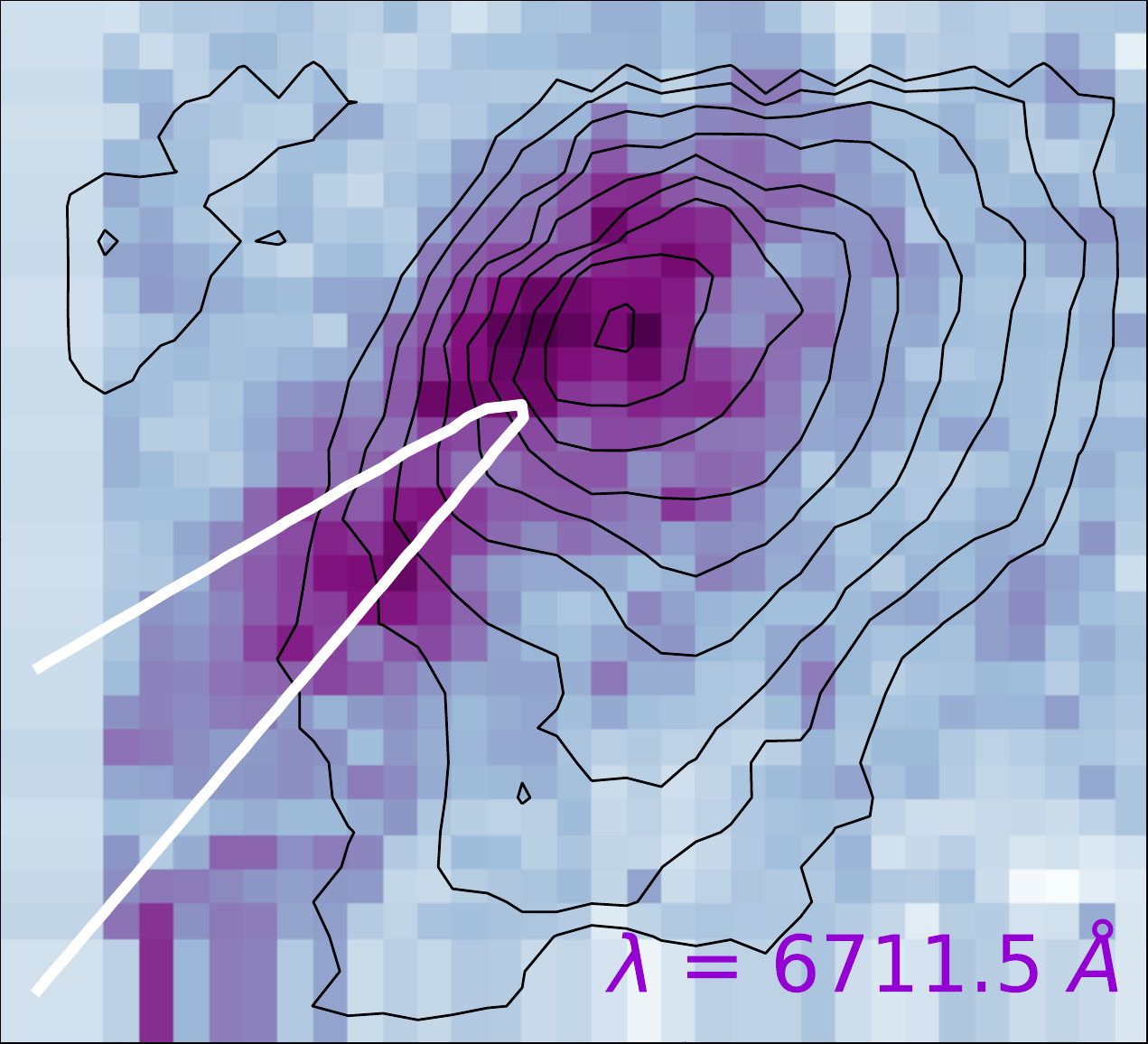} \\

\caption{H$\alpha$ velocity channel maps for the central part of the galaxy UGC 10205 in the velocity range 6363.14 $-$ 6797.40 km\,s$^{-1}$. A velocity bin of 22.86 km\,s$^{-1}$ corresponding to a $\Delta$$\lambda$ $=$ 0.5 \AA\, is used to slice the datacube. The wavelength of each channel is shown in the lower right corner. White lines represent the isovelocity curves for a thin-disk model with inclination of 75$^{\circ}$. Green and orange crosses represent the positions of the Voronoi cells that are best reproduced by a model with inclination of 83$^{\circ}$ as discussed in Section~\ref{dust_lane_section}. Continuum flux contours from the LR$-$V datacube are shown as black solid lines.  \\}
\label{channel_map}
\end{figure*}

\subsubsection{Thin disk model} \label{thin_disk_model}

Both \citet{Garcia_Lorenzo_2015} and \citet{Vega_1997} derived monotonically rising rotation velocity curves for the inner regions of UGC 10205. Following these results, we adopt an arctangent parameterization with a smooth turn-over to recover the shape of the rotation curve, \mbox{V$_{r}$ $=$ (2/$\pi$) v$_{max}$ arctan({\it r/r$_{t}$})} \citep{Courteau_1997}, where v$_{max}$ is the asymptotic maximum rotation speed and r$_{t}$ is the transition radius between the rising and flat part of the rotation curve. 

A suite of 2D kinematic models is built to recover the 2D ionized velocity field of the purely rotating galactic disk. The models contain six free parameters: v$_{max}$, r$_{t}$, position angle ({\it PA}), inclination ({\it i}) and the position of the dynamical center (x0,y0). To define the parameters range we use reported measurements and allow variations consistent with our data. The corresponding ranges are: \mbox{v$_{max}$ $=$ (250 $-$ 350) km\,s$^{-1}$,} \mbox{r$_{t}$ $=$ (0.05 $-$ 0.5) kpc}, \mbox{{\it i} $=$ (20 $-$ 89)$^{\circ}$}, \mbox{{\it PA} $=$ (40 $-$ 170)$^{\circ}$}. For the dynamical center, we use the photometric center as our reference value. The photometric center is defined as the peak of the continuum emission in the region (6540 - 6620)\,\AA\, and we allow a variation of 2 arcsecs around it. A systemic velocity of 6556 km\,s$^{-1}$ is taken from NED.

We utilize a minimization based on a mean-squared error to find the best fit between the model and the data. As the H$\alpha$ channels show a somehow clumpy distribution (see Figure~\ref{channel_map}), we apply the minimization on the voronois that are more relevant to the shape of the rotation curve. The model parameters that best describe the data are \mbox{v$_{max}$ $=$ 292 km\,s$^{-1}$}, \mbox{r$_{t}$ $=$ 0.12 kpc}, \mbox{{\it i} $=$ 75$^{\circ}$} and \mbox{{\it PA} $=$ 130$^{\circ}$}. In the case of the inclination, a wealth of values can be found in the literature. \citet{Garcia_Lorenzo_2015} derived a photometric inclination of 54$^{\circ}$ using the whole galaxy while \citet{Rubin_1985} reported \mbox{{\it i} $=$ 84$^{\circ}$}. Due to the recent interactions suffered by the galaxy \citep[as suggested in][]{Reshetnikov_1999}, the gas disk of the outer parts might be mostly affected by them. Thus, the inclination for the innermost parts might be different. We have obtained \mbox{{\it i} $=$ 75$^{\circ}$} for these regions. The orientation of the line of nodes is in reasonably good agreement with the value inferred by \citet{Barrera-Ballesteros_2015}. They derived a morphological \mbox{{\it PA} $=$ (132.4 $\pm$ 1.6)$^{\circ}$} by fitting an ellipse to an isophote at radius 15 arcsec in the {\it r}$-$band {\it SDSS} image. They also measured kinematic {\it PA} for the ionized component and derived \mbox{{\it PA} $=$ (120.3 $\pm$ 2.7)$^{\circ}$} and \mbox{{\it PA} $=$ (129.4 $\pm$ 1.5)$^{\circ}$} for the approaching and receding sides, respectively.

Figure~\ref{channel_map} shows the H$\alpha$ channel maps of the HR$-$R datacube. Each map represents the line emission at a particular wavelength channel (from 6705.0\,\AA\, to 6709.5\,\AA) with a channel step of $\Delta$$\lambda$ = 0.5\,\AA. From the analysis of this figure, we infer the presence of a highly-inclined gas-rich disk located in the central regions of this galaxy. The isovelocity contours are well reproduced by our model for the central parts (see the white solid lines in the channels maps from 6707.0\,\AA\, to 6708.5\,\AA\, of Figure~\ref{channel_map}). However, some of the ionized gas do not follow the general rotation pattern as indicated by the isovelocity curves (see green and orange crosses in the channel maps from $\lambda$ $=$ 6704.5\,\AA\, to $\lambda$ $=$ 6707.5\,\AA\, in Figure~\ref{channel_map}). This component will be described later.

\begin{figure}
\centering
\includegraphics[trim={1.25cm 0cm 0.5cm 0.52cm}, width=85mm]{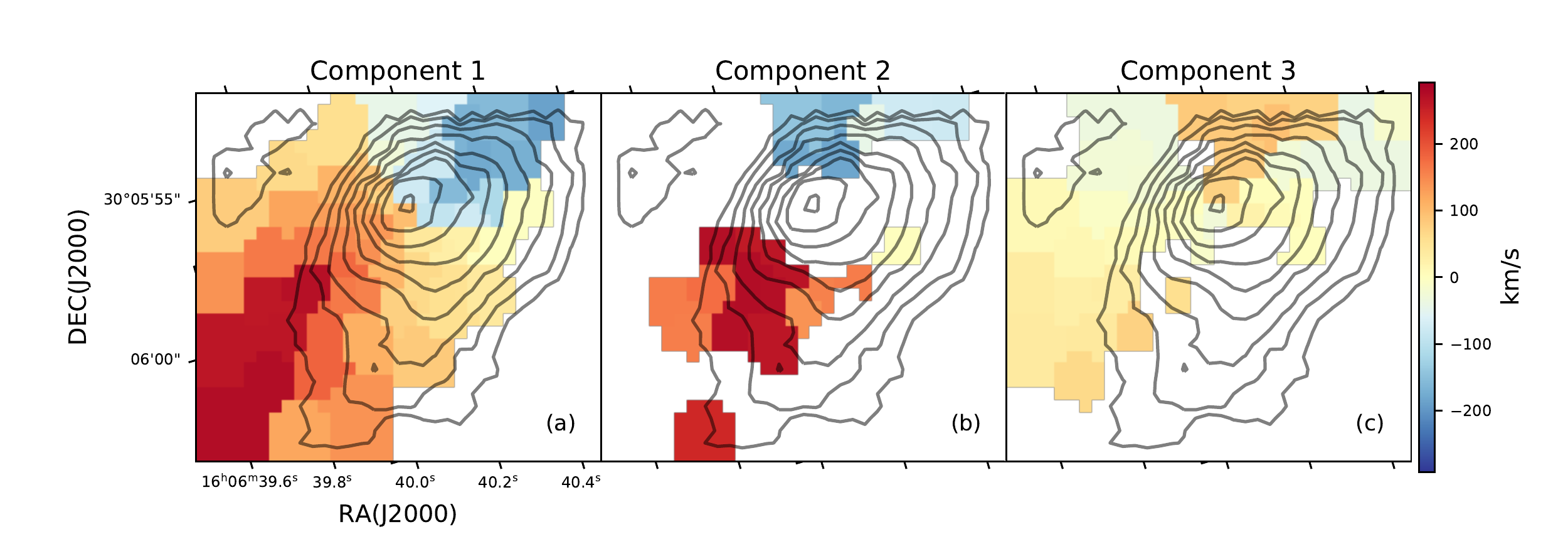} \\
\includegraphics[trim={1.25cm 0cm 0.5cm 0.52cm}, width=85mm]{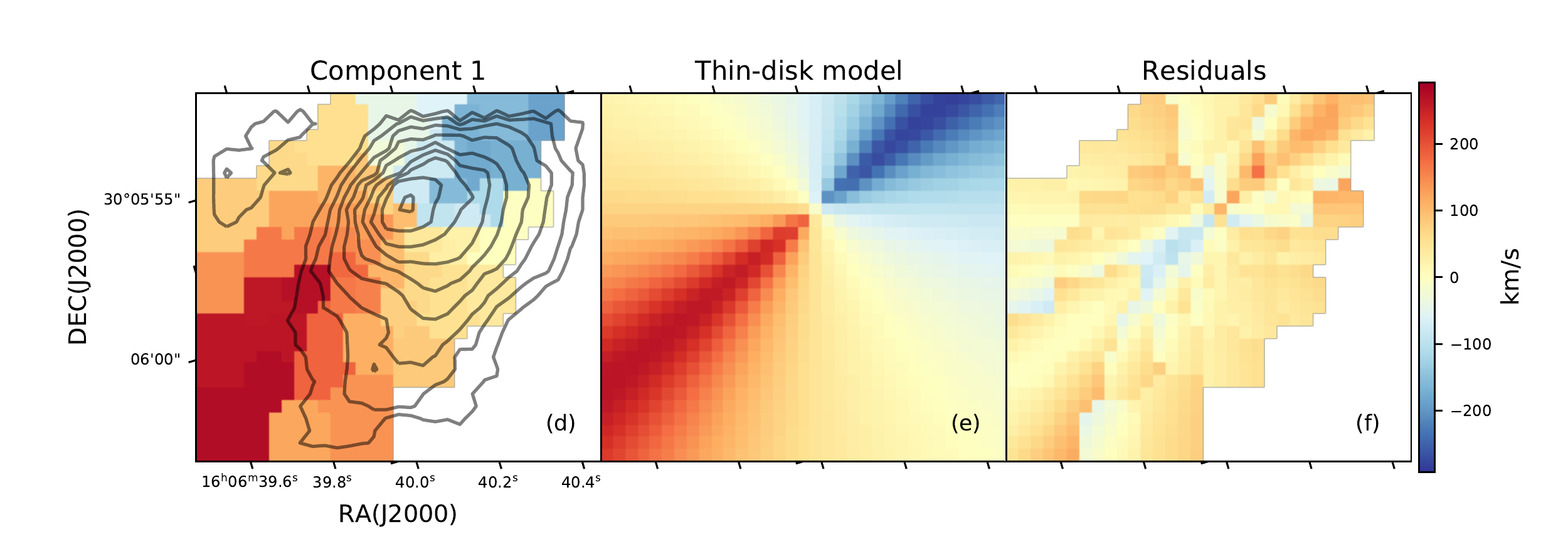} 
\caption{Top panel: velocity maps of the three kinematically distinct gaseous components: (a) component defined by the thin-disk model (arctangent function), (b) second component with a similar rotation pattern than the previous one, (c) component with a velocity close to the systemic value. These components are defined by the blue, red and orange Gaussian fits in the spectra of Figure~\ref{examples_ha_profiles}, respectively. Continuum flux contours from the LR$-$V datacube are shown as gray solid lines. Bottom panel: (d) velocity map for the ionized gas component defined by the thin-disk model, (e) the best thin-disk model for the previous component, (f) the residuals after subtracting the best thin-disk model.}
\label{ionized_maps}
\end{figure}

To recover the velocity maps of the ionized gas components, we need to disentangle the different emission lines in the H$\alpha$ profile. We fit multiple Gaussian components and a polynomial continuum to separate up to three kinematically distinct gaseous components. The corresponding ionized gas kinematic maps are shown in the top panel of Figure~\ref{ionized_maps}.

The first component ({\it component 1}) is the one associated with the thin-disk model. We trace this component based on the velocity associated to the centroid of each Voronoi cell in the disk model, i.\,e., we defined the Gaussian fit of this component as the closest one to the value of the velocity in the model. Panels (e) and (f) in the bottom part of Figure~\ref{ionized_maps} show the best thin-disk model for {\it component 1} and the residuals after subtracting the best thin-disk model. The latter map indicates that the residuals fluctuate around zero randomly which confirms that the thin-disk model is a good approximation. The results by \citet{Vega_1997} pointed out the existence of a second component with a similar rotation pattern to the previous one. The spatial extension of this component ({\it component 2}) is more limited to the central part of the galaxy as it can be seen in panel (b) of Figure~\ref{ionized_maps}. There is also a third component ({\it component 3}) with a velocity close to the systemic, ranging from $-$ 60 km\,s$^{-1}$ to 85 km\,s$^{-1}$ relative to the systemic. This last component is spatially coincident with the regions previously characterized as deviations from the general rotation pattern (green and orange crosses in Figure~\ref{channel_map}).

\begin{figure*}
\centering
\includegraphics[trim={1cm 0cm 0.8cm 0cm}, width=43.5mm]{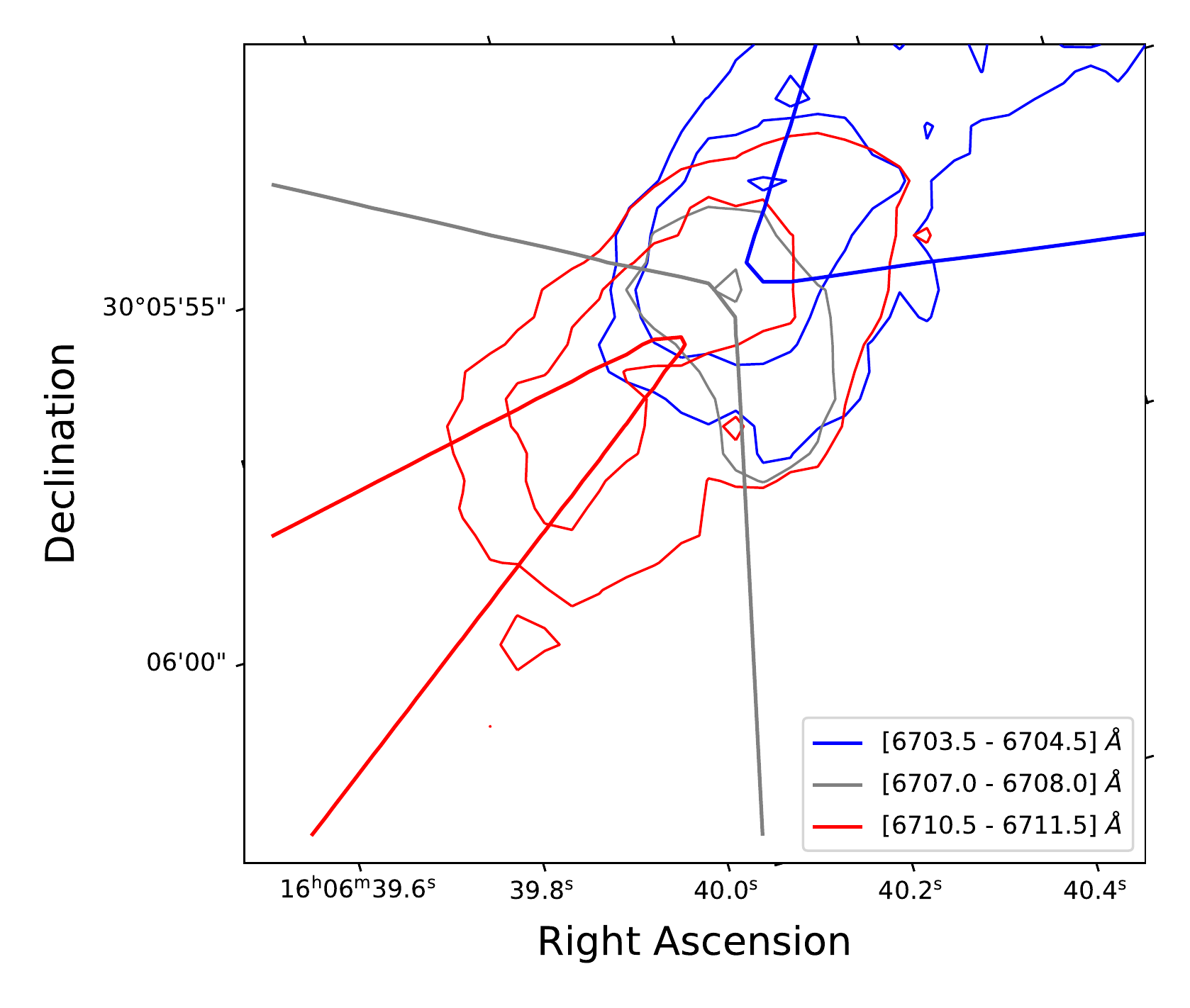}
\includegraphics[trim={0cm 0cm 0.8cm 0cm}, width=36mm]{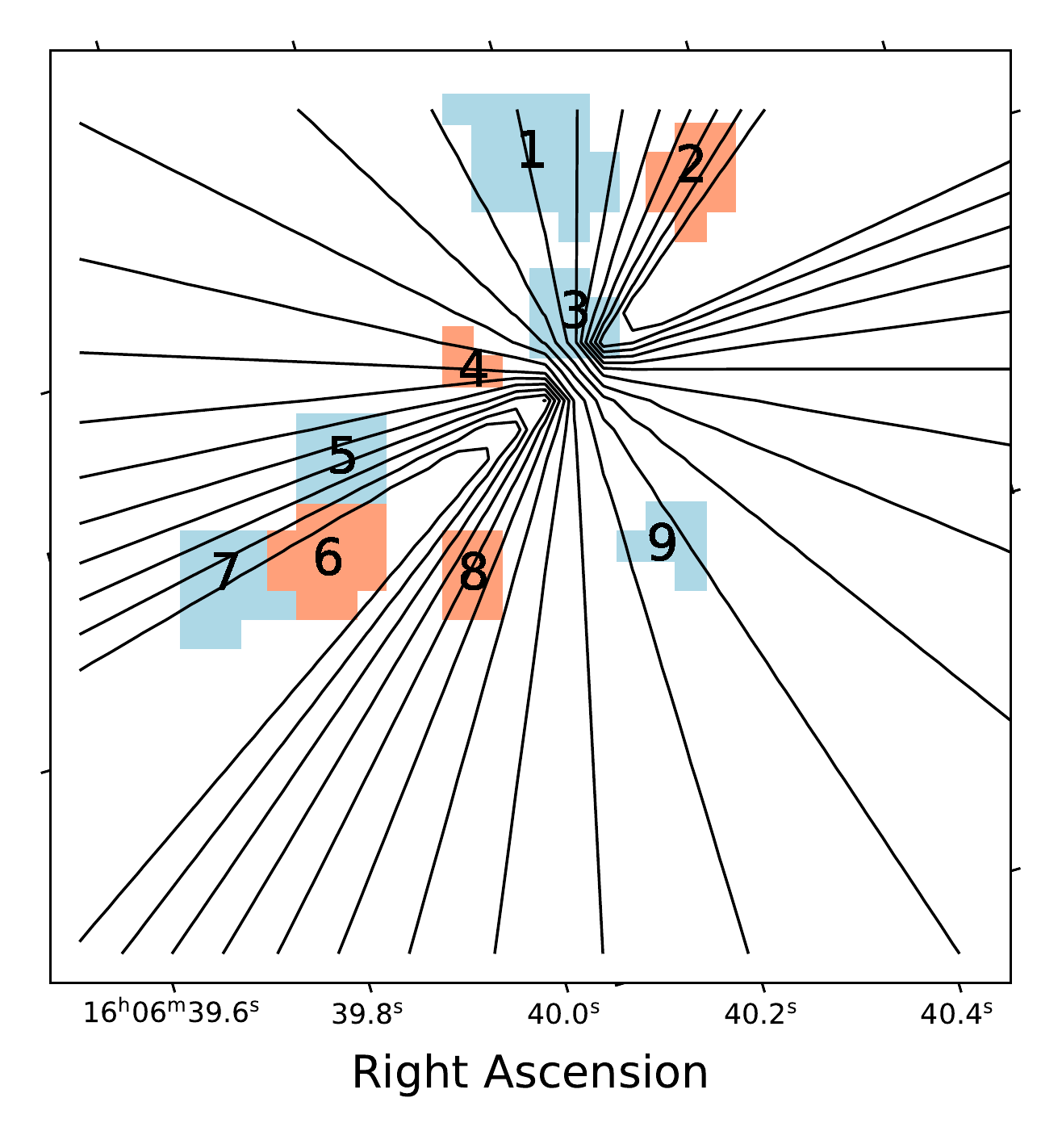} 
\includegraphics[trim={0cm 0cm 0cm 0cm},width=85mm]{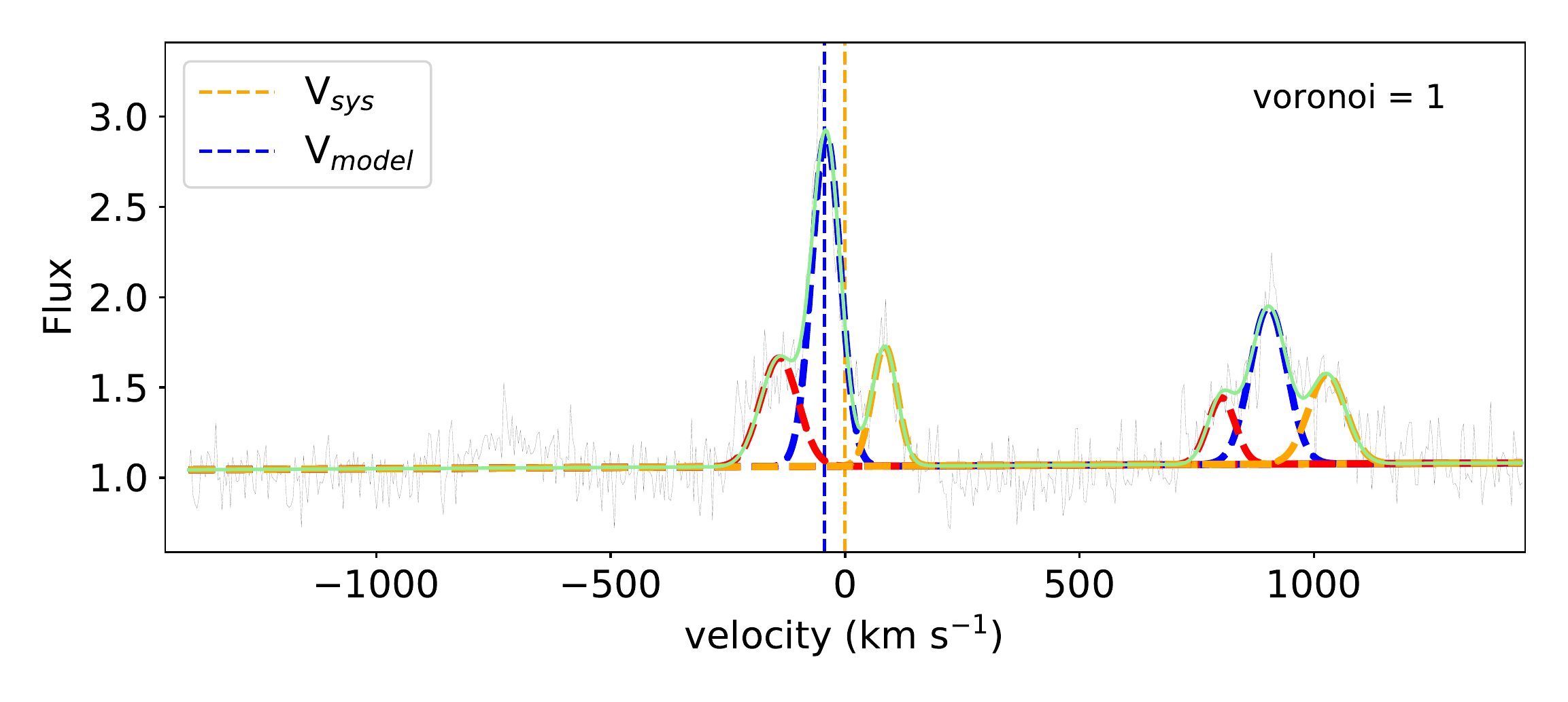} \\

\includegraphics[width=85mm]{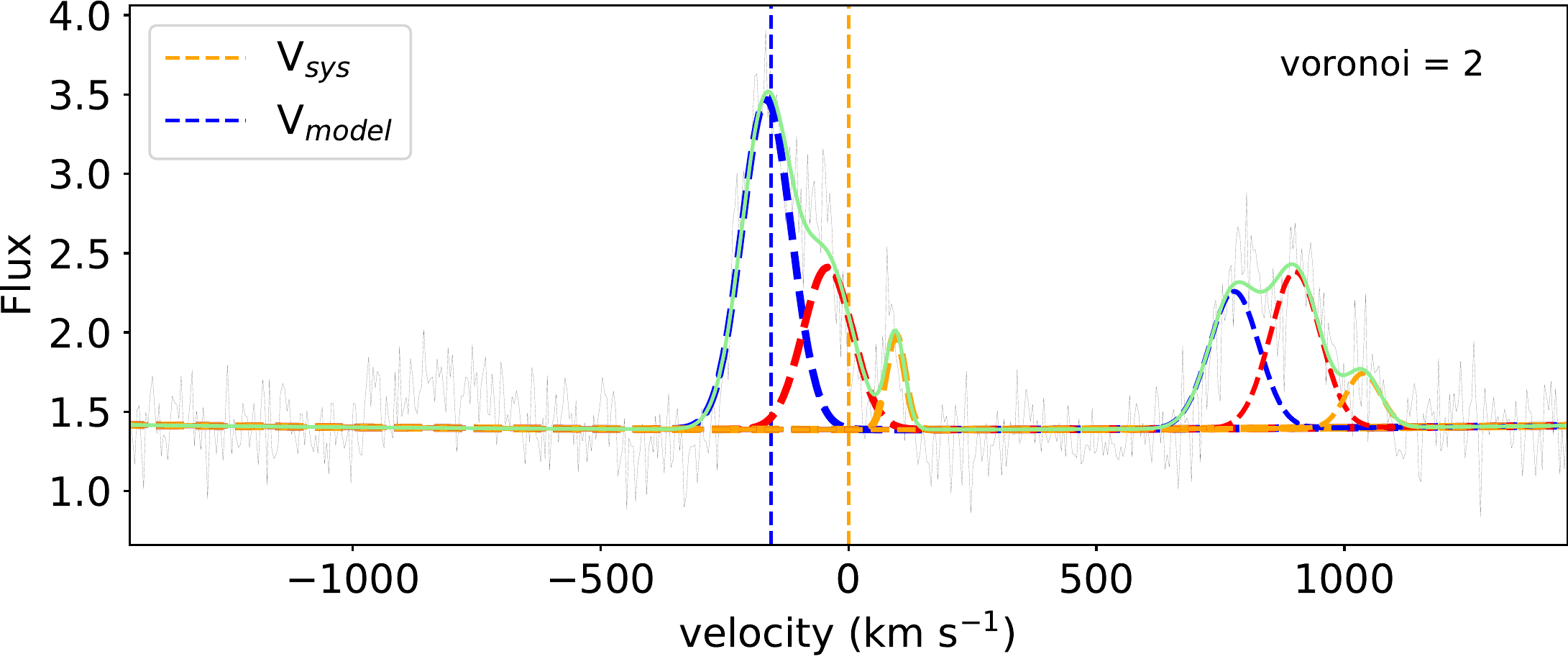} 
\includegraphics[width=85mm]{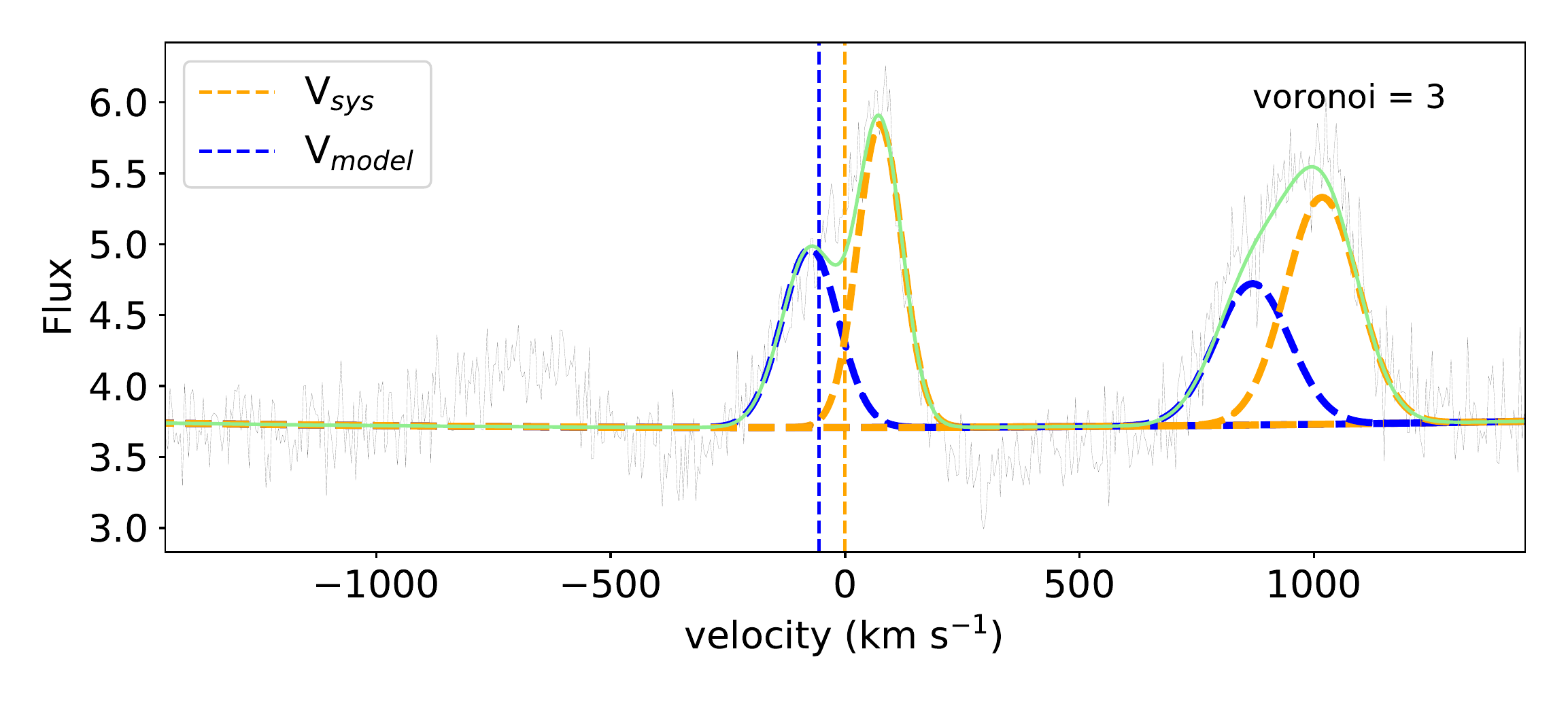} \\

\includegraphics[width=85mm]{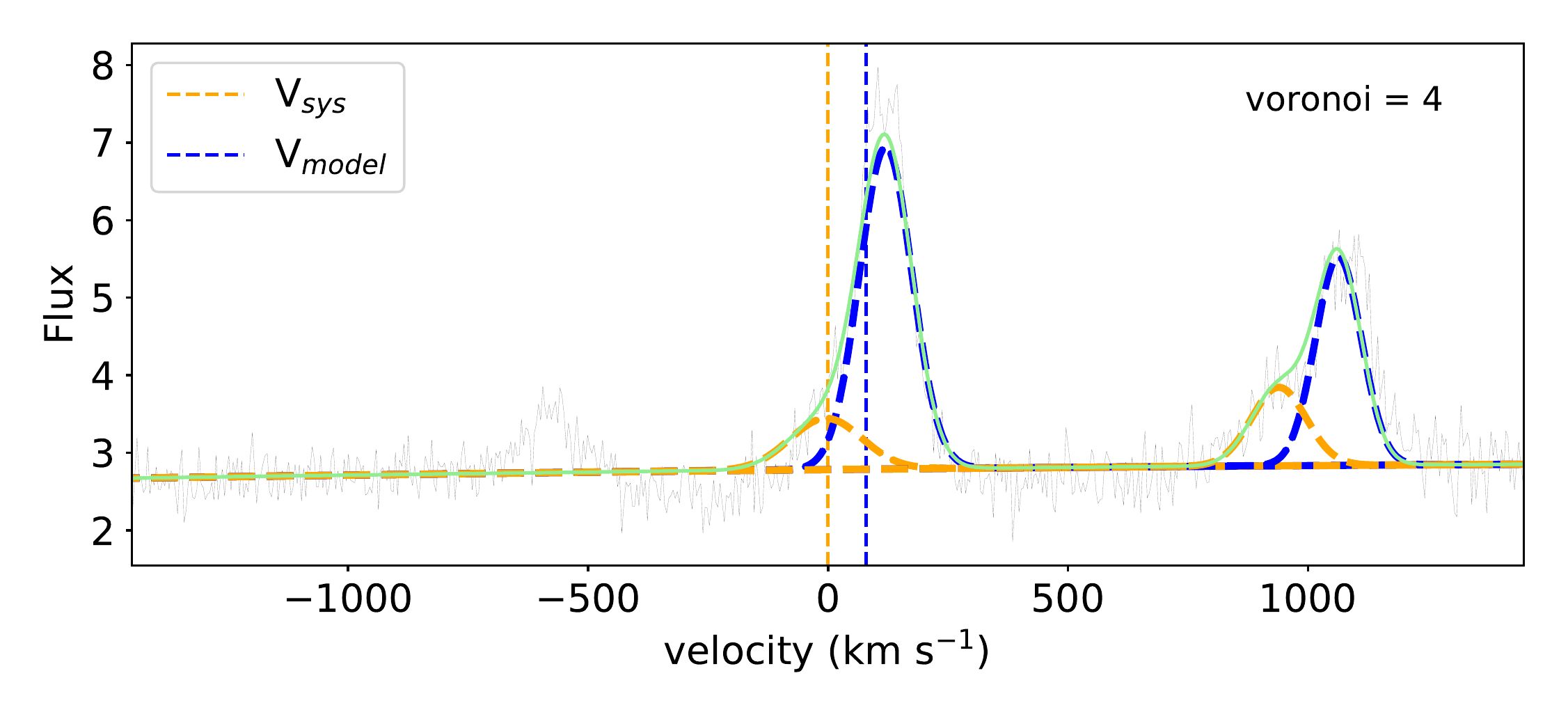} 
\includegraphics[width=85mm]{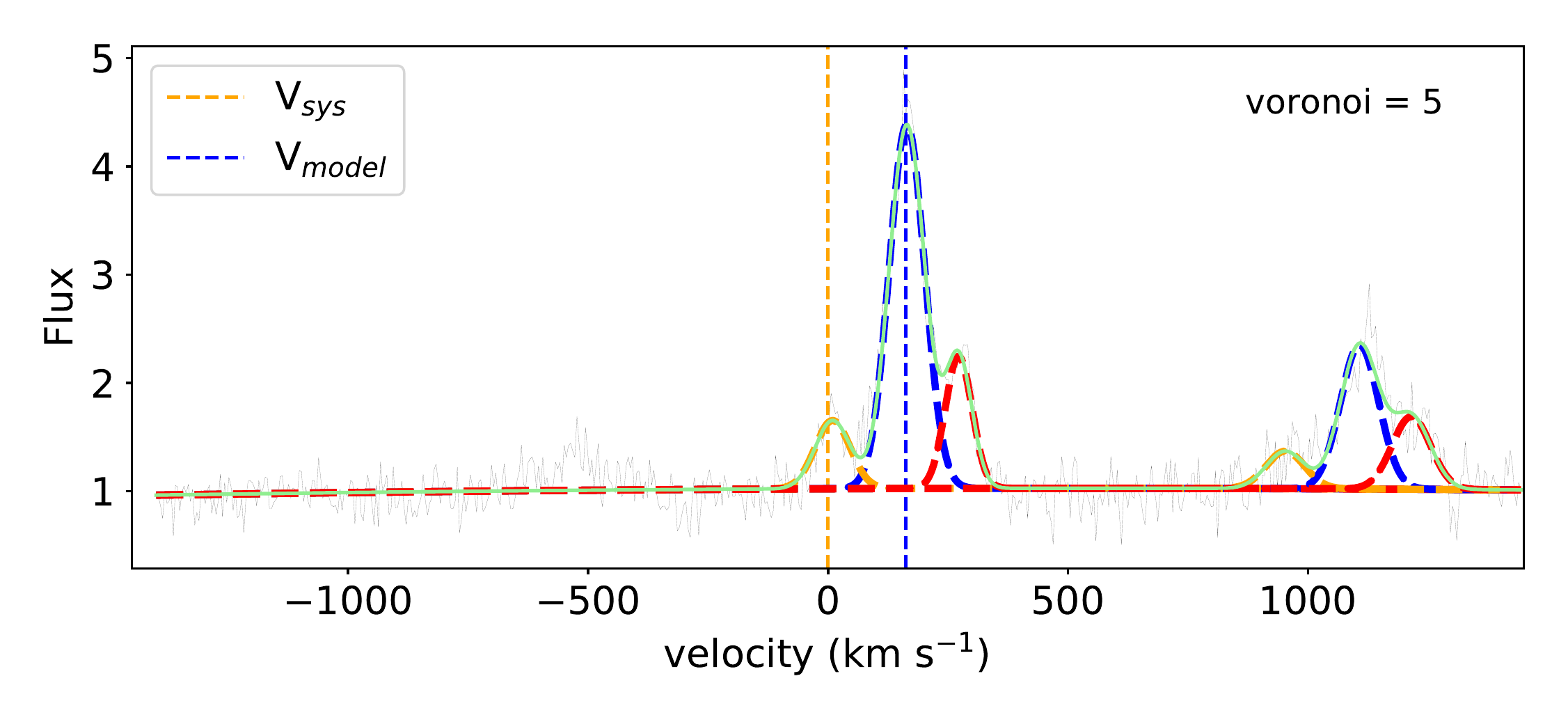} \\

\includegraphics[width=85mm]{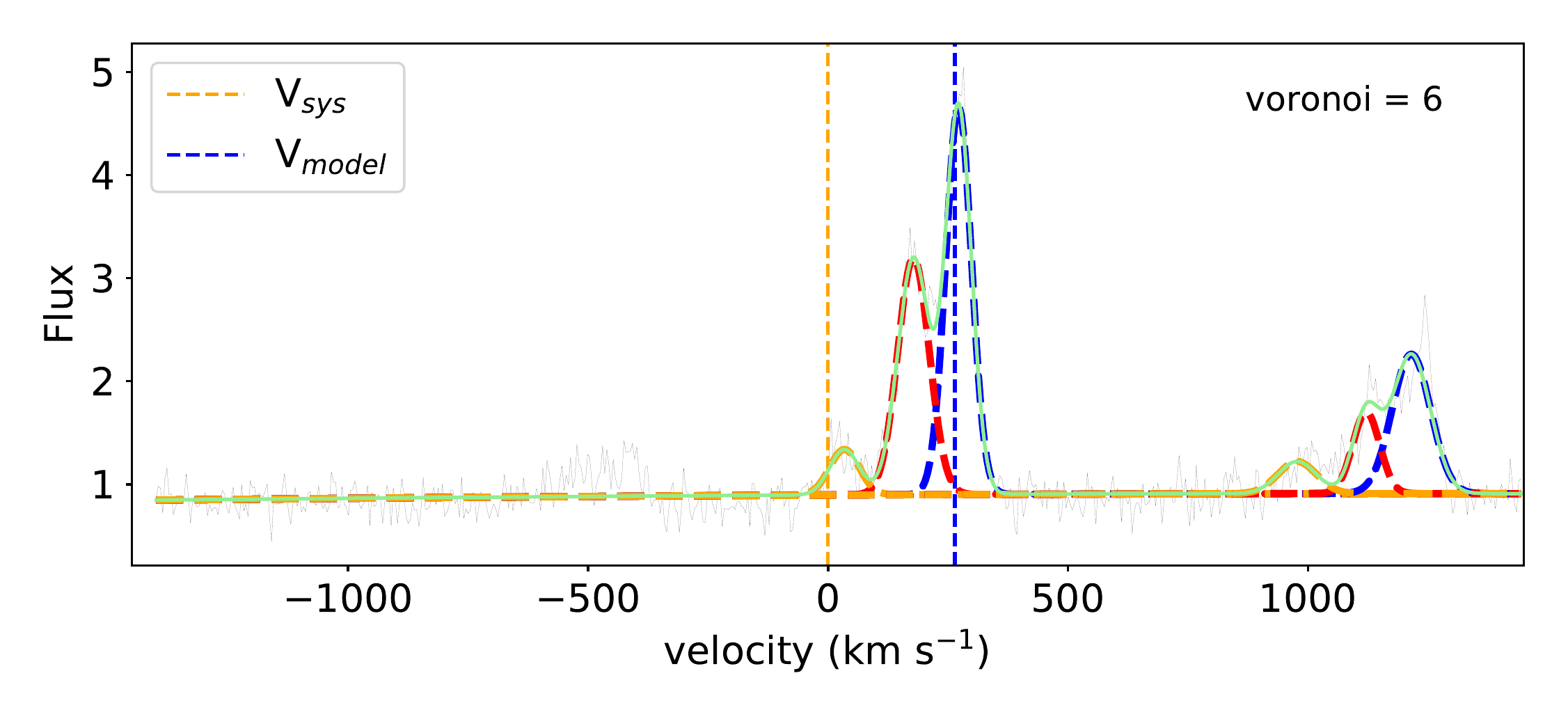} 
\includegraphics[width=85mm]{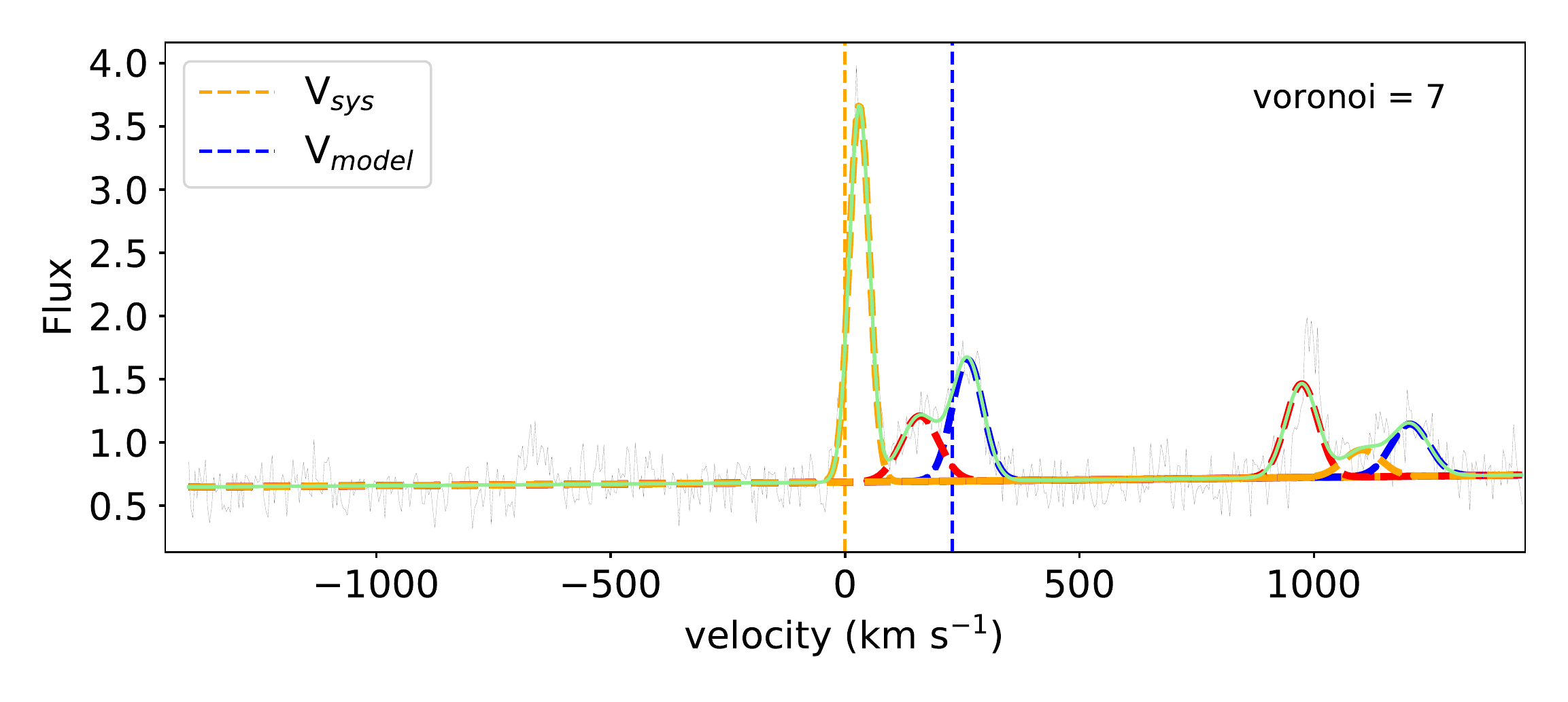} \\

\includegraphics[width=85mm]{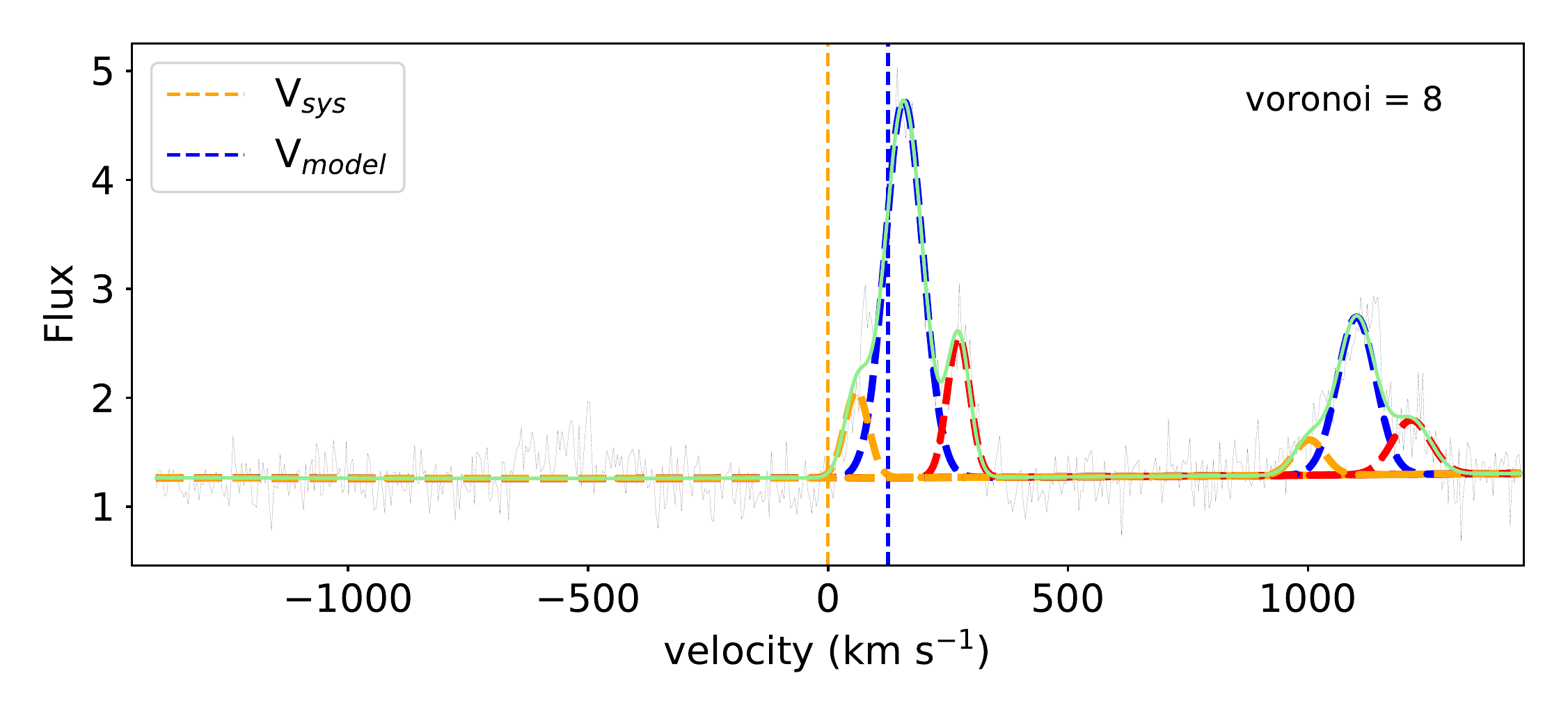} 
\includegraphics[width=85mm]{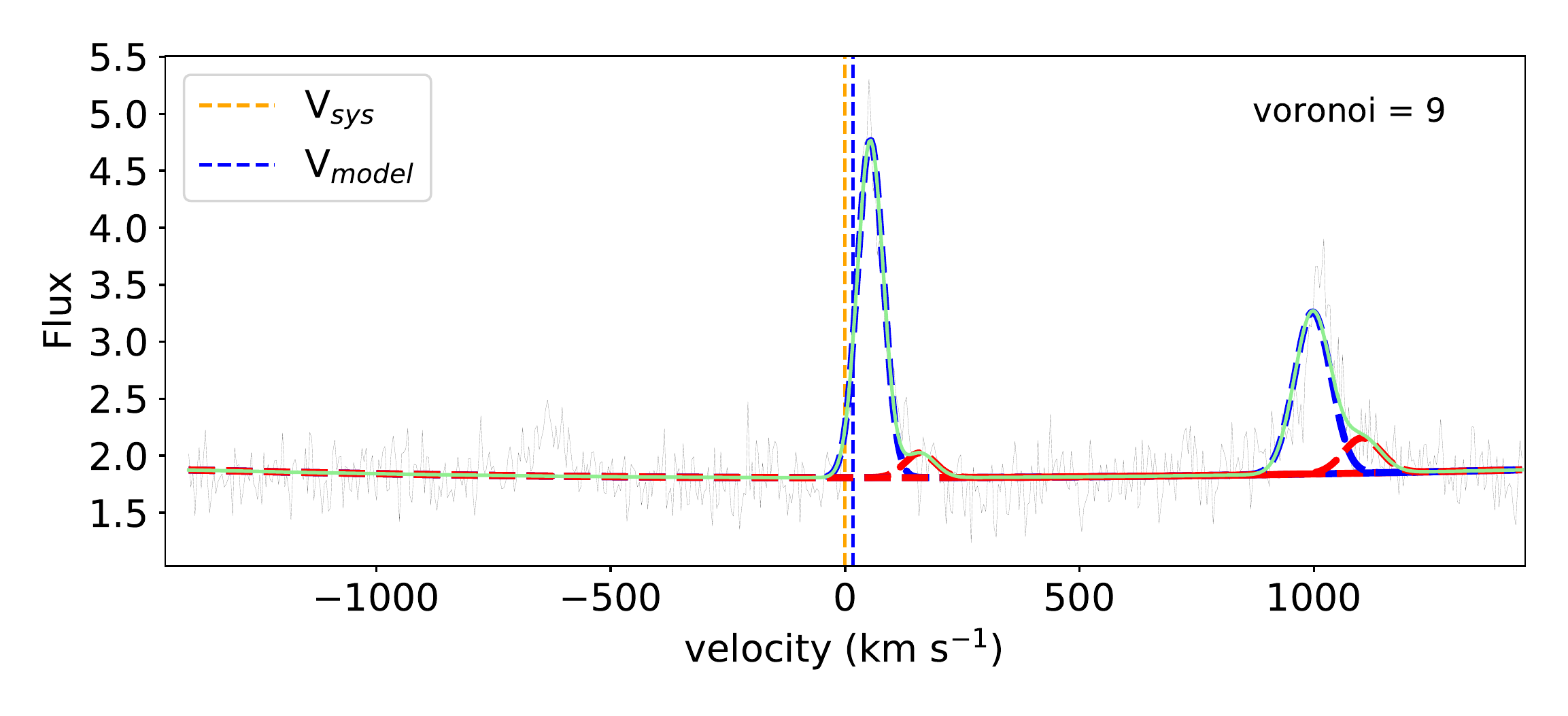} \\

\caption{Left top panel: H$\alpha$ flux contours together with the rotating thin disk model derived for the H$\alpha$ velocity field are shown as blue, grey and red colours for ranges [6703.5 $-$ 6704.5]\,\AA, [6707.0 $-$ 6708.0]\,\AA\,and [6710.5 $-$ 6711.5]\,\AA, respectively. Middle top panel: Isovelocity contours for the thin disk model ({\it i} $=$ 75$^{\circ}$) showing the location of some specific Voronoi cells within the datacube. The color coding assigned to the Voronoi cells is chosen for convenience, allowing a better visual inspection of their spatial extent within the map. Rest panels: Examples showing the presence of kinematically distinct gaseous H$\alpha$ components for the Voronoi cells indicated in the middle top panel. The zero velocity is set at the systemic redshift of the galaxy (orange dashed line) while the velocity of the thin-disk model for each particular Voronoi cell is shown by the blue dashed line. Gaussian fits represent the component associated with the thin-disk model (blue), the component with a similar rotation pattern to the previous one but restricted to a smaller region (red) and the component with a velocity close to the systemic (orange), respectively. The units of the horizontal axis are in km\,s$^{-1}$ relative to the systemic velocity of the galaxy.\\}
\label{examples_ha_profiles}
\end{figure*}

Finally, we select a few characteristic Voronoi cells to show the complexity of the emission line profiles together with the Gaussian fits of the three kinematically distinct components. The position of these Voronoi cells is shown with blue and orange colours in the middle panel of the top row in Figure~\ref{examples_ha_profiles}. Blue, red and orange Gaussian fits in the spectra of this figure correspond to {\it components 1}, {\it 2} and {\it 3} in Figure~\ref{ionized_maps}, respectively. The blue dashed line is the velocity of the thin-disk model while the orange dashed line represents the systemic velocity. Also, a significant rotational broadening for the H$\alpha$ profiles in the central regions can be seen in this figure (see panel corresponding to \mbox{voronoi $=$ 3}). As we will describe later in Section~\ref{adc_literature}, this effect is due to the steep rise of the rotational velocity curve in the innermost regions of this galaxy. We would expect to better constraint the internal dynamics of these regions in nearly face-on or less inclined systems where the $\sigma$ values are expected to be less broaden by the effect of rotation. In spite of the complex structure that follows the ionized gas distribution, our thin disk model is able to reproduce the kinematic pattern found in this object.

\subsubsection{Deviations from the pure thin disk model} \label{dust_lane_section}

As commented before, the distribution of the ionized gas in the low velocity channels (from 6704.5\,\AA\, to 6707.5\,\AA) reveals the presence of a kinematically distinct component that do not follow the rotation pattern of the gas-rich disk. To explain the gas velocity field in those particular regions, two possible interpretations arise.

On one hand, this dynamical component seems to be spatially coincident with the position of the dust lanes that might be driving the motions of the gas (see positions marked as green and orange crosses in Figure~\ref{channel_map}). Dust lanes are in most cases morphologically associated with a gas dense component even in early-type galaxies \citep{Finkelman_2010,Finkelman_2012}. Judging from the large-scale images available for UGC 10205 (see upper panel in Figure~\ref{ugc10205_megara_fov} as an example), the well-defined and prominent dust lanes of this object are highly inclined. Due to the reduced size of the MEGARA FoV, we are sampling a relatively small angle on the sky in comparison with the large-scale size of the object ($\sim$ 1 arcmin). Assuming a close to edge-on orientation for the dust lanes, we expect to recover angles ($\phi$) around 90$^{\circ}$ and 270$^{\circ}$ within this structure. Should that be the case, the difference between the radial velocity component of the gas in each Voronoi cell and the systemic velocity of the thin-disk model ({\it i} $=$ 75$^{\circ}$) should be nearly zero. Possible values for the inclination of the dust lane were investigated to check whether or not the velocities in these regions might be coherent with the structure of a disk model with higher inclination than the main thin disk model ({\it i} $=$ 75$^{\circ}$). An optimal value of {\it i} $=$ 83$^{\circ}$ was found.

\begin{figure}
\centering
\includegraphics[trim={0.15cm 0.15cm 0.15cm 0.15cmm},width=90mm]{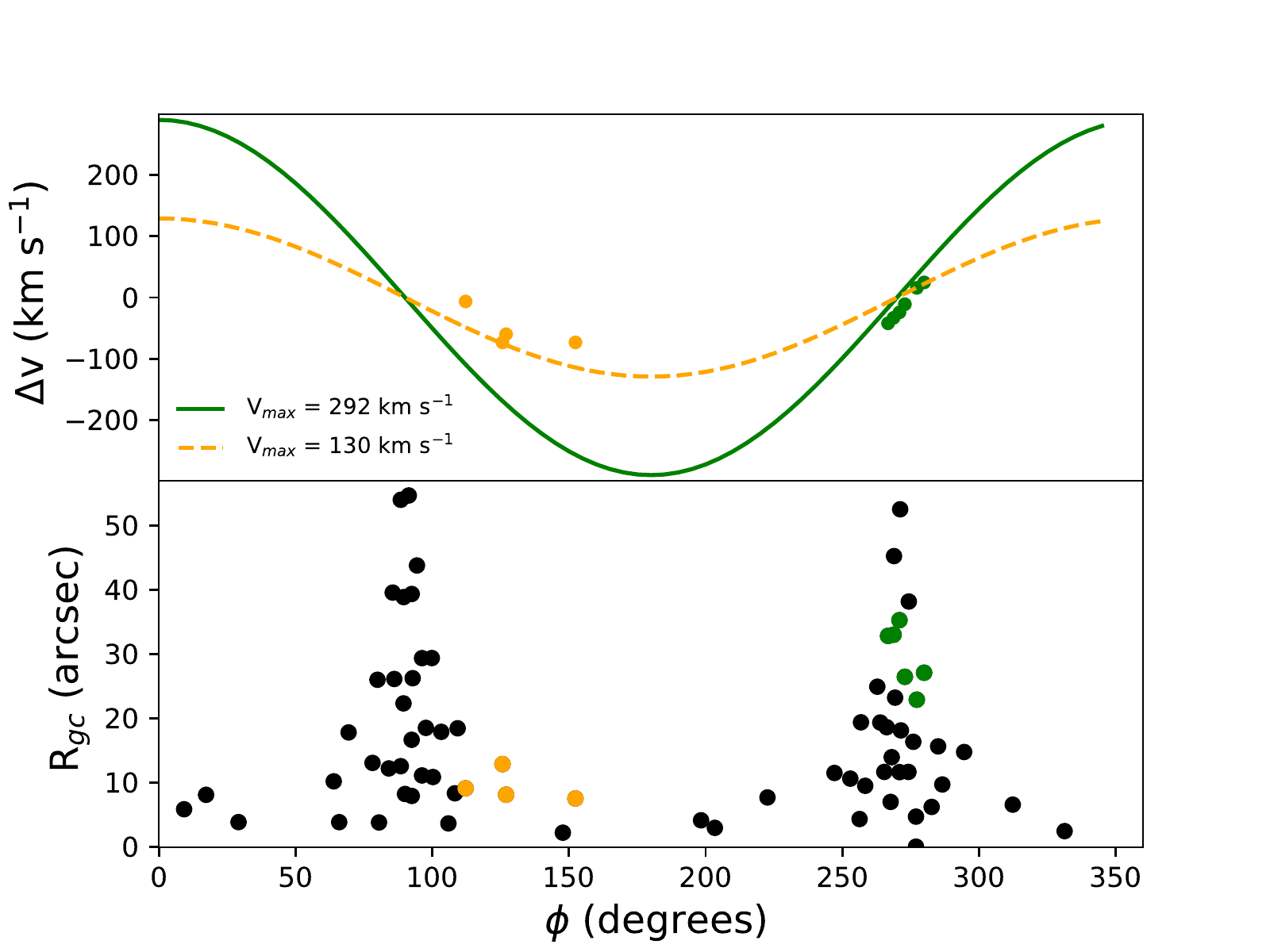}\\
\caption{Top panel: Velocity distribution in the plane of the dust lanes ({\it i} $=$ 83$^{\circ}$). Orange (approaching side) and green (receding side) points represent the difference between the radial velocity component of the gas and the systemic velocity of the thin-disk model ({\it i} $=$ 75$^{\circ}$) computed for each Voronoi cell ($\Delta$v) within the dust lane. Green solid line and orange dashed line are the velocity models with v$_{max}$ $=$ 292 km\,s$^{-1}$ and v$_{max}$ $=$ 130 km\,s$^{-1}$, respectively. Bottom panel: Distribution of the galactocentric distances and angles for the different Voronoi cells. Orange and green points represent the values of the ones located in the dust lane.\\}  
\label{dust_lane_plot}
\end{figure}

Thus, the bottom panel in Figure~\ref{dust_lane_plot} shows the distribution of the distances and angles probed by the Voronoi cells in the plane of the dust lanes assuming {\it i} $=$ 83$^{\circ}$ for this plane. The SE (approaching) side of the dust lane is represented by the orange points while the SW-to-NW (receding) side is marked by the green points. In particular, the points represent the centroid of each Voronoi cell associated with the regions where the line-of-sight passes through the plane of the dust lanes (positions also marked as green and orange crosses in Figure~\ref{channel_map}). 

Additionally, the difference between the velocity of this gas component and the systemic velocity of the thin-disk model ({\it i} $=$ 75$^{\circ}$) computed for each Voronoi cell is shown in the upper panel of Figure~\ref{dust_lane_plot} as green/orange points. As expected, this difference in velocity ($\Delta$v) is nearly zero for $\phi$ $\sim$ 90$^{\circ}$ and $\phi$ $\sim$ 270$^{\circ}$. As an example of a particular Voronoi cell, we include the spectrum corresponding to voronoi = 7 (Figure~\ref{examples_ha_profiles}). In this spectrum, two different features can be clearly seen: (i) a component almost at the systemic velocity of the model and (ii) the component of the thin-disk model. The former one would be the gas component associated with the dust-lane structure. 

At the galactocentric distances corresponding to the positions of the receding side of the dust lane ($\sim$ 30 arcsec), we expect to reach the maximum velocity for the thin disk model (v$_{max}$ $=$ 292 km\,s$^{-1}$, green solid line). \mbox{{\it i} $=$ 83$^{\circ}$} is the inclination of the dust-lane that minimizes the distance between the model and $\Delta$v. For the case of the SE side of the dust lane (orange points), the galactocentric distances are $\sim$ 10 arcsec so the velocity curve has still not reach its maximum velocity. A velocity of 130 km\,s$^{-1}$ (orange dashed line) is the best value consistent with the inclination derived previously. As a conclusion, having a slightly more inclined dust-lane ({\it i} $=$ 83$^{\circ}$) with the same galaxy's potential than the thin disk model ({\it i} $=$ 75$^{\circ}$) seems to be a reasonable good explanation for these regions.

On the other hand, the origin of this component could be associated with the merging scenario that creates a localized enhancement of the star formation with a distinct kinematic pattern and the gas might be tracing non-circular motions in those regions. The numerical model presented by \citet{Reshetnikov_1999} relied on the tidal disruption of a small E/SO object by the massive early-type galaxy. Non-circular gas motions in the nuclear region of UGC 10205 were suggested by these authors. However, the nice match between both the $\Delta$v values and its azimuthal variation with respect to the one of highly-inclined dust lanes seems to favor the former scenario.

\subsection{Resolved stellar kinematics} \label{resolved_stellar_kinematics}

\begin{figure}
\centering
\includegraphics[trim={1.5cm 0cm 0.5cm 0.5cm},width=100mm]{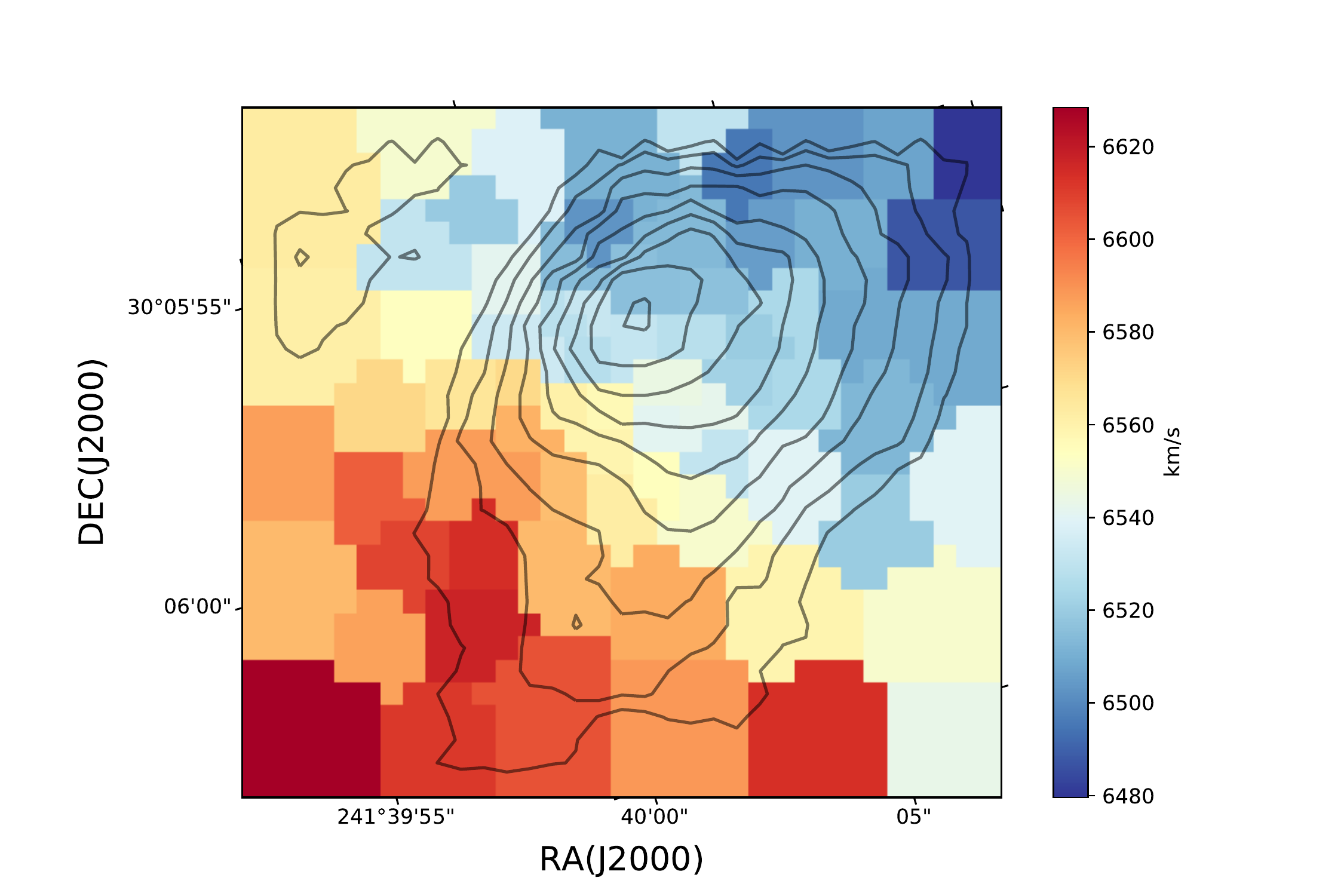}
\includegraphics[trim={1.5cm 0cm 0.5cm 0.5cm},width=100mm]{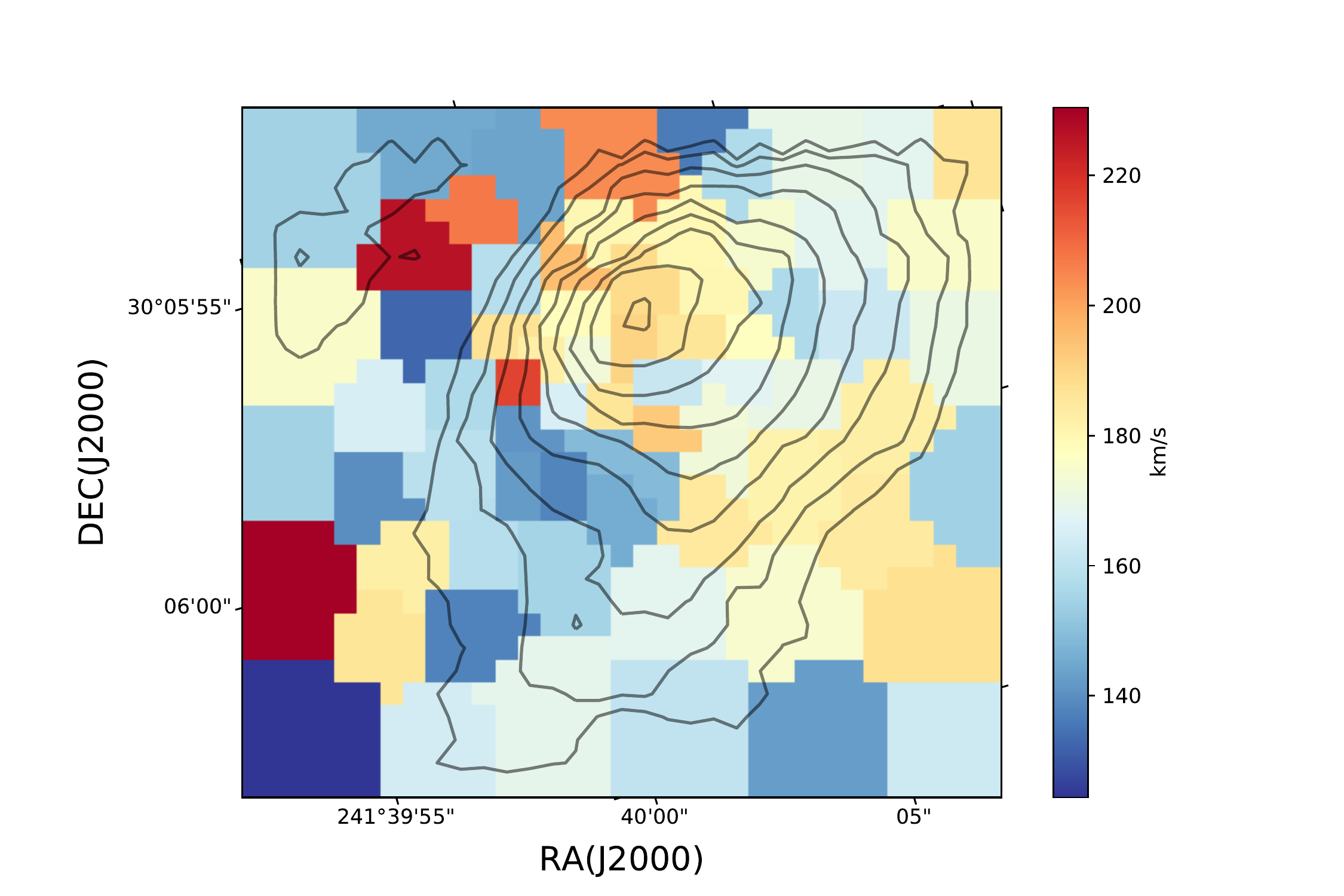} \\
\caption{Top panel: Stellar velocity field for UGC 10205. Note that the galaxy systemic velocity is reported to be 6556 km\,s$^{-1}$. Bottom panel: Stellar velocity dispersion. Contours of the continuum emission (black) are overlaid on both maps. Color scales and units are indicated on the right part of each map.\\} 
\label{velocity_map_LRV}
\end{figure}

The MEGARA absorption-line kinematics for the central regions of the galaxy UGC 10205 are presented in Figure{~\ref{velocity_map_LRV}}. This figure shows both the stellar velocity field and the stellar velocity dispersion maps derived from our pPXF analysis. pPXF yields the values of the stellar mean velocity and stellar velocity dispersion for each Voronoi cell applying a S/N rejection criterion of 20. The reader is referred to Section \ref{LR-V grating} where the main procedure followed to create these maps is explained with detail.

The stellar velocity map (upper panel in Figure{~\ref{velocity_map_LRV}}) shows a clear stellar rotation pattern with a symmetric velocity field respect to the galaxy center and with a maximum radial velocity of $\sim$ 6628 km\,s$^{-1}$. Typical errors in our stellar velocity measurements range from 4 to 23 km\,s$^{-1}$ with mean errors of 10 km\,s$^{-1}$. To check the accuracy of our kinematic analysis a visual inspection is performed using the results presented in \citet{Falcon_Barroso_2017}. These authors found a similar stellar rotation pattern using the mid-resolution CALIFA data (V1200 with R $\sim$ 1650) in the wavelength range 3850 $-$ 4600\,\AA. 

The stellar velocity dispersion map (bottom panel in Figure{~\ref{velocity_map_LRV}}) displays values ranging from 124 to 230 km\,s$^{-1}$, with values in the nuclear region $\sim$ 190 km\,s$^{-1}$. These values are already corrected for the instrumental resolution, which in our case is \mbox{$\sigma$ $\sim$ 21 km\,s$^{-1}$}. Although no signs of kinematically decoupled components appeared to be present, we have performed the analysis of the Gauss-Hermite moment {\it h3} (skewness of the velocity profile) to examine this effect. The {\it h3} map (not shown) indicates that we can safely discard the presence of an anti-correlation between these values and the stellar velocities ones.

\subsection{Estimating the asymmetric drift correction}

By applying an asymmetric drift correction (ADC) it is possible to account for the effect of the stellar random motions within galaxies. Basically, the mean tangential velocity of a stellar population lags the actual circular velocity as defined by the gravitational potential of the galaxy. ADC accounts for this difference. The effect of the ADC phenomenon will be more significant in those cases where the velocity dispersion of different stellar populations is higher. Although this effect was first noted in the solar neighborhood \citep{Stromberg_1924,Stromberg_1925}, asymmetric drift corrections have been extensively used in the literature to reconcile the gas and stellar rotation curves observed in external galaxies. 

The Stellar Velocity Ellipsoid (SVE) is described in cylindrical coordinates by its radial, tangential and vertical components: $\sigma$$_{R}$, $\sigma$$_{\phi}$ and $\sigma$$_{Z}$. The shape of the SVE could be parameterized by the axial ratios $\sigma$$_{Z}$/$\sigma$$_{R}$, $\sigma$$_{\phi}$/$\sigma$$_{R}$ and $\sigma$$_{Z}$/$\sigma$$_{\phi}$. For the purpose of this work, we focus our attention on the $\alpha$ $=$ $\sigma$$_{Z}$/$\sigma$$_{R}$ and $\beta$ $=$ $\sigma$$_{\phi}$/$\sigma$$_{R}$ parameters. In particular, \citet{Mogotsi_2018} found a global vertical-to-radial velocity dispersion ratio of \mbox{$\alpha$ $=$ 0.97 $\pm$ 0.07} for UGC 10205. Similar values of the radial and vertical dispersions point out to an isotropic heating.

Here, we assume that the stellar azimuthal anisotropy ($\beta$$_{\phi}$) can be approximated using the epicycle anisotropy, i.\,e., the stellar orbits are considered nearly circular \citep{Gerssen_1997, Gerssen_2000, Shapiro_2003, Ciardullo_2004, Westfall_2011, Gerssen_2012, Blanc_2013, Gentile_2015}. Thus, we measure $\beta$$_{EA}$ (hereafter $\beta$ for simplicity) using the following expression:
\begin{equation}
\beta = \sqrt{ \frac{1}{2}  \left( \frac{\partial ln(v_{\phi})}{\partial ln(R)} + 1 \right)}
\end{equation}
where v$_{\phi}$ is the tangential speed of the stars. We found a mean value of $\beta$ $=$ 0.86 $\pm$ 0.12. Due to the small dispersion found and in order to simplify the calculations, we assume the previous value for all the Voronoi cells.

Once $\alpha$ and $\beta$ are known, we infer $\sigma$$_{R}$ in each Voronoi cell using the values of $\sigma$$_{LOS}$ previously obtained (bottom panel of Figure~\ref{velocity_map_LRV}):
\begin{equation}
\sigma_{LOS}^{2} = \sigma_{R}^{2} \sin \phi^{2} \sin i^{2} + \sigma_{\phi}^{2} \cos \phi^{2} \sin i^{2} + \sigma_{Z}^{2} \cos i^{2}
\end{equation}
where $\phi$ is the angle of the central position of each Voronoi cell in the plane of the galaxy measured from the line of nodes and {\it i} refers to the inclination. We assume the same inclination for each Voronoi cell, i.\,e., in this case we adopt the same inclination for the most inner regions of the galaxy. Due to the agreement between the ionized gas rotation axis and the stellar rotation axis with a difference of \mbox{(6 $\pm$ 3)$^{\circ}$}, we use an inclination of 75$^{\circ}$ for the ADC.

To estimate the ADC in each Voronoi cell, we apply the following expression as reproduced from equation (4-33) of \citet{Binney_1987}:
\begin{equation}
\small
V_{C}^{2} = v_{\phi}^{2} - \sigma_{R}^{2} \left[\frac{\partial ln(\nu_{R})}{\partial ln(R)} +  \frac{\partial ln(\sigma_{R}^{2})}{\partial ln(R)} + 1 - \frac{\sigma_{\phi}^{2}}{\sigma_{R}^{2}} 
+ \frac{R}{\sigma_{R}^{2}}  \frac{\partial (v_{R} v_{Z})}{\partial Z} \right]
\label{adc_equation}
\end{equation}
This expression shows the difference between the true velocity (i.e., V$_{C}$ is the circular velocity of a test particle in the potential) and the observed circular velocity. If the SVE is aligned with the cylindrical coordinate system (R,$\phi$,z), the last term in equation~\ref{adc_equation} could be neglected.

The term $\nu$$_{R}$ in equation~\ref{adc_equation} is proportional to the stellar mass surface density ($\Sigma$) under the assumption that $\Sigma$ is well traced by the surface brightness profile of the galaxy. In this case, to trace the radial dependence of $\nu$ we apply a one dimensional (1D) photometric decomposition over the Panoramic Survey Telescope and Rapid Response System \citep[Pan-STARRS,][]{Panstarrs_2016} {\it y}-band light profile. Among all the Pan-STARRS bands, the near-infrared one is considered the best tracer of the stellar mass as the light is mainly dominated by old MS and RGB stars. Also, problems associated with the interstellar dust (as shown by the presence of prominent dust lanes in UGC 10205) are partly avoided. The fit was performed following the procedure explained in \citet{Dullo_2019}. We discard non-symmetric features due to bars or spiral arms in the fitting process. The bulge component is parameterized using a S\'ersic profile \citep{Sersic_1968} with a best-fitting S\'ersic index {\it n} $=$ 1.48 $\pm$ 0.22 and a half-light radius {\it r$_{e}$} $=$ 4.10 $\pm$ 0.81 arcsec. The galaxy disk component is described with an exponential profile with scale length {\it h} $=$ 10.8 $\pm$ 1.3 arcsec.

\subsection{Asymmetric drift correction: comparison with previous data in the literature} \label{adc_literature}

\begin{figure}
\centering
\includegraphics[trim={1cm 0cm 1cm 0cm}, width=90mm]{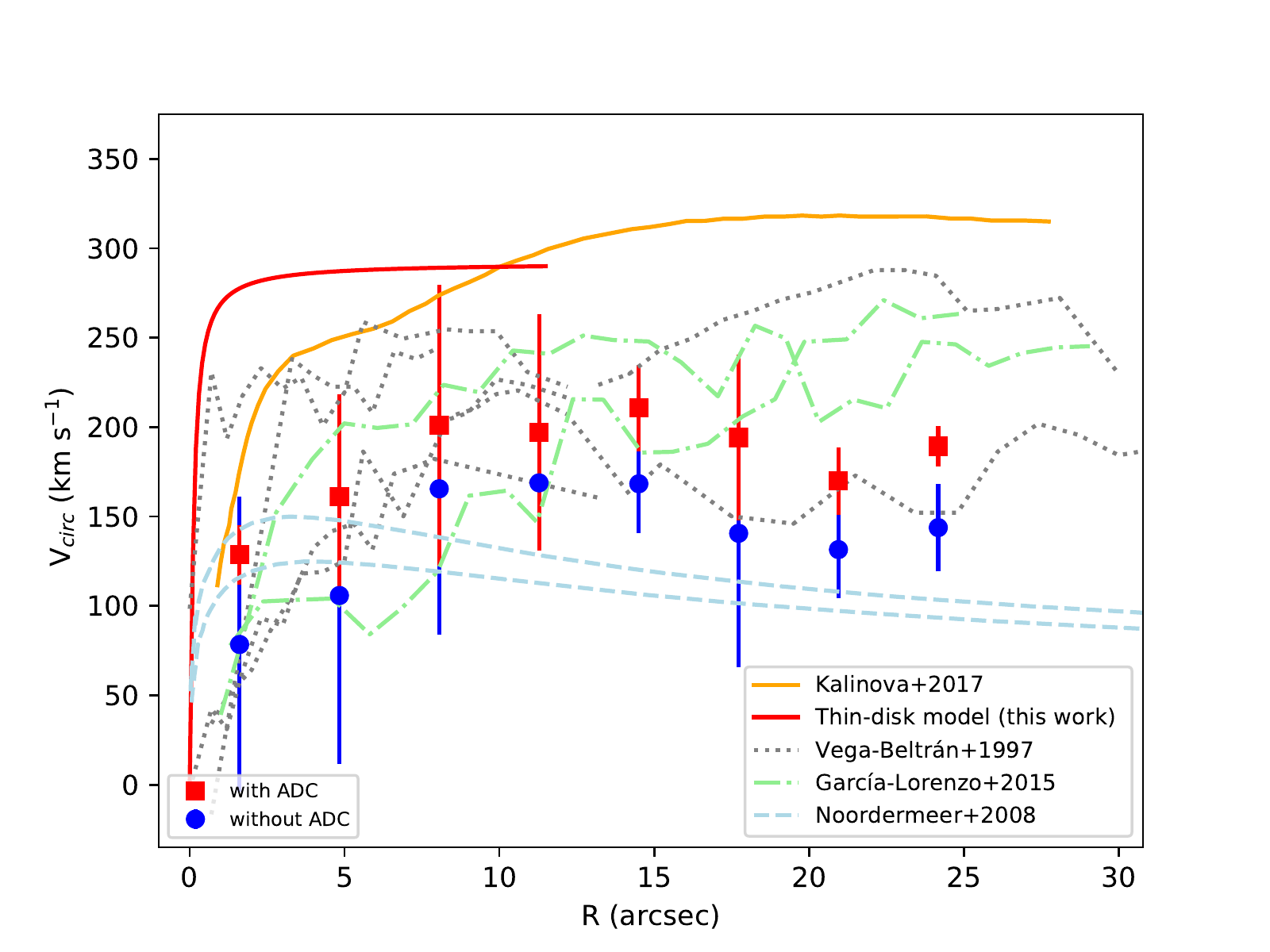}
\caption{Stellar (blue circles) rotation curve for UGC 10205. The ionized gas rotation curves (red squares) after applying the ADC under the assumption of an inclination {\it i} $=$ 75$^{\circ}$. The light-blue dashed lines are the predictions for the rotation curves of a S\'ersic bulge with {\it n} $=$ 2  and different values of the axis ratio {\it q} $=$ 0.2 (top) and {\it q} $=$ 1 (bottom) derived by \citet{Noordermeer_2008} and scaled up to match the mass and the effective radius of our bulge component. The best rotation thin disk model is represented by the red solid line. The observed velocity curve values from \citet{Vega_1997} and \citet{Garcia_Lorenzo_2015} are shown in dotted gray lines and dotted-dashed green lines, respectively. The velocity curve profile measured by \citet{Kalinova_2017} is also shown (orange line).\\}  
\label{ADC_graph_final}
\end{figure}

Figure~\ref{ADC_graph_final} shows the results obtained after applying the ADC to the stellar component. For comparison, the predictions for the rotation curves of S\'ersic bulges obtained by \citet{Noordermeer_2008} are also shown. These curves have been scaled up to take into account the mass and the effective radius of the bulge component. The effective radius, obtained directly from our 1D decomposition, is {\it r$_{e}$} $=$ 4.10 $\pm$ 0.81 arcsec ({\it r$_{e}$} $\sim$ 1.97 kpc). For the stellar mass derivation, we apply the bulge-to-total flux ratio (B/T $=$ 0.19 $\pm$ 0.3) obtained as a result of the 1D decomposition and the total stellar mass of the galaxy derived in \citet{Walcher_2014} (log(M$_{\star}$) $=$ 10.997 $\pm$ 0.111 M$_{\sun}$). As the 1D decomposition yields {\it n} $=$ 1.48 $\pm$ 0.22, we represent the rotation curve for a fiducial bulge with a S\'ersic concentration parameter {\it n} $=$ 2 as the best approximation in our case. For the same {\it n} $=$ 2, blue dashed line curves correspond to variations of the axis ratio {\it q} $=$ 0.2 (top) and {\it q} $=$ 1 (bottom) in Figure~\ref{ADC_graph_final}. It is expected that the rotation curves from \citet{Noordermeer_2008} are a lower limit as they only consider the bulge mass component but our V$_{C}$ measurements also trace the mass in the inner disk of the galaxy.

We have also included in Figure~\ref{ADC_graph_final} some observational data found in the literature for this galaxy. In particular, the stellar rotation curve derived in \citet{Kalinova_2017} is shown in orange. In the new classification proposed by these authors, the rotation curve of UGC 10205 is classified as {\it round-peaked} meaning that {\it the circular velocity rises steeply and has a round peak, gradually changing to the flat part of the circular velocity curve (CVC) at radius $\sim$ 0.5 R$_{e}$}, where R$_{e}$ is the value they derived for the whole galaxy. They also found that this class is the most common one among early-type galaxies (E4$-$S0a) and spiral galaxies classified as Sa$-$Sbc. The typical value expected for the average amplitude of the CVC is around 300 km\,s$^{-1}$, similar to the v$_{max}$ value derived in Section~\ref{gas_kinematics_section} for the thin-disk model.

\begin{figure*}
\centering

\includegraphics[trim={1.5cm 0cm 1cm 0cm}, width=85mm]{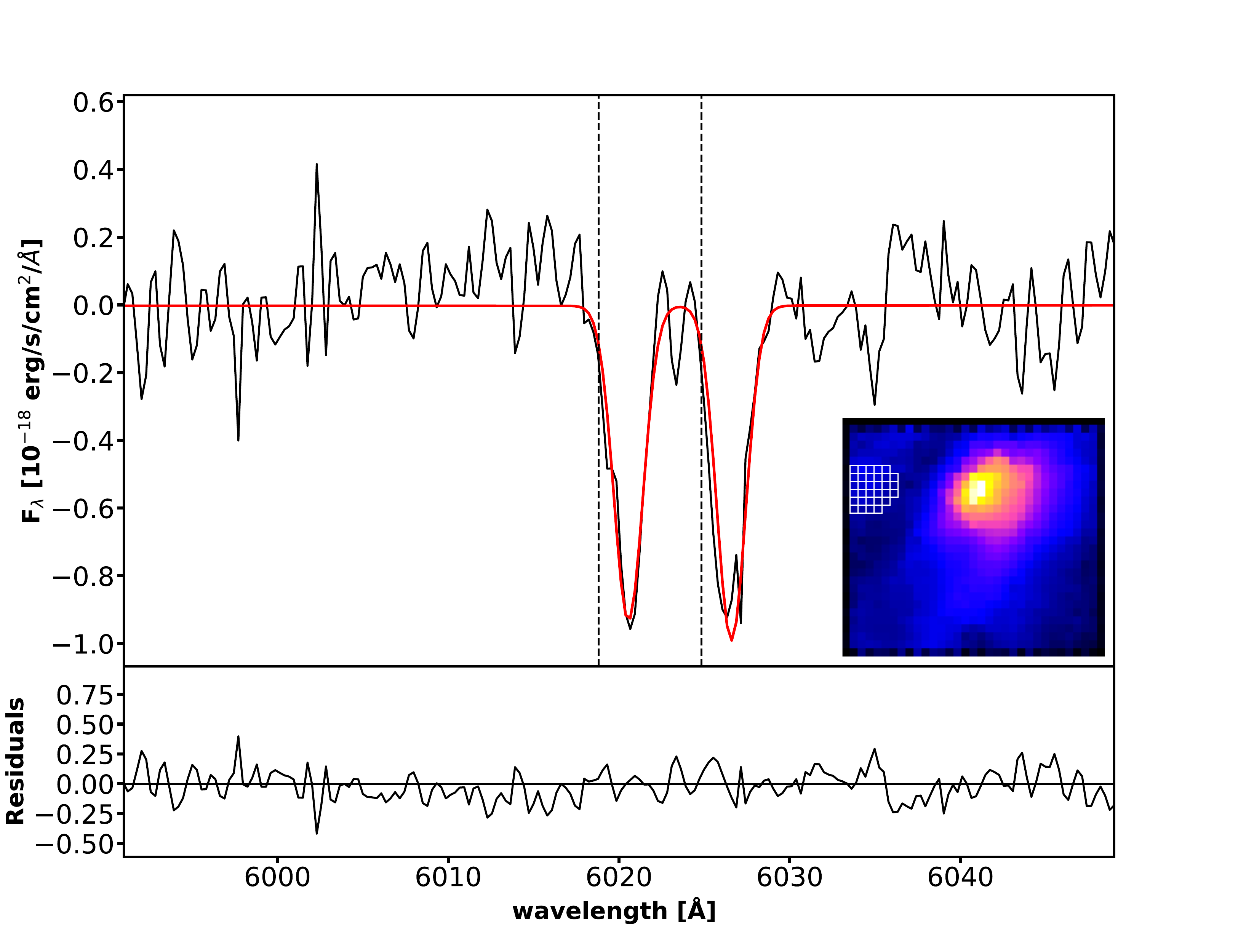}
\includegraphics[trim={1cm 0cm 1.5cm 0cm}, width=85mm]{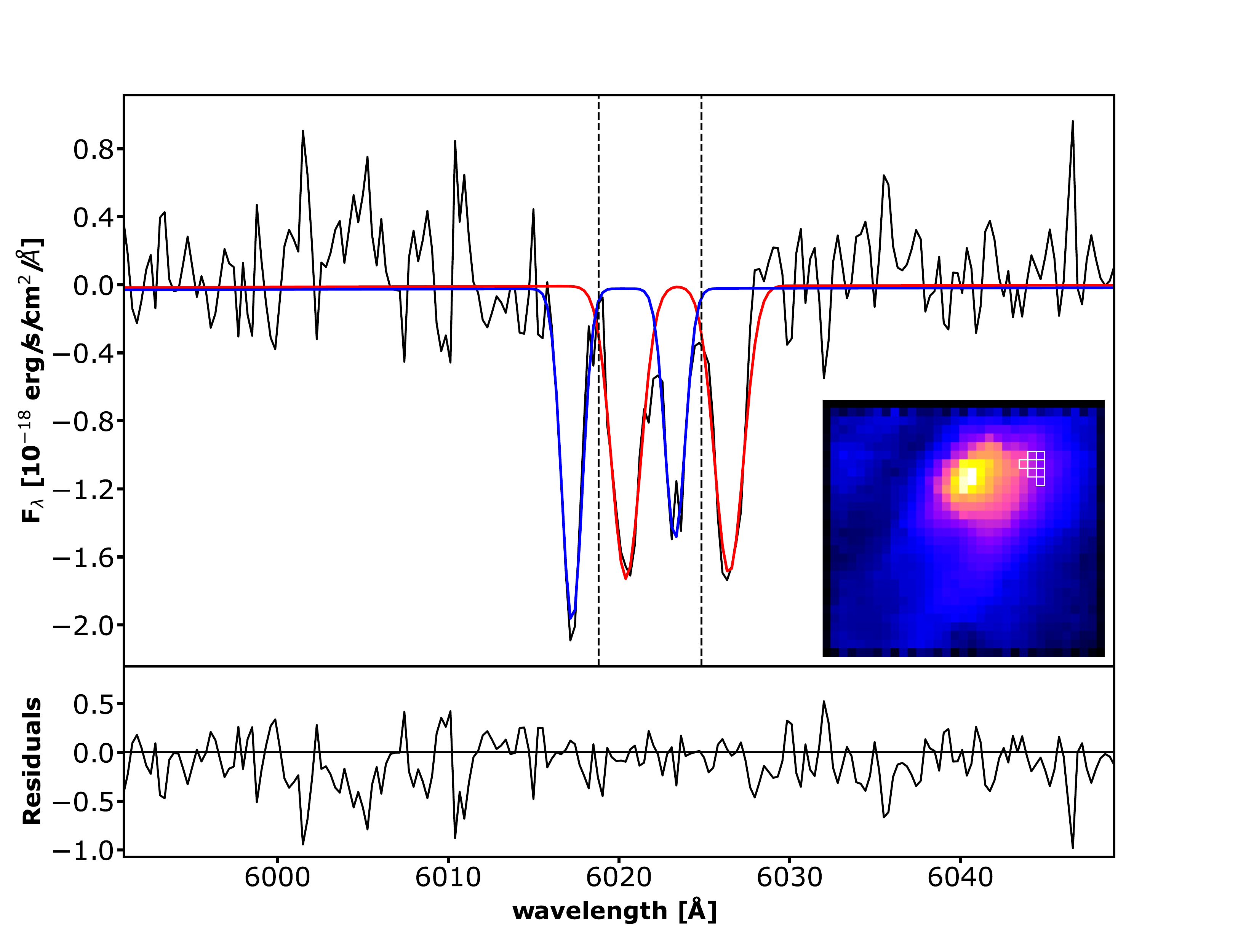} \\

\caption{Examples of the ISM absorption \mbox{Na {\scriptsize I} D} line profiles for UGC 10205. The insets show the position of the Voronoi cells used to obtain each spectrum. The vertical dashed lines represent the wavelengths corresponding to the \mbox{Na {\scriptsize I}} $\lambda$$\lambda$\,5890, 5896 at the systemic velocity of the galaxy. In the right panel, the thick blue line shows the best fit to the individual narrow Gaussian components for the blue-shifted component of the \mbox{Na {\scriptsize I} D} doublet that traces the outflow. The red line is the fit to the red-shifted \mbox{Na {\scriptsize I} D} absorption signatures that appears as an indicator of gas moving towards the galaxy as an infalling gas. In the left panel only the presence of the red-shifted \mbox{Na {\scriptsize I} D} absorption could be seen.\\}  
\label{nad_profiles}
\end{figure*}

On the other hand, measurements of the ionized gas velocity rotation curve are also included. As the thin-disk model derived in Section~\ref{gas_kinematics_section} was recovered using mainly the inner regions of the galaxy, we include here other literature data that cover the whole galaxy. The velocity curve based on the H$\alpha$ + [NII] emission-line data obtained by \citet{Garcia_Lorenzo_2015} within CALIFA appears as the green dot-dashed lines in Figure~\ref{ADC_graph_final}. 
The UGC 10205 major-axis kinematic curves derived by \citet{Vega_1997} for the multiple H$\alpha$ components found on this object using long-slit spectroscopy data are shown by the gray dotted lines. Both \citet{Garcia_Lorenzo_2015} and \citet{Vega_1997} derived monotonically rising rotation velocity curves for the inner regions of UGC 10205. These data have been homogenized to {\it i} $=$ 75$^{\circ}$. Our stellar data points once corrected by ADC effects are in good agreement with the values obtained by the previous authors up to a galactocentric distance of R $\sim$ 17 arcsec. For R $>$ 17 arcsec, our measurements match the ones by \citet{Vega_1997} along the NW side of the major axis of the galaxy. These results suggest that assuming an inclination of 75$^{\circ}$ for the stellar component in the central part of this galaxy was a reasonable approach. Also, the adoption of a tangential parameterization with a smooth turn-over to recover the shape of the thin disk model in Section~\ref{gas_kinematics_section} is justified by the shape of the previous rotation curves.

Average values of the ADC in local disk galaxies are usually 10\,$\%$ $-$ 20\,$\%$ in late-type systems \citep[see][]{Ciardullo_2004, Merrett_2006, Noordermeer_2008_ADC, Herrmann_2009, Westfall_2011, Martinsson_2013}. Higher values for the ADC are expected for E/S0 and early-type spirals. Here, we are analyzing the central part of a S0/Sa galaxy. We found a mean value of V$_{rot}$(stars)/V$_{rot}$(gas) $=$ 0.75 $\pm$ 0.08. This result is consistent with the one by \citet{Cortese_2014}. These authors explored only the central parts of a sample mainly composed by early-type spirals finding an average ratio of V$_{rot}$(stars)/V$_{rot}$(gas) $=$ 0.75. These results point out the necessity of having high values of the ADC in order to reconcile the discrepancy between the ionized and the stellar velocity rotation curves for these morphological types.

\subsection{Nature of the absorbing gas through ISM \mbox{Na {\small I} D}  measurements} \label{nature_sodium}

Galactic winds are an essential element that determines the evolution of galaxies through cosmic time. In particular, as explained in the introduction, galactic outflows are thought to play a key role in self-regulating star formation in galaxies as well as they constitute necessary mechanisms to explain the chemical evolution and metal enrichment of the circumgalactic medium \citep{Oppenheimer_2006,Oppenheimer_2008}.

At optical wavelengths, the ISM \mbox{Na {\small I} D} absorption doublet is commonly used to trace the properties of neutral atomic gas inflows and outflows in the interstellar medium \citep{Heckman_2000,Schwartz_2004,Veilleux_2005,Rupke_2005,Martin_2005,Martin_2006,Chen_2010,Sarzi_2016,Roberts_Borsani_2019}. Due to the high inclination of this object the \mbox{Na {\small I} D} absorption line profiles are confused with multiple components along the line of sight. Searching for signatures of inflowing/outflowing material in highly inclined systems require having high spatial and spectral-resolution observations to break up the absorption profile into different components. Here, the observations provided by MEGARA show the variety of the profiles and the distinct kinematic components for the \mbox{Na {\small I} D} excess. Some examples of the ISM absorption \mbox{Na {\small I} D} lines that will be explained later (Sections~\ref{inflow} and~\ref{outflow}) are provided in Figure~\ref{nad_profiles}.

For the discussion about inflows/outflows that appear in the following sections, we assume that foreground gas is the dominant source of the absorptions found in the \mbox{Na {\small I} D} profiles. In this scenario, blue-shifted absorption is a signature of gas moving towards the observer (outflow) while red-shifted absorption is indicative of gas moving towards the galaxy (inflow). We do not detect re-emission of absorbed photons from the background gas. For more details about this topic, the reader is referred to the detailed discussion found in \citet{Roberts_Borsani_2019}.

\subsubsection{Infalling neutral gas} \label{inflow}

The analysis of the previous profiles allows us to capture the presence of a red-shifted high-velocity component (see left panel in Figure~\ref{nad_profiles}) that might be associated with material moving towards the galaxy center. In particular, the $\Delta$V distribution of the red-shifted narrow \mbox{Na {\small I} D} continuum absorption features shows a positive mean value of $\Delta$V $=$ (108.35 $\pm$ 18.05) km\,s$^{-1}$ relative to the systemic velocity of the galaxy. This red-shifted component is located all over the MEGARA FoV. Also, the small variation of the $\Delta$V distribution among all the Voronoi cells is an indication that the infalling neutral gas extends beyond the spatial coverage of the FoV. 

The previous value suggests a scenario for the ISM \mbox{Na {\small I} D} absorption lines in which the red-shifted \mbox{Na {\small I} D} interstellar gas is moving inward toward the central regions of UGC 10205. Thus, the absorbing neutral material would be located in a shell-like spatial distribution region moving towards the galaxy as an inflowing gas. 
We consider this structure to be a shell-like type rather than a stream based mainly in the following two arguments. From a morphological point, there is a complete 2D coverage of the FoV. In the case of the shell a 2D coverage is expected while the stream structure would be seen as a strip of stars or gas. From a kinematical point of view, in the current scenario where the bulk velocity of the inflowing gas presents almost a constant value (approximately 108 km\,s$^{-1}$ relative to the systemic velocity of the galaxy) instead of showing a velocity gradient, we consider that the most simple explanation would be to have roughly spherically symmetric gas inflows. The current stage of this galaxy should be highly affected by the previous minor merger suffered in the past as proposed in \citet{Reshetnikov_1999}. The presence of these gas clouds that are inflowing toward the center of the galaxy might be a consequence of such scenario.

\subsubsection{Presence of outflow signatures: mass and outflow rate of the wind} \label{outflow}

\begin{figure}
\centering
\includegraphics[trim={1cm 0cm 1cm 0cm}, width=90mm]{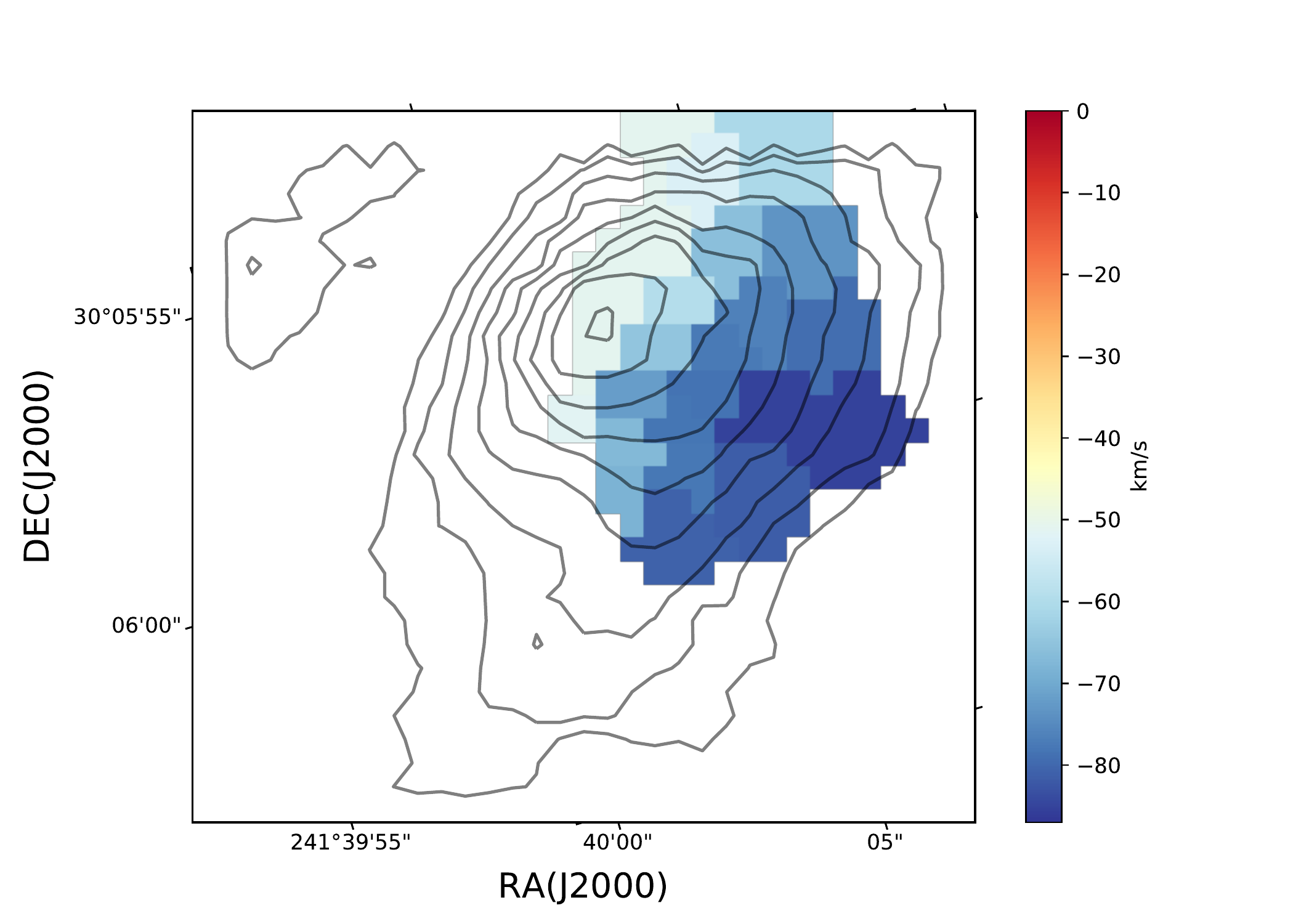}\\ 
\caption{Neutral gas kinematics derived for the blue-shifted absorption component of the \mbox{Na {\small I} D} doublet with respect to the systemic velocity of the galaxy as inferred from the MEGARA data. Contours indicate the distribution of the continuum flux from the LR$-$V datacube.\\}  
\label{outflow_plot}
\end{figure}

Major galaxy mergers and ULIRGS in the local universe exhibit outflows with relatively large velocities and a cone shape \citep{Rupke_2002,Rupke_2005_2,Westmoquette_2012,Soto_2012,Cazzoli_2014,Perna_2019}. Here, we present results from a less active star-forming galaxy. The right panel in Figure~\ref{nad_profiles} shows that clear blue-shifted Na I D absorption signatures are present in this galaxy. The spectra clearly show the presence of double-peaked components in the Na I D absorption line profiles. High spectral resolution is needed to distinct these components. The analysis of the spatially-resolved optical spectroscopy of the blue-shifted NaI D suggests the presence of a wide-angle conical outflow emerging from the central regions of UGC 10205 and perpendicular to the galactic plane (Figure~\ref{outflow_plot}). The outflow shows a maximum projected velocity of up to \mbox{$-$87 km s$^{-1}$}, blue-shifted with respect to the systemic. 

To explore to what extent this neutral gas outflow might remove relatively large amounts of gas from the galaxy, we estimate both the total mass in the wind and the outflow mass rate applying the thin-shell free wind model \citep[equations 13 and 14 in][]{Rupke_2005}. Since we have information on the spatial distribution of the wind thanks to the IFS data, we follow the approach of \citet{Shih_2010,Rupke_2013,Cazzoli_2014,Cazzoli_2016} using a Voronoi-by-Voronoi basis and assuming a series of thin shells to make it a valid approach for this case. Following the previous prescription, the outflow mass rate is proportional to the deprojected velocity, the HI column density and the deprojected solid angle measured in each Voronoi cell. The prescriptions applied to convert Na column densities into HI column densities are largely affected by the assumptions made on the ionization fraction, the depletion onto dust grains and the Na abundance. To avoid the previous uncertainties in the estimation of the column density of HI atoms, we adopt here the same approach as in \citet{Cazzoli_2014,Cazzoli_2016}. These authors make use of both the relation between the HI and the color excess proposed by \citet{Bohlin_1978} \mbox{(N$_{HI}$ $=$ 4.8 $\times$ E$_{(B-V)}$ $\times$ 10$^{21}$ cm$^{-2}$)} and the relation between the color excess and the equivalent width as in \citet{Turatto_2003} \mbox{(E$_{(B-V)}$ $=$ $-$0.04 + 0.51 $\times$ EW$_{NaD}$)}. We calculate a mean value of \mbox{1.65 $\pm$ 0.07 $\times$ 10$^{21}$ cm$^{-2}$} for the HI column density which is in the range of the expected values for low-z galaxies as reported in \citet{Heckman_2015} and \citet{Martin_Fernandez_2016}. The cold gas mass outflow rate is \mbox{0.78 $\pm$ 0.03 M$_{\sun}$ yr$^{-1}$} and the total mass in the wind is equal to \mbox{4.55 $\pm$ 0.06 $\times$ 10$^{7}$ M$_{\sun}$}. These values are consistent with the ones derived using the formalism of the thin-shell model in galaxies with lower star formation rates as in \citet{Roberts_Borsani_2019} where the authors used galaxies with SFR in the range \mbox{1.45 $<$ SFR [M$_{\sun}$ yr$^{-1}$] $<$ 16.98} obtaining outflow mass rates ranging from \mbox{0.15 $<$  $\dot{M}$$_{out}$ [M$_{\sun}$ yr$^{-1}$] $<$ 1.74}.

A significant mass loss has been postulated to regulate the star formation processes within galaxies. The mass loading factor, expressed as the ratio \mbox{$\eta$ $=$ $\dot{M}$$_{out}$ / SFR}, is used to describe the strength of the winds. This factor has important consequences for the predictions in the theoretical models with some findings suggesting that $\eta$ increases with redshift \citep{Barai_2015, Muratov_2015, Sugahara_2017}. To calculate the SFR, we have derived dust-corrected H$\alpha$ growth curves using CALIFA data \citep[following the methodology explained in detail in][]{Catalan_Torrecilla_2017}. In the region covered by the MEGARA FoV and applying the conversion \mbox{SFR(H$\alpha$) = 5.5 $\times$ 10$^{-42}$ L(H$\alpha$$_{corr}$)} \citep{Kennicutt_2009}, we obtained \mbox{SFR(H$\alpha$$_{corr}$) $=$ 0.49 M$_{\sun}$ yr$^{-1}$} and \mbox{$\eta$ $=$ 1.59}. Due to the large amount of dust present in this galaxy, the attenuation correction based on Balmer Decrement measurements might be underestimating the SFR and affecting the real value of $\eta$. To mitigate this problem, we make use of hybrid calibrations combining luminosities measured directly with that of the light emitted by dust after being heated by young massive stars \citep{gordon_2000, inoue_2001, iglesias_paramo_2006, Calzetti_2007, Hao_2011, Kennicutt_Evans_2012, Catalan_Torrecilla_2015}. We adopt the H$\alpha$$_{obs}$, FUV$_{obs}$, 22$\mu$m and TIR luminosities calculated for this object in \citet{Catalan_Torrecilla_2015} as our best estimations. Using the hybrid SFR prescriptions for the different combinations of \mbox{H$\alpha$ + a$\times$IR} and \mbox{FUV + b$\times$IR}, we derived a mean SFR of \mbox{SFR $=$ 1.32 M$_{\sun}$ yr$^{-1}$} and a loading factor of \mbox{$\eta$ $=$ 0.59}. As these SFR measurements are for the whole galaxy while the effect of the outflow has been computed at a restricted distance limited to our current FoV, the estimated value of $\eta$ serves here as a lower limit. Mass loading factors below unity are typically found on galaxies similar to UGC 10205 \citep[see][]{Roberts_Borsani_2019}. Large-scale observations that trace the wind extension outside the disk are needed to explore the effect in the neighboring ISM and circumgalactic medium.

\section{Discussion: Proposed scenario to understand the complex kinematics in UGC 10205} \label{scenario}

UGC 10205 was proposed as a possible candidate for being a polar-ring galaxy \citep{Whitmore_1990,van_Driel_2000}. Further observations made with \mbox{CALIFA} revealed that there is not a clear misalignment between the velocity maps of the gas \citep{Garcia_Lorenzo_2015} and the stellar components \citep{Falcon_Barroso_2017}. As seen in the 2D stellar kinematics maps of Figure~\ref{velocity_map_LRV}, the approaching radial velocities are in the SE side of the galaxy which means the rotation of the disk is counterclockwise.

Dust lanes appeared roughly parallel to the major axis of the galaxy. In particular, the one that is crossing the South region is more clearly visible in front of the bulge (i.\,e., on the near side of the galaxy) while the one in the North seems to be hidden behind the bulge. This location within the galaxy is consistent with the orientation of the reported outflow, which is seen over the near side of the galaxy. As the cold neutral gas traced by \mbox{Na {\small I} D} needs to be observed against the stellar background, the outflow is detected in the approaching cone while the receding part of the cone is not observed. The dust lane that is crossing the galaxy in the South position would prevent us for seeing the outflow component arising in this region which suggest that the dust lane is located on the near side of the galaxy from our position. On the contrary, the dust lane in the North would be located on the far side of the galaxy.

To elucidate the possible mechanism that originated the outflow, it is important to discriminate between different ionization sources in the nuclear region. In \citet{Catalan_Torrecilla_2015} we applied the standard Baldwin, Phillips $\&$ Terlevich (BPT) diagram \citep{Baldwin_1981} with the demarcation lines of \citet{Kauffmann_2003} and \citet{Kewley_2001}. In particular, we estimated the [OIII]/H$\beta$ and the [NII]/H$\alpha$ line flux ratio for the central 3'' spectra using the CALIFA IFS data. The results pointed out that the center of this galaxy could be classified as a mix SF/AGN region, i.\,e., the points lied on the composite area of the diagram. This analysis discards the presence of a strong AGN contribution. This is consistent with the studies by \citet{Sarzi_2016} where the authors examined a sample composed mostly of massive early-type galaxies suggesting that the cold-gas outflows typically occur in objects dominated either by central starbursts or composite AGN/star formation activity. The spatial distribution of the H$\alpha$ emission shows that there is a strong resolved H$\alpha$ contribution in the nuclear region (see the channel maps presented in Figure~\ref{channel_map}). This might favor a scenario where star formation activity could be the potential mechanism triggering the galactic wind. The presence of the outflow is compatible with being created at the very center of the galaxy where most of the H$\alpha$ emission is found. Also, AGN-driven outflows are expected to display higher velocities than the ones observed here \citep{Rupke_2013,Rupke_2017}. Although further investigations need to be done, it seems plausible to discard the AGN-driven as the potential mechanism that generate the outflow in this case.

Finally, the analysis of the ionized and the cold phase of the ISM shows that they have different locations within the galaxy. The H$\alpha$ emission is confined to a disk structure supported by rotation while the \mbox{Na {\small I} D} is associated with (i) a shell-like structure moving infalling towards the galaxy center and (ii) a blue-shifted \mbox{Na {\small I} D} absorption in the near side of the galaxy that reflects the presence of a galactic outflow along the NE semi-minor axis. The ionized gas in the central region shares the kinematic pattern of the thin-disk model proposed to explain the motion of the gas. Thus, the analysis of the ionized H$\alpha$ emission line shows the presence of three different dynamical components, with one of them associated with a highly-inclined ({\it i} $=$ 75$^{\circ}$) rotating disk. Also, a second dynamical component associated with the position of the dust lanes is present. The gas is rotating in this dusty structure surrounding the central regions with a higher inclination, {\it i} $=$ 83$^{\circ}$.

\section{Summary and Conclusions} \label{final_conclusions}

In this paper, we present high resolution optical IFS observations of the nearby galaxy UGC 10205, made with MEGARA on the GTC 10.4m telescope at La Palma Observatory. The combination of the LR$-$V and HR$-$R set-ups provides a wealth of information on the state of the multi-phase interstellar medium and the stellar kinematics in UGC 10205 and allows us to study the kinematics of the cool and the ionized gas in detail. MEGARA IFS observations have shown an intriguing and complex structure in the central regions of UGC 10205. The main results obtained in this study can be summarized as the following:

\begin{itemize}

\item We present fundamental parameters of the galaxy kinematics by means of creating 2D stellar kinematics maps for the first two order moments of the line of sight velocity distribution (V and $\sigma$).

\item In an attempt to reconciliate the ionized and the stellar velocity rotation curves, ADC corrections are applied. Having high quality IFS data is fundamental to properly estimate the ADC in each Voronoi cell and to recover the shape of the SVE. A mean value of \mbox{V$_{rot}$(stars)/V$_{rot}$(gas) $=$ 0.75 $\pm$ 0.08} is found. Applying this correction, there is a significant agreement between the derived stellar velocity rotation curve and the gaseous ionized H$\alpha$ one previously derived in the literature by both \citet{Vega_1997} using long-slit spectroscopy data and \citet{Garcia_Lorenzo_2015} with the CALIFA data. These results suggest that the innermost regions of early-type spirals needed large values of the asymmetric drift correction as proposed by \citet{Cortese_2014}.

\item We also explore the presence of interstellar \mbox{Na I $\lambda$$\lambda$ 5890,5896} doublet absorption that is commonly used to trace the properties of neutral gas inflows/outflows. The complexity of the neutral gas kinematics is revealed by the presence of multiple narrow components. An overall shift of \mbox{$\Delta$V $=$ (114 $\pm$ 5) km\,s$^{-1}$} for the foreground red-shifted component of the neutral gas is detected through the \mbox{Na {\small I} D} absorption. This suggests the presence of mild inflows towards the inner regions. The inflowing gas detection might be triggered by a past minor merger suffered by the galaxy \citep{Reshetnikov_1999}. 

\item The characterization of outflows is essential to place constraints on the models of galaxy formation and evolution. The majority of works in the literature has focused on understanding the complexity of outflows in extreme objects such as ULIRGS. Here, we point out the necessity to extend these studies including less active galaxies. To elucidate the role that this outflow might play in removing gas from UGC 10205, we calculate physical properties of the cold gas entrained in the wind applying the formalism of the thin-shell model. We obtained a total mass in the wind \mbox{M$_{out}$ $=$ 4.55 $\pm$ 0.06 $\times$ 10$^{7}$ M$_{\sun}$} and a cold gas mass outflow rate \mbox{$\dot{M}$$_{out}$ $=$ 0.78 $\pm$ 0.03 M$_{\sun}$ yr$^{-1}$}. The calculation of the mass loading factor yield values of \mbox{0.59 $<$ $\eta$ $<$ 1.59}, although more accuracy SFR measurements are needed to constraint this value as well as large-scale observations to explore the effect of the wind in the neighboring ISM and circumgalactic medium.

\item Finally, assuming that the gas describes circular orbits in the galaxy's plane, we model the H$\alpha$ gas distribution in the inner regions using a thin-disk model with \mbox{({\it i} $=$ 75$^{\circ}$)}. The primary goal of our modeling is to characterize the kinematics and the ionized gas geometry. Despite the complexity of the H$\alpha$ emission line profile where up to three kinematically distinct gaseous components appeared, our simple thin-disk model still can reproduce most of the kinematical features seen in the central regions of UGC 10205. Also, a distinct kinematic component of the gas with a higher inclination \mbox{({\it i} $=$ 83$^{\circ}$)} pattern is detected in a position coincident with the dust lane structures. 

\end{itemize}

Finally, we emphasize the potential of MEGARA to make meaningful comparisons between the neutral and the ionized gas phase components in the central regions of nearby galaxies. 

UGC 10205 is a remarkable example of an object that has revealed an active current phase, reinforcing the importance of spatially-resolved high-resolution analyses in less active galaxies. A systematic exploration of galactic winds in normal star-forming objects and in its multiple phases is required to provide a complete picture on the role that these phenomena might play to shape galaxies and its influence on the surrounding environments.

\acknowledgements

Authors acknowledge financial support from the Spanish Ministry of Economy and Competitiveness (MINECO) under grant number AYA2016-75808-R which is partly funded by the European Regional Development Fund (ERDF), AYA2017-90589-REDT, RTI2018-096188-B-I00, S2018/NMT-4291 and ESP2017-83197P projects. C.C.-T. gratefully acknowledges the support of the European Youth Employment Initiative (YEI) by means of the Postdoctoral Fellowship Program. GB acknowledges financial support from DGAPA- UNAM through PAPIIT project IG100319.

\bibliographystyle{apj} 
\bibliography{ugc10205_bibliography} 

\twocolumngrid

\end{document}